\documentclass[twocolumn]{aastex63}

\graphicspath{{./}{figures/}} %%to tell LateX to also look for figures in the figures folder

\newcommand\sersic{S\'ersic}

\newcommand\gamornet{G\textsc{a}M\textsc{or}N\textsc{et}}

\shorttitle{\gamornet{} employed to study galaxy morphology and quenching}
\shortauthors{Ghosh et al.}

\usepackage[normalem]{ulem}
\usepackage{enumitem}
\usepackage{bm}
\usepackage{amsmath}
\usepackage{gensymb}
\usepackage{subfigure}
\usepackage{multirow}
\usepackage{array}
\usepackage{hhline}
\usepackage{hyperref}
\PassOptionsToPackage{hyphens}{url}\usepackage{hyperref}

\begin{document}

\title{Galaxy Morphology Network: A Convolutional Neural Network Used to Study Morphology and Quenching in $\sim 100,000$ SDSS and $\sim 20,000$ CANDELS Galaxies}

\author[0000-0002-2525-9647]{Aritra Ghosh}
\affil{Yale Center for Astronomy and Astrophysics, and Department of Astronomy, \\ Yale University, New Haven, CT, USA}
\email{aritra.ghosh@yale.edu}

\author[0000-0002-0745-9792]{C. Megan Urry}
\affil{Yale Center for Astronomy and Astrophysics, and Department of Physics, \\Yale University, New Haven, CT, USA}

\author[0000-0002-9490-6326]{Zhengdong Wang}
\affiliation{Department of Computer Science, Yale University, New Haven, CT, USA}

\author{Kevin Schawinski}
\affiliation{Modulos AG, Technoparkstr. 1, CH-8005, Zurich, Switzerland}

\author{Dennis Turp}
\affiliation{Modulos AG, Technoparkstr. 1, CH-8005, Zurich, Switzerland}

\author[0000-0003-2284-8603]{Meredith C. Powell}
\affil{Yale Center for Astronomy and Astrophysics, and Department of Physics, \\Yale University, New Haven, CT, USA}
\affil{Kavli Institute for Particle Astrophysics and Cosmology, Stanford University, Stanford, CA, USA}

\begin{abstract}
We examine morphology-separated color-mass diagrams to study the quenching of star formation in $\sim 100,000$ ($z\sim0$) Sloan Digital Sky Survey (SDSS) and $\sim 20,000$ ($z\sim1$) Cosmic Assembly Near-Infrared Deep Extragalactic Legacy Survey (CANDELS) galaxies. To classify galaxies morphologically, we developed Galaxy Morphology Network (\gamornet{}), a convolutional neural network that classifies galaxies according to their bulge-to-total light ratio. \gamornet{} does not need a large training set of real data and can be applied to data sets with a range of signal-to-noise ratios and spatial resolutions. \gamornet{}'s source code as well as the trained models are made public as part of this work (\href{http://www.astro.yale.edu/aghosh/gamornet.html}{Link\,1}$\vert$\href{http://gamornet.ghosharitra.com}{Link\,2}). We first trained \gamornet{} on simulations of galaxies with a bulge and a disk component and then transfer learned using $\sim25\%$ of each data set to achieve misclassification rates of $\lesssim5\%$. The misclassified sample of galaxies is dominated by small galaxies with low signal-to-noise ratios. Using the \gamornet{} classifications, we find that bulge- and disk-dominated galaxies have distinct color-mass diagrams, in agreement with previous studies. For both SDSS and CANDELS galaxies, disk-dominated galaxies peak in the blue cloud, across a broad range of masses, consistent with the slow exhaustion of star-forming gas with no rapid quenching. A small population of red disks is found at high mass ($\sim14$\% of disks at $z\sim0$ and 2\% of disks at $z \sim 1$). In contrast, bulge-dominated galaxies are mostly red, with much smaller numbers down toward the blue cloud, suggesting rapid quenching and fast evolution across the green valley. This inferred difference in quenching mechanism is in agreement with previous studies that used other morphology classification techniques on much smaller samples at $z\sim0$ and $z\sim1$.
\end{abstract}

\keywords{Galaxies (573), Galaxy classification systems (582), Galaxy evolution (594), Galaxy quenching (2040), Astronomy data analysis (1858), Neural networks (1933), Convolutional neural networks (1938)}

\section{Introduction} \label{sec:intro}

We know from large-scale surveys that both local and high-redshift galaxies show a bimodal distribution in the galaxy color-mass space\,\citep{strateva_01,baldry_04,baldry_06,brammer_09} with a ``blue cloud,'' a ``red sequence'' and a ``green valley.'' Galaxy color-mass diagrams are useful for studying galactic evolution, as the stellar mass of a galaxy indicates its growth over time, and the color tracks its rate of star formation. The standard interpretation of the bimodal color-mass distribution is that, because there are few galaxies in the green valley, star formation in blue cloud galaxies must be quenched rapidly, perhaps aided by emission from an active galactic nucleus (AGN; \citealp{bell_04,faber_07}). Direct evidence of this AGN feedback remains murky, however \citep{harrison_17}.

Galaxy morphology adds a third interesting dimension to the color-mass space. Because elliptical galaxies typically form in major mergers, and galactic disks usually do not survive them, morphology can be used as a tracer of the recent merger history of a galaxy. The observed bimodality in the color-mass diagram (as well as interpretations therefrom) comes from superposing distinct populations with different morphological types, as first shown by \citet{schawinski_14_green_herring}, who used Galaxy Zoo morphological classifications to study local ($z\sim0$) galaxies. They suggested that there are two separate evolutionary tracks for galaxies: (1) major mergers forming ellipticals from disk-dominated galaxies, accompanied by AGN triggering and rapid quenching of star formation, and (2) slow, secular growth of disk-dominated galaxies, until they reach a critical halo mass, after which the remaining cold gas is slowly consumed and the stellar population gradually reddens. At $z\sim0$, the latter population is an order of magnitude larger than the merger-created ellipticals.

\begin{figure*}[htbp]
	\begin{center}
	\begin{tabular}{cccc}
	\hline 
	\hline
	Disk-Dominated & Bulge-Dominated & Indeterminate & \\[0.15cm]
	\hline \\
    \includegraphics[width=3.25cm,height=3.25cm]{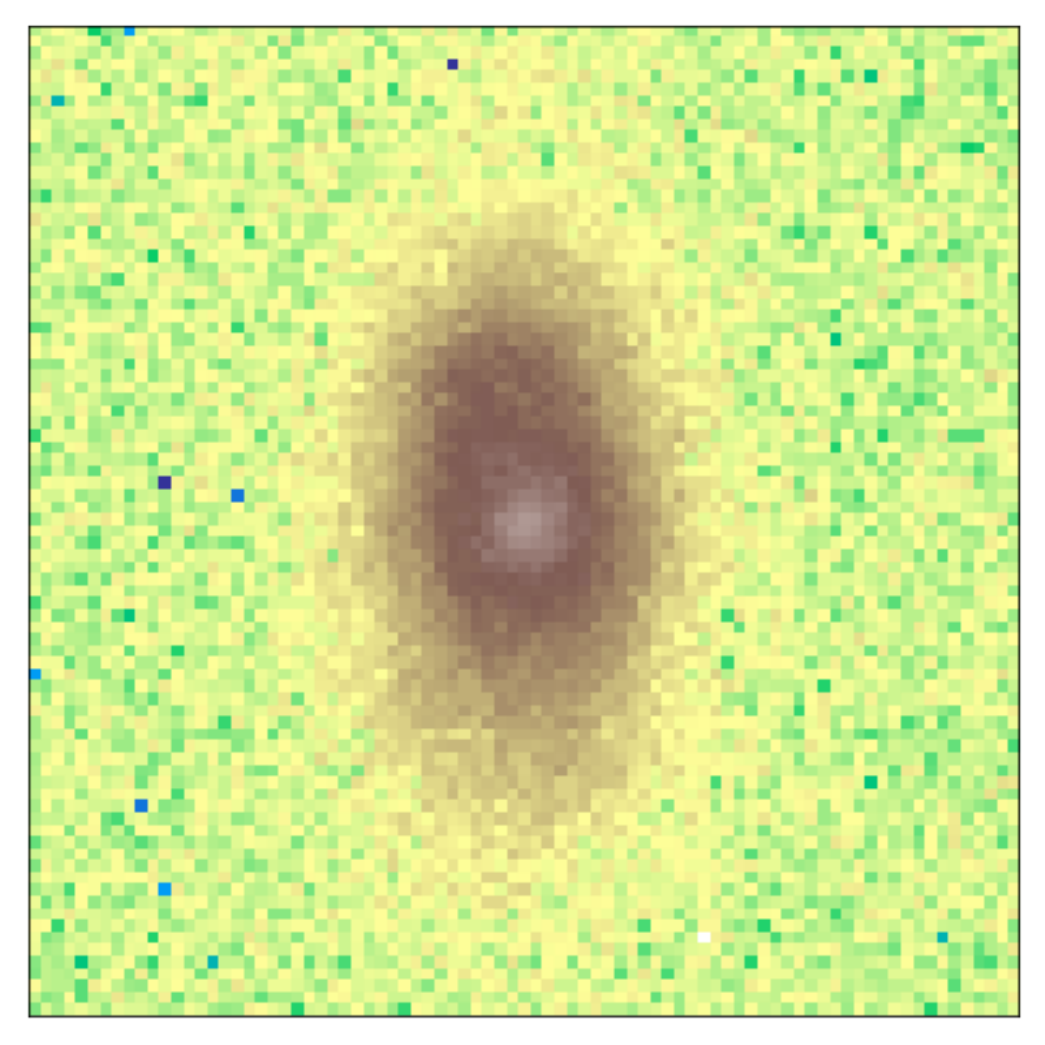} & \includegraphics[width=3.25cm,height=3.25cm]{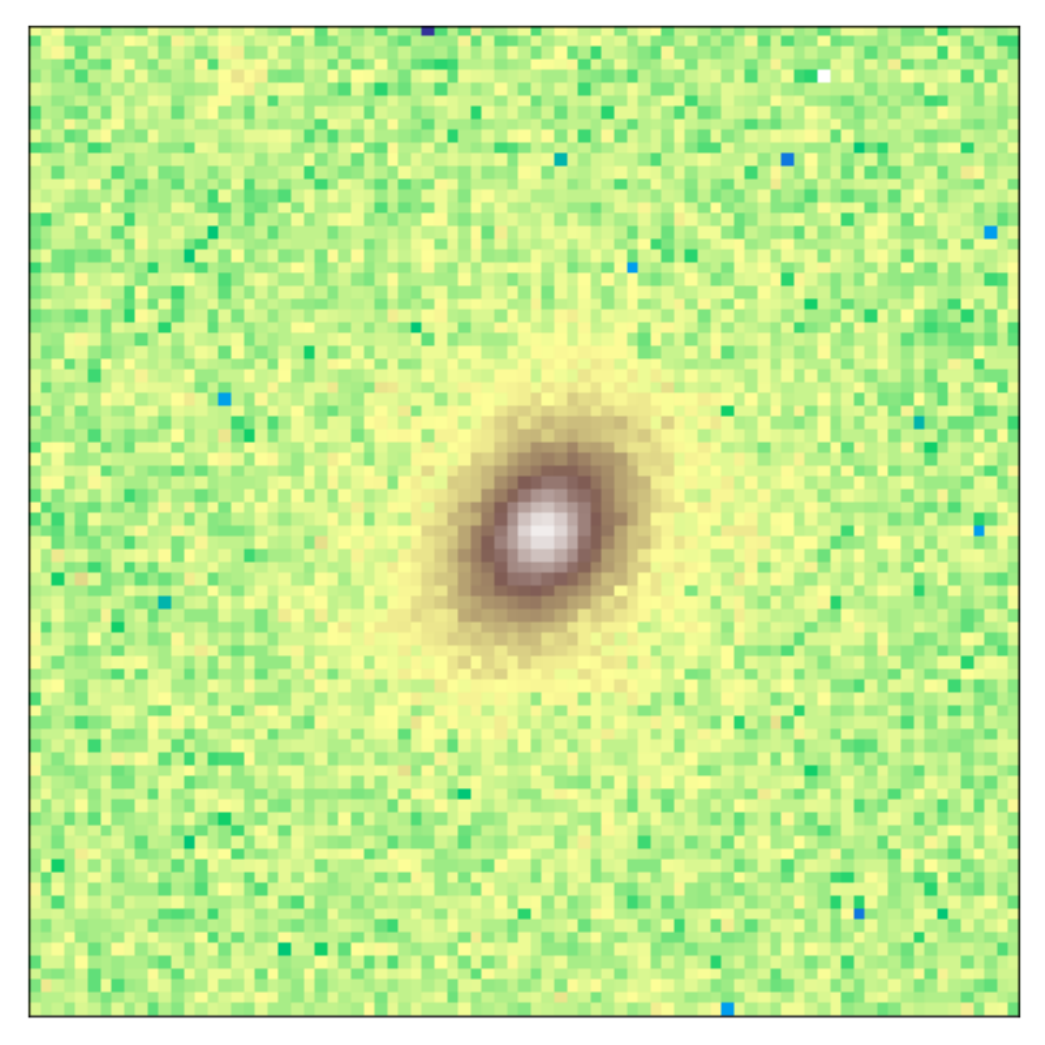} & \includegraphics[width=3.25cm,height=3.25cm]{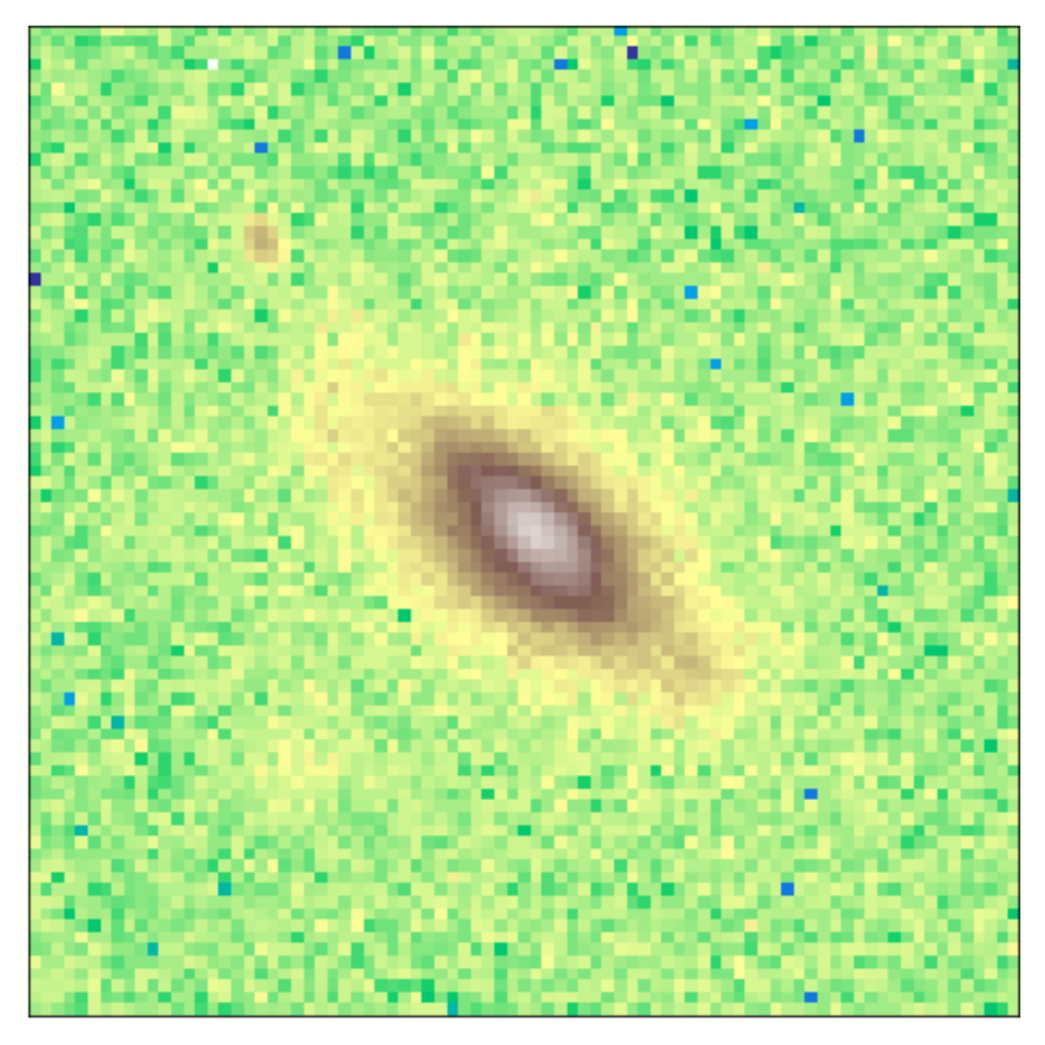} &
     \multirow{2}{*}[2.5cm]{\includegraphics[height=6cm]{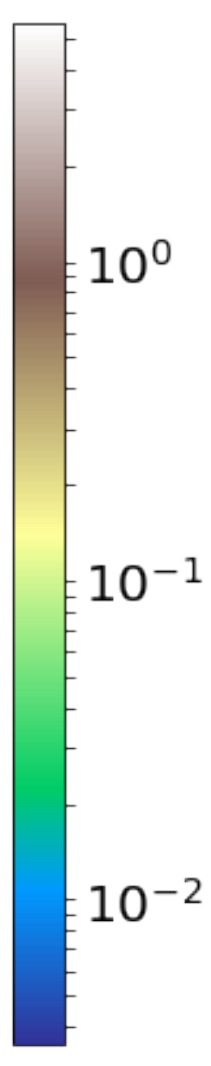}}
    \\[0.3cm]
     \includegraphics[width=3.25cm,height=3.25cm]{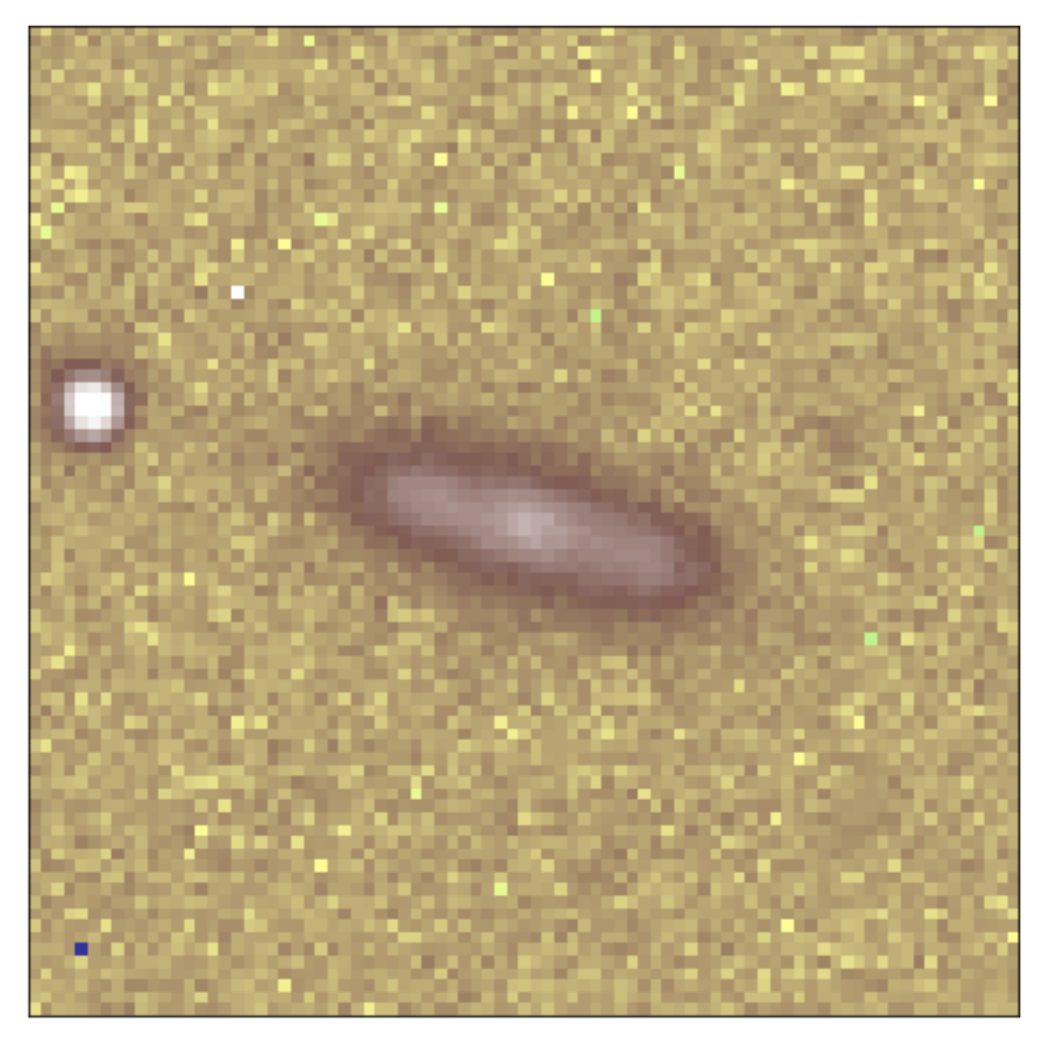} & \includegraphics[width=3.25cm,height=3.25cm]{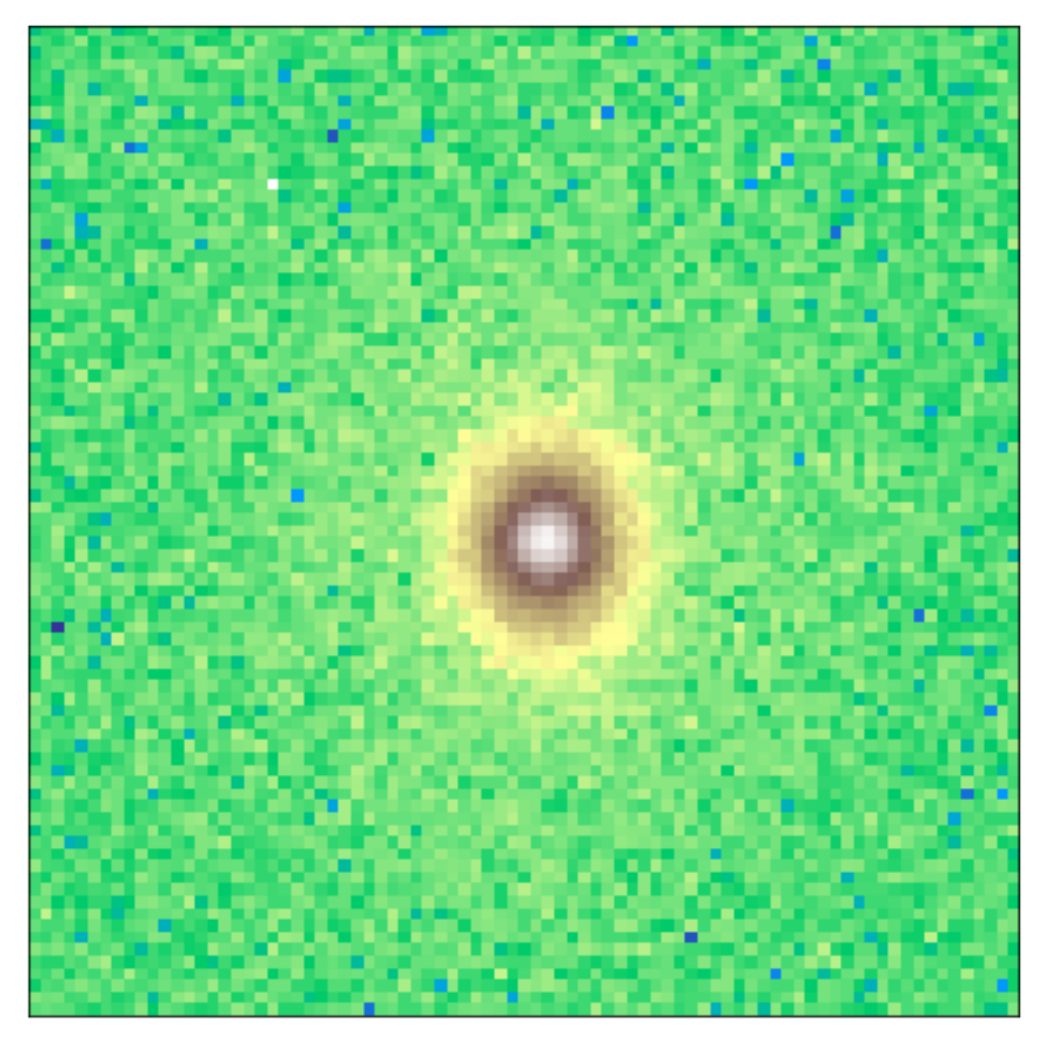} & \includegraphics[width=3.25cm,height=3.25cm]{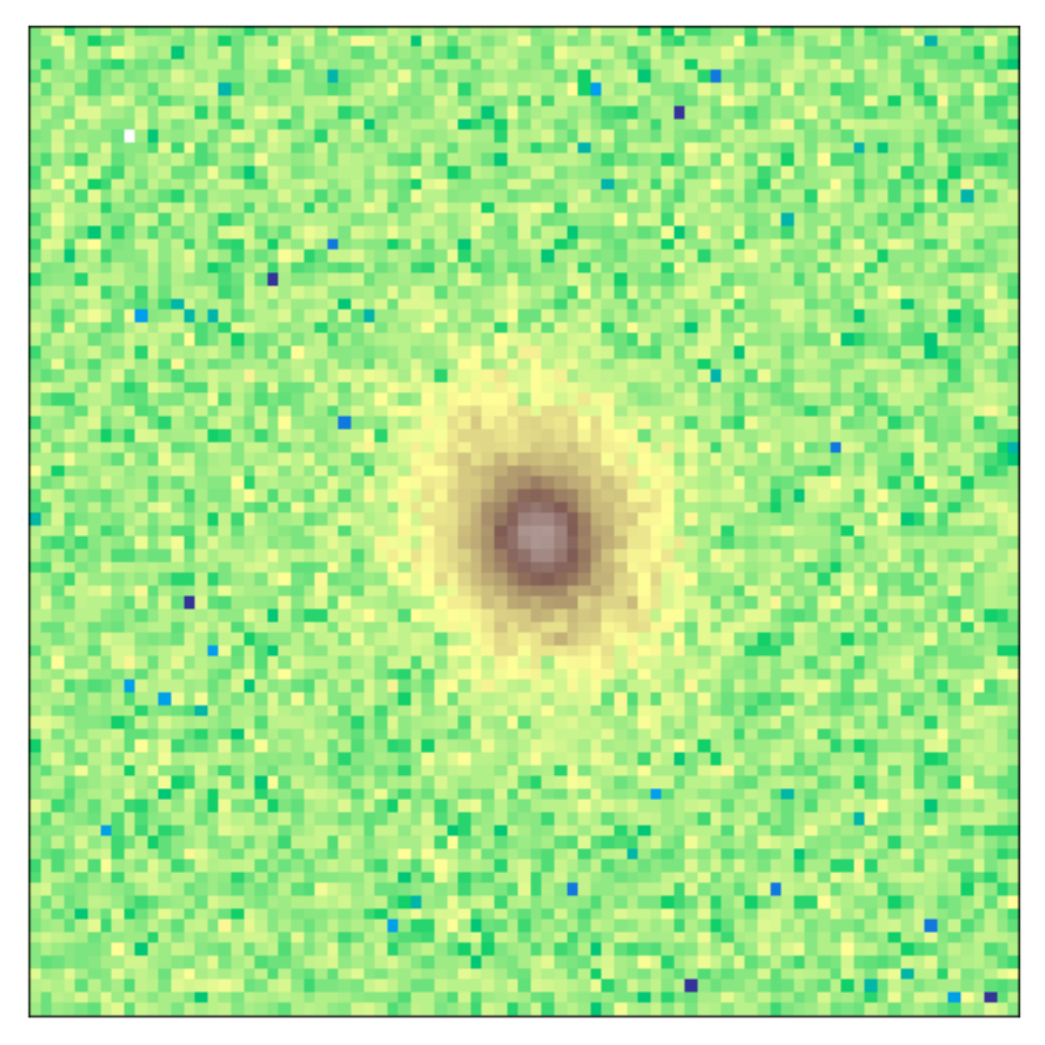}
     &
     \\[0.3cm]
     \hline
     \\
      \includegraphics[width=3.25cm,height=3.25cm]{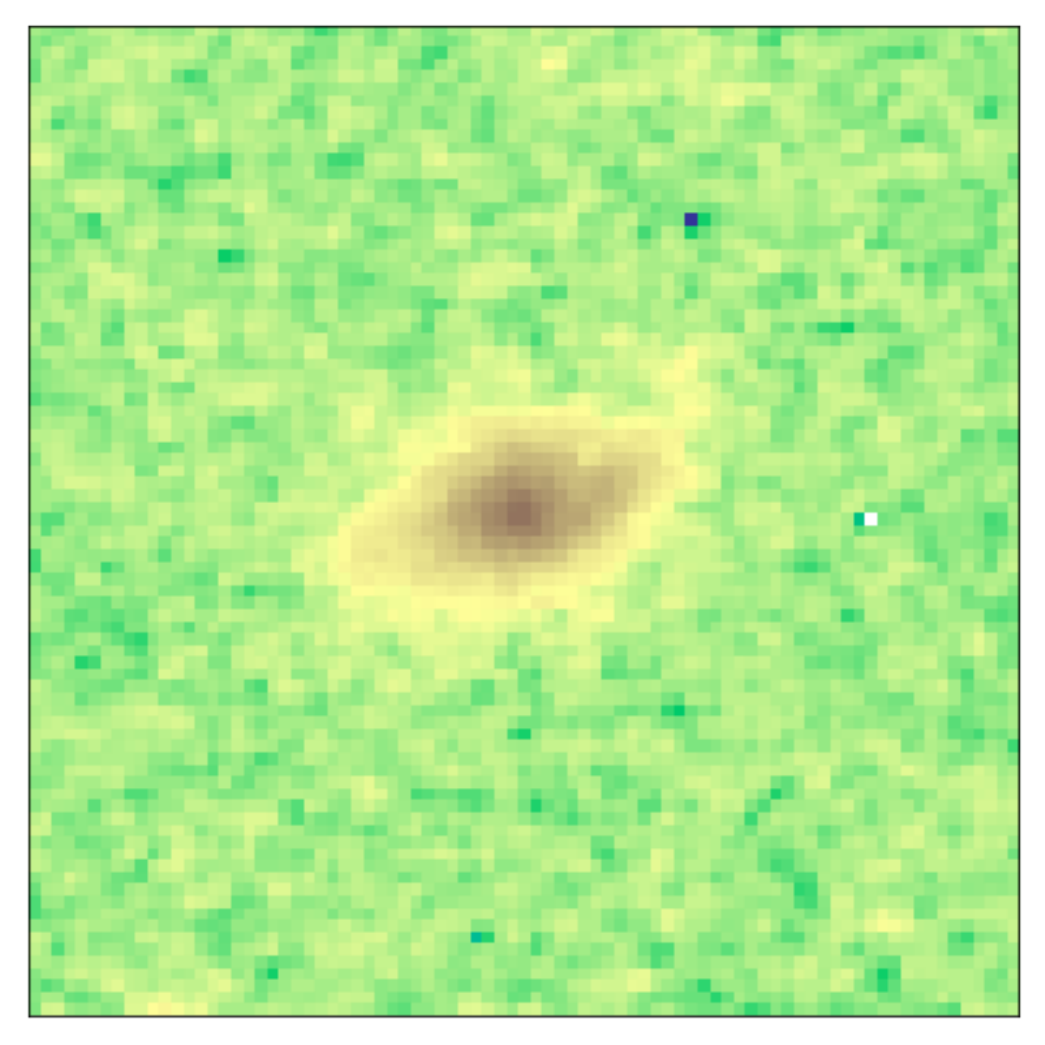} & \includegraphics[width=3.25cm,height=3.25cm]{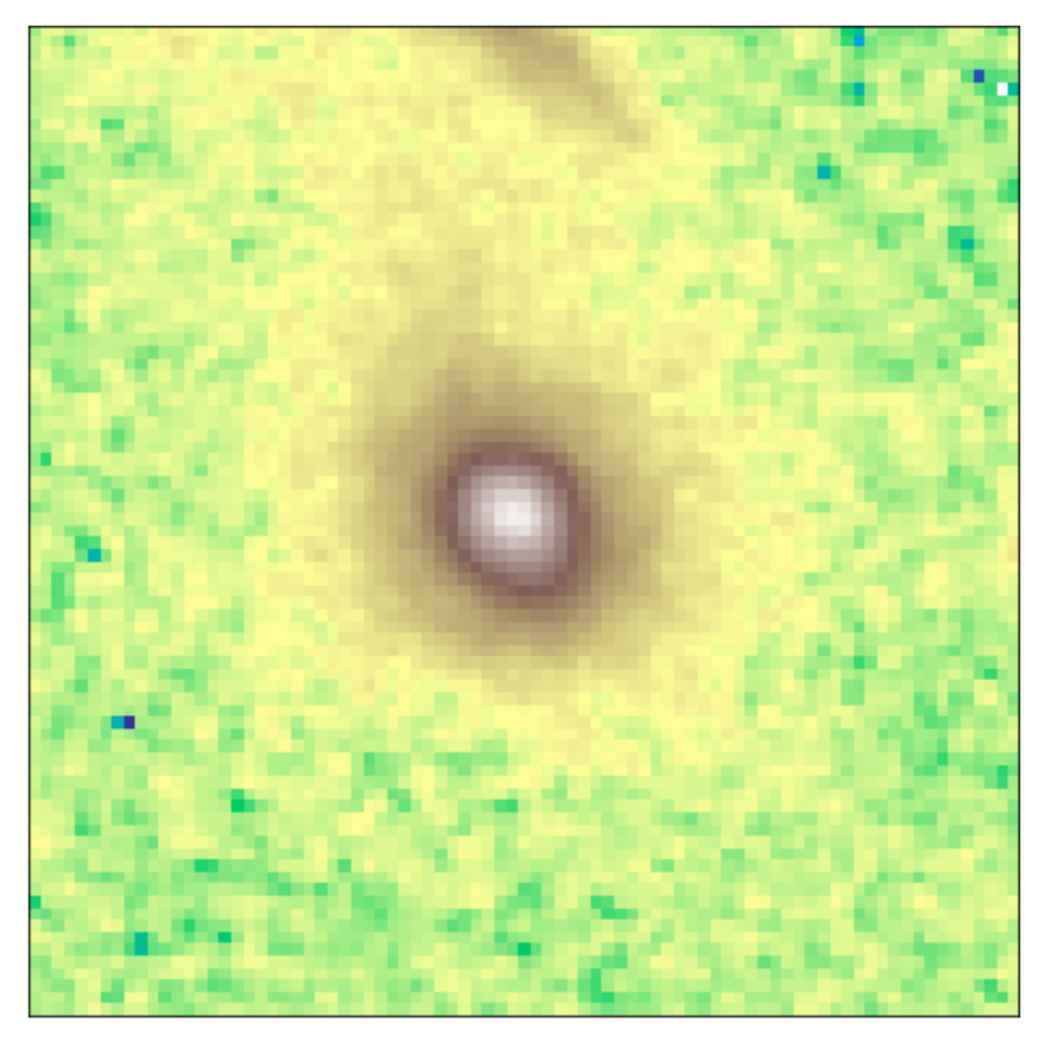} & \includegraphics[width=3.25cm,height=3.25cm]{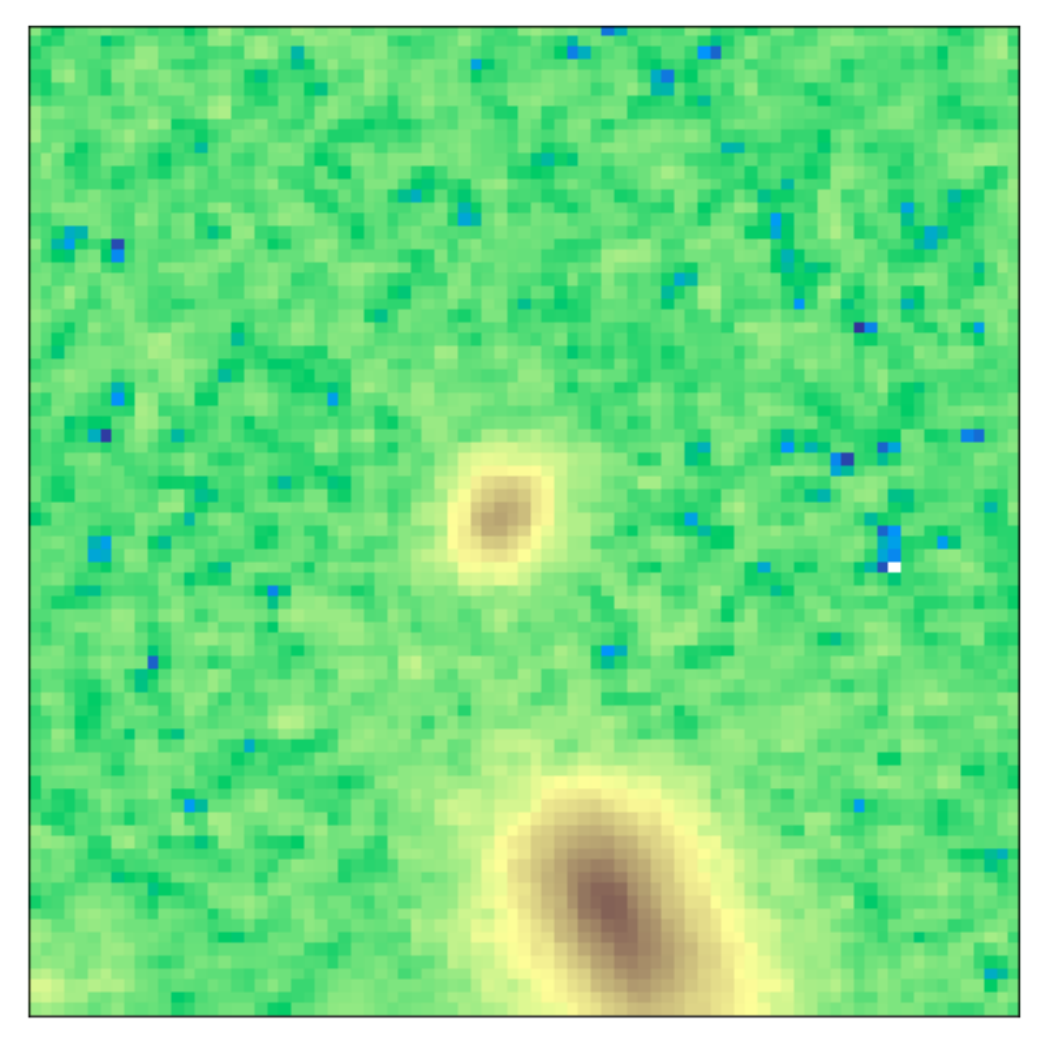} &
     \multirow{2}{*}[2.5cm]{\includegraphics[height=6cm]{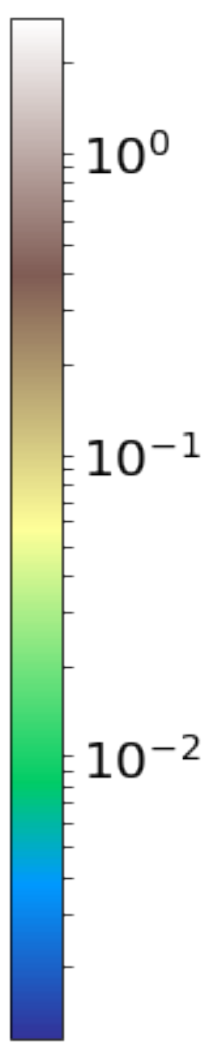}}
     \\[0.3cm]
     \includegraphics[width=3.25cm,height=3.25cm]{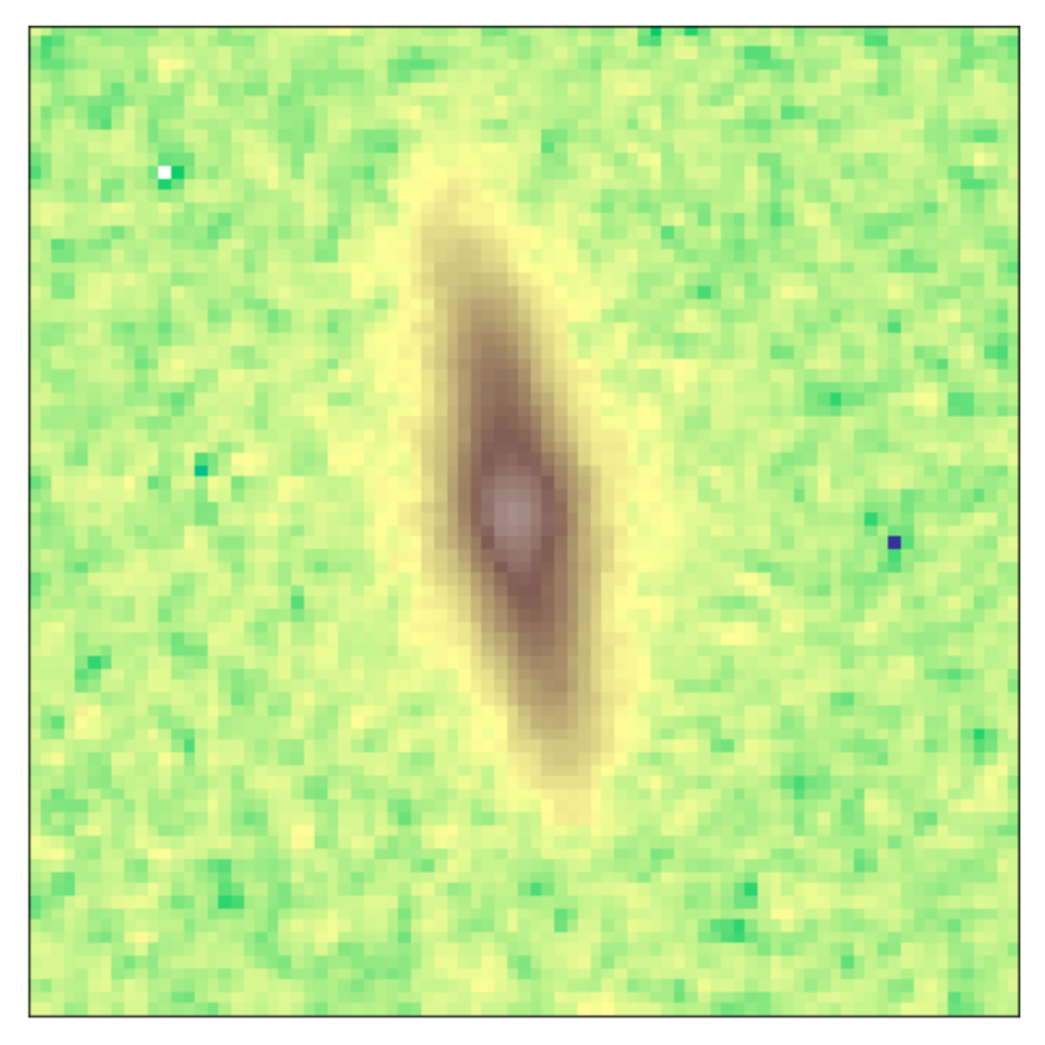} & \includegraphics[width=3.25cm,height=3.25cm]{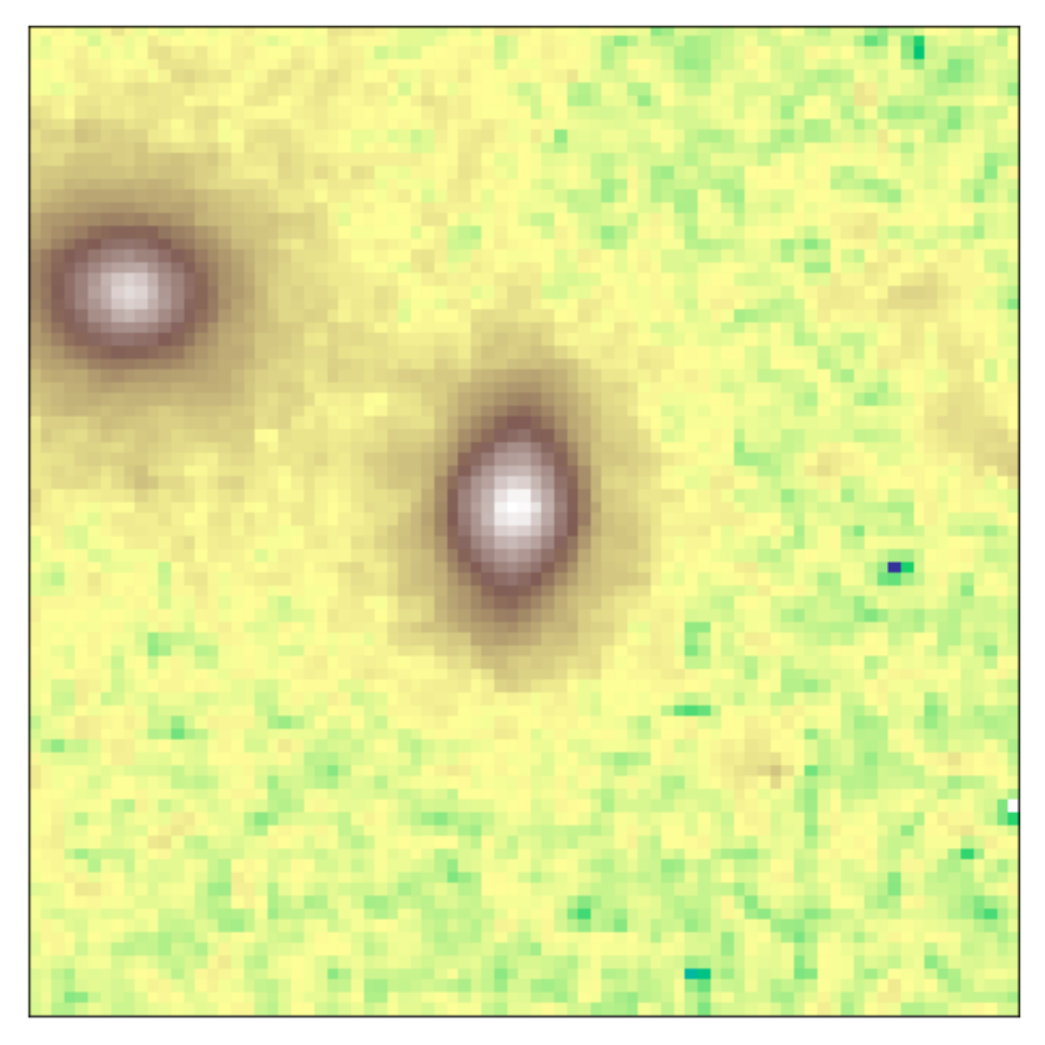} & \includegraphics[width=3.25cm,height=3.25cm]{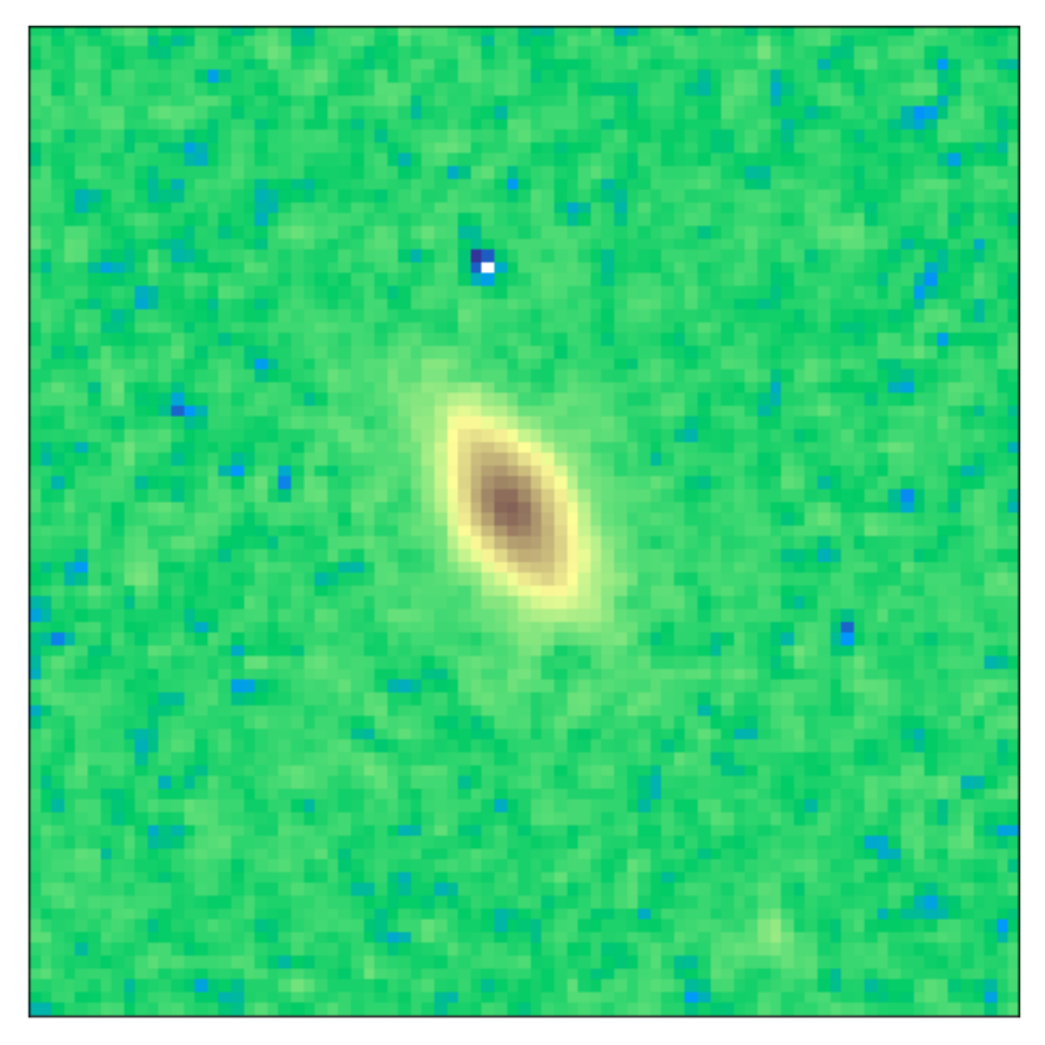} 
     & 
     \\[0.3cm]
     \hline
    \end{tabular}
  \end{center}
  \caption{The above figure contains randomly chosen galaxies from both our data sets classified by \gamornet{} as being disk-dominated\,(left column panels), bulge-dominated\,(middle column panels) or indeterminate\,(right column panels). Refer to \S~\ref{sec:morph_results} for the definitions of these categories. The top two rows show SDSS cutouts, which are 33.07" $\times$ 33.07" (83 pixels $\times$ 83 pixels) and the bottom two rows show CANDELS cutouts, which are 4.98" $\times 4.98"$ (83 pixels $\times$ 83 pixels). During training, \gamornet{} focuses on galaxies located at the center of the image and, thus, can process cutouts with other objects in the frame besides the central galaxy, as is evident from the images above.}
  \label{fig:eg_images}
\end{figure*}

Still, most star formation and the most pronounced galaxy evolution happen not locally but at $z\sim1$ and above. Thus, it is important to investigate the galaxy color-mass diagram at $z\gtrsim1$. \citet{powell_17} studied galaxies from The Great Observatories Origins Deep Survey\,(GOODS)-N and GOODS-S at $z\sim1$ and found that disks and spheroids have distinct color-mass distributions in rough agreement with the results at $z\sim0$. From the distribution of X-ray-selected AGN hosts in this sample, they concluded that AGN feedback may quench star formation in galaxies that undergo major mergers, but these are still less than half the galaxy population. However, this study was done with a sample of only $2651$ disks and $126$ spheroids. Much larger studies, across a broader redshift range, will better illuminate the effect of mergers and AGN on galaxy evolution. 

The two traditional ways of obtaining morphological classifications --- visual classification and fitting light profiles --- are not easily scalable to the large data volumes expected from The Large Synoptic Survey Telescope\,(LSST), the Wide Field Infrared Survey Telescope\,(WFIRST) and Euclid. The most popular galaxy light profile fitting program, GALFIT\,\citep{galfit}, and automated versions of it like GALAPAGOS\,\citep{galapagos}, suffer from the fact that the quality of the fit depends heavily on the input parameters, and when dealing with hundreds of thousands of galaxies, such hand-refinement of input parameters is an impossible task. There have been attempts to employ visual classifications on large galaxy samples via citizen science projects like Galaxy Zoo\,\citep{gzoo_p1,gzoo_p2}, but even these will fail to keep up with the coming data volume. Moreover, reliable visual classifications require a decent signal-to-noise ratio (S/N), take time to set up and execute, and require an extremely careful de-biasing of the vote shares obtained \citep{gzoo_p1,gzoo_candels}. 

For these reasons, using machine learning to classify galaxy morphology is particularly attractive. Data available from the Sloan Digital Sky Survey\,(SDSS) inspired early attempts at using machine learning to classify galaxies morphologically on a large scale\,\citep[e.g.,][]{Ball2004GalaxyNetworks,Kelly2004MorphologicalSurvey,banerji_10}. These methods required the user to select proxies for morphology (such as color, concentration index, and spectral features) as inputs to the models. However, as the proxies could have an unknown and biased relation with galaxy morphology, these early networks were not ideal substitutes for the traditional classification methods. 

In the last few years, convolutional neural networks\,(CNNs) have revolutionized the field of image processing\,\citep{dl_1,dl_2}. They are ideal for galaxy morphology classification as they do not require selection of morphological proxies by hand and the network itself decides on which features of the image best discriminate among the different classes. The first serious attempt at using a CNN to classify galaxies morphologically came out of the ``Galaxy Challenge" organized by Galaxy Zoo, where teams competed to reproduce the vote shares of each question in Galaxy Zoo 2 using a CNN (the top entry was by \citealp{dieleman_15}). This was followed by the work of \citet{company_15}, who used a CNN to reproduce visual classifications for galaxies in the Cosmic Assembly Near-Infrared Deep Extragalactic Legacy Survey\,(CANDELS). \citet{tuccillo_18} used domain adaptation combined with one-component Se\'rsic simulations to reproduce morphological classifications for $\sim 5000$ CANDELS galaxies. There have also been attempts at using CNNs for measuring photometric redshifts from galaxy images\,\citep{ml_pz}, doing star/galaxy separation\,\citep{ml_sz}, detecting bars in galaxies\,\citep{ml_bars} and detecting mergers\,\citep{ml_mergers}. 

Most of the previous work involving the use of CNNs to study galaxy morphology has depended on the availability of a large training set of galaxies with known properties. However, if CNNs are to truly replace traditional methods for morphology classification, then there needs to be a single prescription/network that works across multiple data sets and does not require an already classified large training set. 

In this paper, we introduce Galaxy Morphology Network (\gamornet{}), a CNN that can classify galaxies according to their bulge-to-total ratio ($L_B/L_T$) for very different data sets without the need for a large,  pre-classified training set of real galaxies. We first trained our network on simulated galaxies with both bulge and disk components and then transfer learned on a small part of our real sample to produce bulge/disk classifications for $\sim80,000$ ($z \sim 0$) SDSS \textit{g}-band galaxies and $\sim20,000$ CANDELS ($z \sim 1$) \textit{H}-band galaxies. A collection of 12 randomly chosen galaxy image cutouts from both data sets with their \gamornet{} classifications is shown in Figure~\ref{fig:eg_images}. Using the morphology classifications, we then examine the color-mass diagrams of the two samples, separated by morphology, in order to study the quenching of star formation at $z\sim0$ and $1$. 

We describe the details of the SDSS and CANDELS data that we use in \S~\ref{sec:data}. In \S~\ref{sec:methods}, we describe our simulations, the CNNs we use, and our transfer learning algorithm. In \S~\ref{sec:results}, we present the results of the morphology classification, including the color-mass diagrams, and in \S~\ref{sec:disc}, we summarize our results and discuss future applications of \gamornet{}.

We make all of the source code used in this work public along with the trained CNN models. We also release the \gamornet{} morphological predictions for all of the SDSS and CANDELS galaxies in our data sets. All of the code is being made available under a GNU General Public License v3.0 and more details of the public data release are summarized in Appendix\,\ref{sec:ap:public_data_release}. 

\section{Data Sets Used} \label{sec:data}

One of the primary aims of this paper is to demonstrate how \gamornet{} can be used to identify bulge- and disk-dominated galaxies in different data sets without requiring extensive training on real data. Here, we work with two data sets: the SDSS \citep{sdss_1_2_des}, for nearby galaxies ($z\sim0$), and CANDELS \citep{candels_1,candels_2}, for galaxies at $z\sim1$. Together, these data allow us to probe galaxy evolution at different epochs of star formation and black hole growth.

We first created galaxy samples with which we train and test \gamornet{}. Specifically, we identified galaxies in each survey for which bulge/disk decomposition had already been done or which had already been morphologically classified in some other way. 

For the SDSS sample, we used $112,547$ galaxies in the redshift range $ 0.02 \leq z \leq 0.07$ that were imaged in the $g$ band and had bulge fractions determined by \citet{simard_11}, who fitted double \sersic{} profiles with fixed indices $n = 4$ (pure bulge) and $n=1$ (pure disk). For each galaxy, we prepared square cutouts of 167 pixels on a side, centered on the galaxy, with a resolution of $0.396''$ per pixel. We used 30,000 of these for the process of transfer learning, described in \S\,\ref{sec:tf_intro} and the remaining 82,547 galaxies to test the performance of the network. In order to calculate the \textit{u-r} color for each galaxy, we used extinction-corrected model SDSS magnitudes from the NYU-VAGC\,\citep{nyu_vagc_des} and adopted K corrections to $z=0.0$. We obtained aperture and extinction-corrected specific star formation rates (sSFR) and stellar masses from the MPA-JHU DR7 catalog\,\citep{mpa_jhu_2,mpa_jhu_1}, which are calculated using SDSS spectra and broadband photometry. 

For CANDELS reference data, we used \sersic{} indices from \citet{vdw_12}, who fitted the galaxy surface brightness profiles using GALFIT \citep{galfit} with a single (free) \sersic{} component. From this catalog, we selected galaxies with redshifts $0.7 \leq z \leq 1.3$ and ``good" fits (defined by \citealp{vdw_12} as matching the galaxy total magnitude, and having fits that converged, with parameters within an acceptable range). The ensuing sample of 28,946 $z\sim1$ galaxies from the five CANDELS fields includes 6276 from the Great Observatories Origins Deep Survey--North (GOODS-N), 3942 from the Great Observatories Origins Deep Survey--South (GOODS-S), 7425 from the Cosmic Evolution Survey (COSMOS), 4911 from the Ultra Deep Survey (UDS) and 6392 from the All Wavelength Extended Growth Strip International Survey (AEGIS). We downloaded WFC3/IR F160W(H) mosaics from the CANDELS website\footnote{\href{http://arcoiris.ucolick.org/candels/data\textunderscore access/Latest\textunderscore Release.html}{http://arcoiris.ucolick.org/candels/data\textunderscore access/Latest\textunderscore Release.html}}, then for each galaxy, we made square cutouts of $83$\,pixels$\times83$\,pixels with a resolution of $0.06''$ per pixel. We used 7200 galaxy images for transfer learning and the remaining 21,746 for testing the performance of \gamornet{}. We took the rest-frame \textit{U-R} color, stellar mass, and sSFR of each galaxy from the 3D-HST catalog \citep{3dhst}; the stellar masses are based on spectral energy distribution\,(SED) fits to stellar population models with the FAST code \citep{fast} as described in \citet{skelton_14}. The star formation rates used are from \cite{whitaker_14} and assume that UV light from massive stars is re-radiated in the far-infrared.

\begin{deluxetable*}{cccccc}[htbp]
%\tablenum{2}
\tablecaption{Parameter Ranges for Simulated Galaxies  \label{tab:sim_para}}
\tablecolumns{6}
\tablehead{
\colhead{Component Name} & \colhead{\sersic{} Index} & \colhead{Half-Light Radius} & \colhead{Magnitude} & \colhead{Axis Ratio} & \colhead{Position Angle} \\ 
\colhead{} & \colhead{} & \colhead{(Pixels)} & \colhead{(AB)} & \colhead{} & \colhead{(degrees)}
}
\startdata
    \hline
    \hline
    \multicolumn{6}{c}{SDSS sample at $z \sim 0$} \\
    \hline
    Disk & 1.0 & 10.0 - 30.0 & 15.0 - 22.0 & 0.3 - 1.0 & -90.0 - 90.0 \\
    %\hline
    Bulge & 4.0 & 4.0 - 17.0 & Disk Comp. $\pm\,(0,3.2)$\tablenotemark{a} & 0.3 - 1.0 & Disk Comp. $\pm\,(0,15)$\tablenotemark{b}\\
    \hline
    \hline
    \multicolumn{6}{c}{CANDELS sample at $z \sim 1$} \\
    \hline
    Disk & 1.0 & 12.0 - 25.0 & 17.0 - 27.8 & 0.3 - 1.0 & -90.0 - 90.0\\
    %\hline
    Bulge & 4.0 & 4.0 - 14.0 & Disk Comp. $\pm\,(0,3.2)$\tablenotemark{a} & 0.3 - 1.0 &  Disk Comp. $\pm\,(0,15)$\tablenotemark{b} \\
\enddata
\tablenotetext{a}{The bulge magnitude differs from the disk magnitude by a randomly chosen value between $-3.2$ and $+3.2$}
\tablenotetext{b}{The bulge position angle differs from the disk position angle by a randomly chosen value between $-15$ and $+15$}
\tablecomments{The above table shows the ranges of the various \sersic{} profile parameters used to simulate the training data. Each simulated galaxy has an $n=1$ disk and an $n = 4 $ bulge component, where n is the \sersic{} index. The distributions of all the simulated parameters are uniform except those for the bulge magnitude and bulge position angle. See \S\,\ref{sec:simulation_code} for more details.}
\end{deluxetable*}

It is well known that dust extinction can redden galaxies, and significant reddening has been observed for high-redshift galaxies\,\citep{brammer_09,williams_09,cardamone_10}. For the SDSS sample, we make no reddening correction since \citet{schawinski_14_green_herring} showed that dust correction has a negligible effect on the color-mass diagram for local galaxies. However, for the higher redshift CANDELS sample, we corrected the \textit{U-R} colors using the \citet{calzetti_00} extinction law:

\begin{equation}
    \Delta (U-R) = 0.65 A_V  .
\end{equation}

\noindent
The $A_V$ values, taken from the 3D-HST catalog, come from SED fits to stellar population models \citep{3dhst}. 

For both data sets, we used only a fraction of the available sample for transfer learning, leaving a much larger fraction for testing the performance of \gamornet{}. This demonstrates that \gamornet{} can effectively be trained initially on (more extensive) simulations, then re-trained using a small set of real data. Thereafter, \gamornet{} can successfully classify a much larger set of real images because it learns to generalize beyond the training galaxies.

\section{Training our Convolutional Neural Network --- \gamornet{}} \label{sec:methods}

The first hurdle in training a neural network to do morphological classifications is finding a large data set that has already been accurately classified. However, if neural networks are to be used widely for astronomical analysis, we need a more flexible approach --- one that does not require extensive analysis by old, slow (legacy) methods during the training phase and that can be adapted easily to new data sets. Here, we describe how to use simulated galaxies for the initial training of the classification network, followed by the application of a machine learning technique known as ``transfer learning'', wherein a much smaller set of galaxies, classified using a legacy method, is used to fine tune a partially trained network. This ensures that the network becomes adept at classifying real galaxies without requiring too many of them for the training process.

The process of training \gamornet{} to classify galaxy morphologies consists of the following steps:
\begin{enumerate}[noitemsep]
\item Simulating galaxies corresponding to the desired data set (here, SDSS or CANDELS).
\item Initial training of the neural network on those simulated images.
\item Retraining the neural network using a small part of the real data at hand; this process is known as transfer learning.
\item Testing a similar amount of real data to validate the results.
\item Processing the remainder of the real data through the trained network to obtain morphological classifications. 
\end{enumerate}
 
The galaxy simulations are described in \S\,\ref{sec:simulation_code}. \S\,\ref{sec:network_description} contains a brief introduction to CNNs and describes the architecture of \gamornet{}, while \S\,\ref{sec:initial_training} describes the initial training of \gamornet{} on the simulations. In \S\,\ref{sec:tf_intro}, we describe how we perform transfer learning to produce the final trained state of \gamornet{}.

\subsection{Simulations} \label{sec:simulation_code}

\begin{figure*}[hbt]
	\begin{center}
    \subfigure[Initial Simulated Image]{\includegraphics[width=5cm,height=5cm]{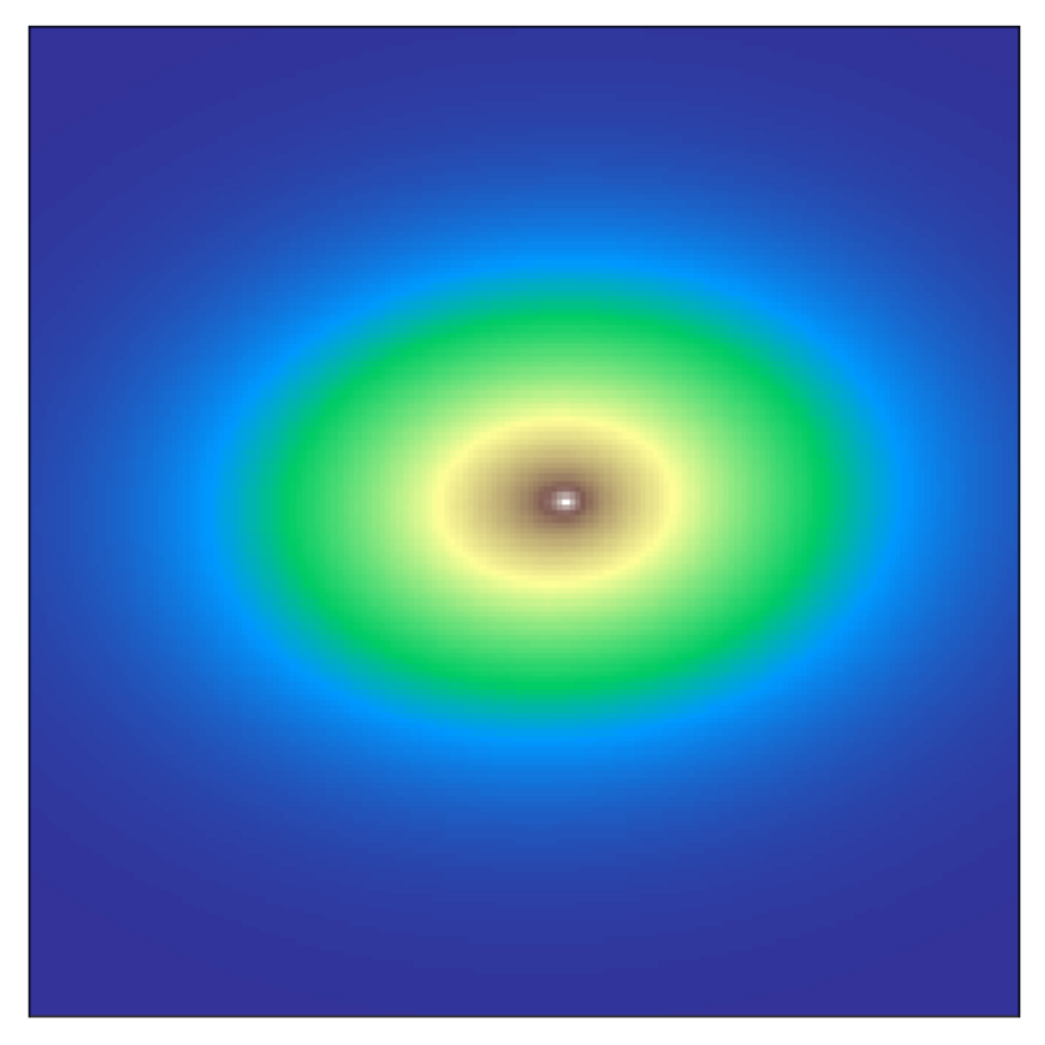}} 
    \subfigure[Convolved with PSF]{\includegraphics[width=5.0cm,height=5cm]{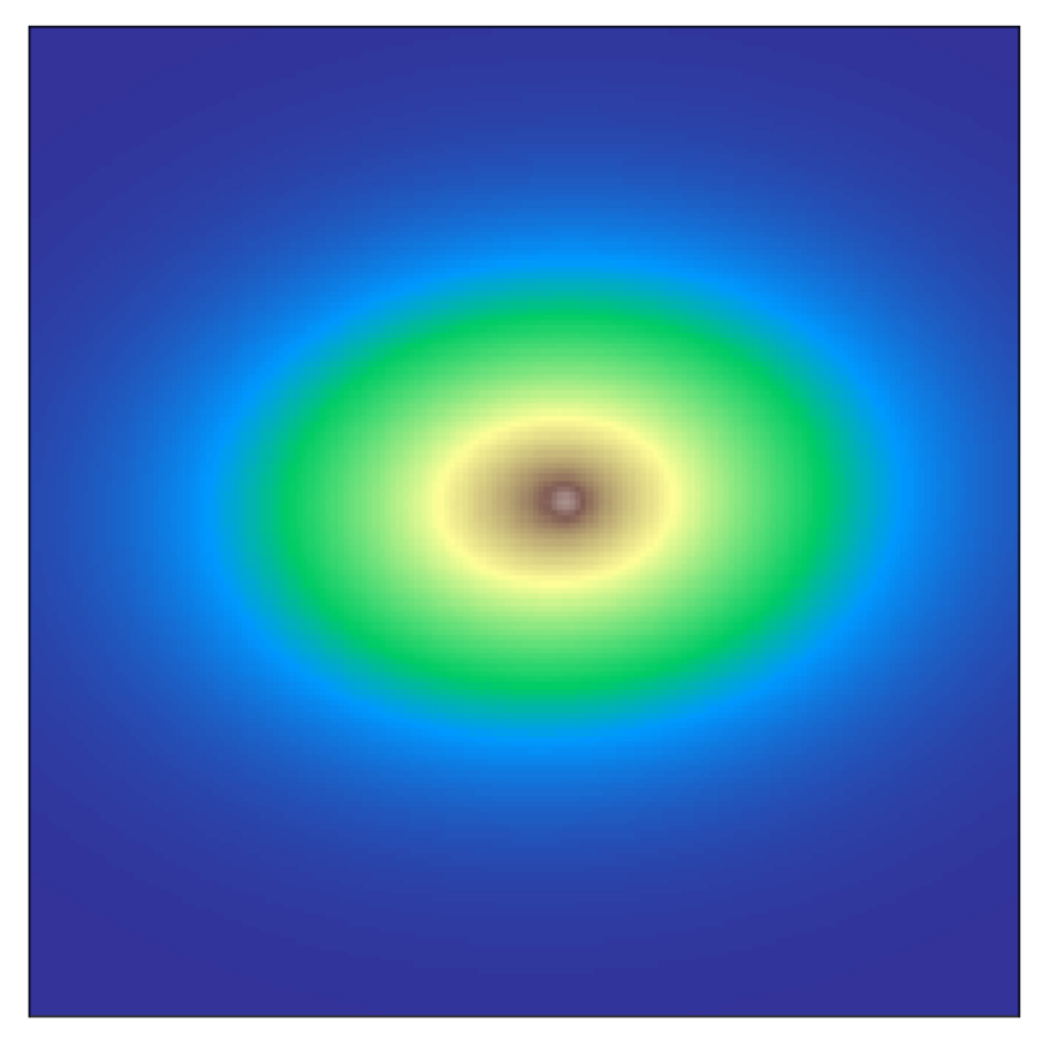}} 
    \subfigure[Noise Added]{\includegraphics[width=5cm,height=5cm]{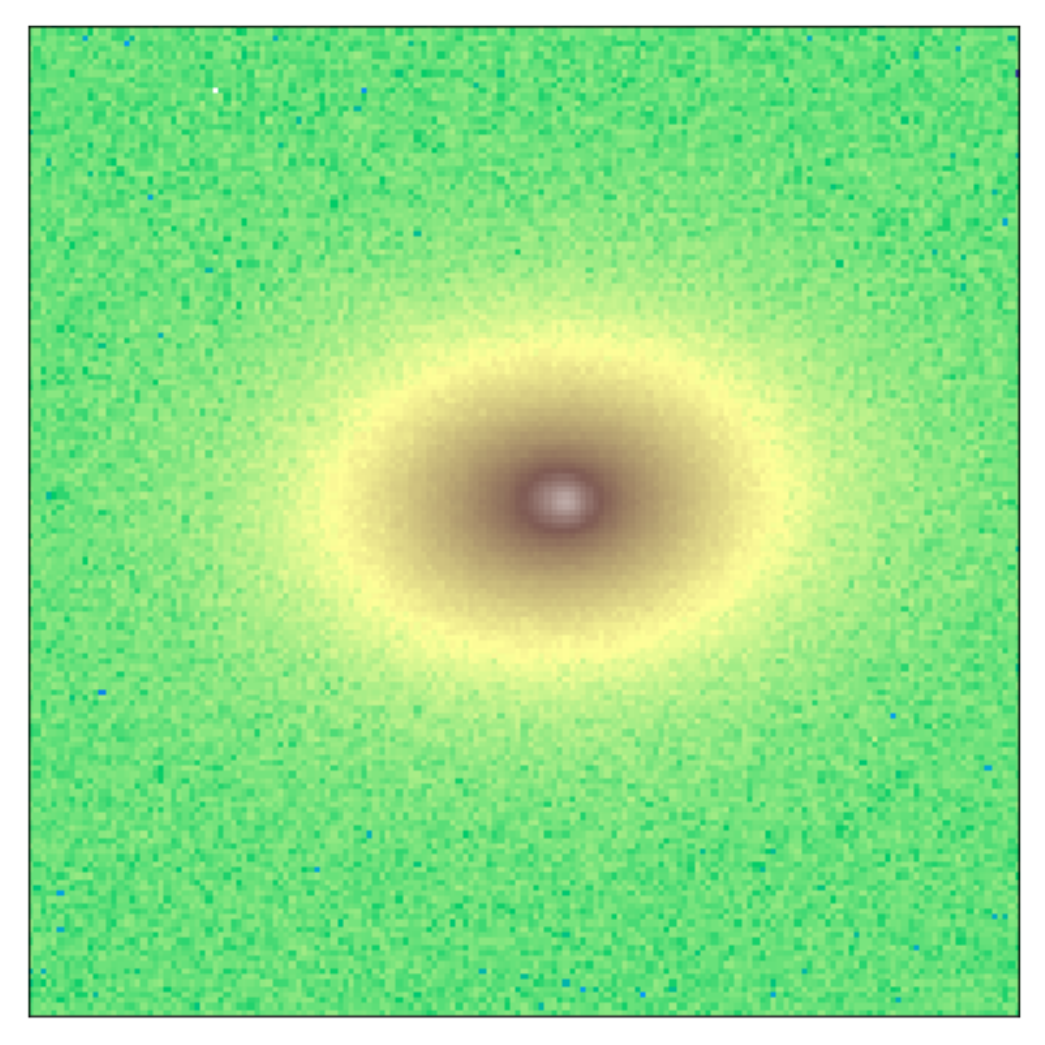}} 
     \subfigure{\includegraphics[height=5cm]{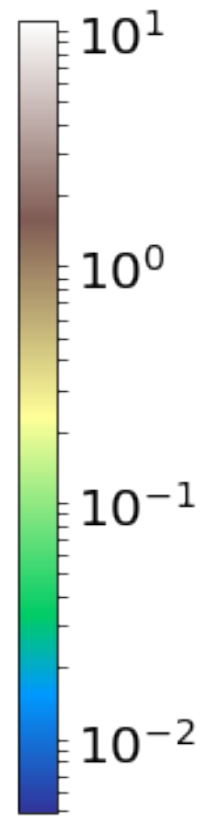}} 
  \end{center}
  \caption{Three stages in simulating an SDSS galaxy. Left (a): Light profile generated by GALFIT with a bulge-to-disk ratio of 0.24. Center (b): The left image convolved with the SDSS PSF. Right (c): SDSS noise added to the middle image. See \S\,\ref{sec:simulation_code} for details of the PSF convolution and noise addition.}
  \label{fig:psf_noise}
\end{figure*}

We simulated galaxies using the GALFIT program \citep{galfit}, which is usually used to fit two-dimensional light profiles of galaxies. Here, we use it instead to create two-dimensional light profiles appropriate for the data sets we are interested in analyzing with \gamornet{}.

For each data set, we simulated 100,000 galaxies consisting of a bulge (\sersic{} component with fixed index $n=4$;\,\citet{de_vac_48}) and disk (\sersic{} index $n=1$). The surface brightness for a galaxy with a \sersic{} profile is given by

\begin{equation}
\label{eq:sersic_fn}
\Sigma(r) = \Sigma_e \exp \left[ -\kappa \left( \left( \frac{r}{r_e}\right)^{1/n} - 1 \right) \right] ,
\end{equation}
where $\Sigma_e$ is the pixel surface brightness at the effective radius $r_e$, $n$ is the Se\'rsic index, which controls the concentration of the light profile, and $\kappa$ is a parameter coupled to $n$ that ensures that half of the total flux is enclosed within $r_e$. 

The parameters required to generate the \sersic{} profiles are drawn from uniform distributions (except the bulge magnitude and position angle) and the ranges of the distributions used for both sets of simulations are summarized in Table~\ref{tab:sim_para}. The galaxy size parameters were chosen to be representative of bright, local galaxies \citep{binney_and_merrifield}; bulges were chosen to have a half-light radius between $3.0$\,kpc and $6.0$\,kpc and disks were assigned half-light radii between $6.0$\,kpc and $10.0$\,kpc. To obtain the corresponding pixel sizes, we placed the samples at $ z = 0.05$ and $z = 1.0$ (corresponding to the mean redshifts of the two samples described in \S\,\ref{sec:data}) using WMAP7 cosmology \citep{wmap7} and using the pixel scale for the appropriate data set. We ensured that the number of simulated galaxies was sufficiently large such that even when we consider subsets of galaxies with similar sizes, they not only span the entire range of $L_B/L_T$ values but also mimic the overall bulge-to-total light ratio distribution.  

The disk magnitudes were drawn from a uniform distribution chosen so as to include most galaxies at these redshifts, and the magnitude of each corresponding bulge is such that it differs from the disk magnitude by a randomly chosen value between $-3.2$ and $3.2$. This was done to ensure that the bulge-to-total ratio varies between $\sim 5\% - 95\%$. Not enforcing this condition and allowing the bulge magnitude to be independent of the disk magnitude causes most galaxies in the training set to have a very high or a very low bulge-to-total ratio, which is not the case for most galaxies and, in any case, is not detectable. Instead, we want to train the network on a sufficient number of galaxies with intermediate bulge-to-total ratios.  

To make the two-dimensional light profiles generated by GALFIT more closely resemble the actual data, we convolved them with a representative point-spread function (PSF), then added noise. For the SDSS simulations, we selected the coordinates of one of the real galaxies in our sample -- R.A.: $213.26064353$, Decl.: $0.14637573$ and then reconstructed the PSF at the corresponding location in the detector using the PSF information stored in the relevant psField file that we obtained from SDSS. To generate the representative noise, we randomly selected 1000 cutouts from our SDSS sample, masked the sources in each cutout using SourceExtractor \citep{s_extract} and then read-in the non-masked pixel values to generate a large sample of noise pixels. We sampled this collection of noise pixels randomly to make two-dimensional arrays of the same size as that of the simulated images and then added them to the images. To make sure that the PSF chosen is representative, we reconstructed the PSFs for 12 more randomly chosen galaxies in our sample and convolved each one with a simulated SDSS galaxy, before adding noise. By inspecting the difference images between each image created using one of the new PSFs and the image created using the originally used PSF, we found the average pixel value of each of these difference images to be at least three orders of magnitude lower than the average pixel value of the galaxy image created using the original PSF.

For the CANDELS sample, we used the model PSF generated by \citet{vdw_12} for the COSMOS field and added noise following the same method as for the SDSS simulations. To make sure that the COSMOS PSF is representative, we followed a procedure similar to what we did for SDSS using the GOODS-S and UDS PSFs. We again found the average pixel value of the difference images to be at least three orders of magnitude lower than the average pixel value of the galaxy image created using the original PSF. 

The effect of convolving the simulated galaxies with PSF and adding noise is depicted in Figure~\ref{fig:psf_noise}.

The goal behind convolving with the PSF and adding noise is not to recreate perfect replicas of the real galaxies in our samples but rather to train the network on realistic simulated images for which we know the intrinsic morphologies. This is why we arbitrarily selected the COSMOS PSF instead of making simulations for each field separately and used only one random SDSS PSF. If we were to make more of an effort to recreate exactly the real data in our sample, then the whole purpose having a CNN is lost. In that case, the neural network ends up having a low variance but an extremely high bias, as it is too closely tied to the training set. Instead, here, the CNN learns to generalize from fewer examples.  

Since the galaxies were independently simulated, the simulation code could be trivially parallelized, and we make the simulation code available as a part of our public data release (see Appendix \ref{sec:ap:galsim}).

\subsection{The Network} \label{sec:network_description}
\begin{figure}[htbp]
	\begin{center}
		\includegraphics[width=0.45\textwidth]{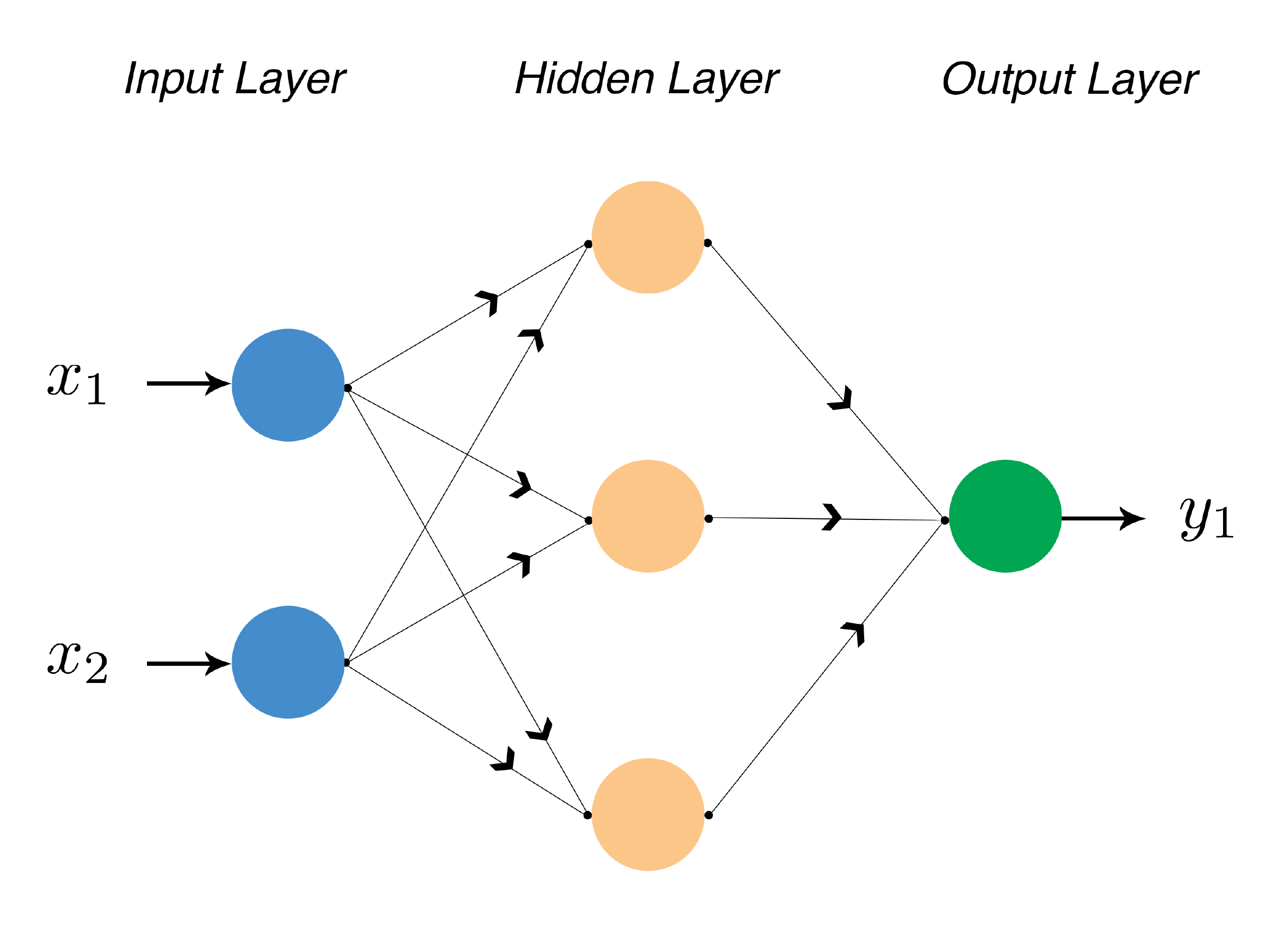}
	\end{center}
	\caption{A schematic diagram showing a simple artificial neural network with a single hidden layer.}
	\label{fig:schematic_network}
\end{figure}

Artificial neural networks, consisting of many connected units called artificial neurons, have been studied for more than five decades now. The neurons are arranged in multiple layers as shown in the schematic representation in Figure~\ref{fig:schematic_network}; each network has an input layer via which the data is fed into the network and an output layer that contains the result of propagating the data through the network, with additional hidden layer(s) in between. Each neuron is characterized by a weight vector $\bm{w}=(w_1,w_2,\ldots,w_n)$ and a bias $b$. The input to a neuron (coming from the outputs in the previous layer) is usually written as $\bm{x}=(x_1,x_2,\ldots,x_n)$ and the output of the neuron is given by 

\begin{equation}
y = \sigma(\bm{w} \cdot \bm{x} + b) ,
\label{eq:neuron_output}
\end{equation}
where $\sigma$ is the chosen activation function of the neuron. The process of ``training" an artificial neural network involves finding out the optimum set of weights and biases of all the neurons such that for a given vector of inputs, the output vector from the network, $\bm{y} = (y_1,y_2,\ldots,y_n)$, resembles the desired output vector $\bm{\hat{y}}=(\hat{y_1},\hat{y_2},\ldots,\hat{y_n})$ as closely as possible. The process of optimization is usually performed by minimizing a loss function, such as the popular cross-entropy loss function,

\begin{equation}
L = -\frac{1}{N}\sum_{j=1}^N \sum_{c=1}^M I_{j,c} \log(p_{j,c}) 
\label{eq:cross_entr_loss}
\end{equation}
where $I_{j,c}$ is a binary indicator function depicting whether class label $c$ is the correct classification for the $j^{th}$ observation. Also, $p$ is the predicted probability (by the network) that observation $j$ is from class $c$, $M$ is the total number of classes and $N$ is the total number of samples. 

Out of the various algorithms available to minimize the loss function, one that is used very widely is stochastic gradient descent\,(SGD) and its different variants \citep{nielsen}. In SGD, we estimate the gradient of $L$ using a mini-batch of training samples and update the weights and biases according to

\begin{equation}
\begin{aligned}
w' = w -\eta \frac{\partial L}{\partial w} \\
b' = b - \eta \frac{\partial L}{\partial b}
\end{aligned}
\label{eq:sgd}
\end{equation}
where $\eta$ is a small positive constant known as the learning rate. Calculation of the gradient is done using the back-propagation algorithm, and we refer the interested reader to \citet{rumelhart_88} for details.

\begin{deluxetable*}{cccc}[htbp]
%\tablenum{3}
\tablecaption{Structure of \gamornet{} \label{tab:network_layers}}
\tablecolumns{4}
\tablehead{
\colhead{Order} & \colhead{Type of Layer} & \colhead{Layer Description} & \colhead{Activation Function}
}
\startdata
    \hline
    1 & Input & Size: $167\times167$(SDSS) $\vert$ $83\times83$(CANDELS) & --  \\
    \hline
    2 & Convolutional & No. of Filters: 96 $\vert$ Filter Size: 11 $\vert$ Strides: 4 & ReLU\tablenotemark{a} \\
    \hline   
    3 & Max-Pooling & Kernel Size: 3 $\vert$ Strides: 2 & -- \\
    \hline
    4 & Local Response Normalization & -- & -- \\
    \hline
    5 & Convolutional & No. of Filters: 256 $\vert$ Filter Size: 5 $\vert$ Strides: 1 & ReLU\tablenotemark{a} \\
    \hline   
    6 & Max-Pooling & Kernel Size: 3 $\vert$ Strides: 2 & -- \\
    \hline
    7 & Local Response Normalization & -- & -- \\
    \hline
    8 & Convolutional & No. of Filters: 384 $\vert$ Filter Size: 3 $\vert$ Strides: 1 & ReLU\tablenotemark{a} \\
    \hline
    9 & Convolutional & No. of Filters: 384 $\vert$ Filter Size: 3 $\vert$ Strides: 1 & ReLU\tablenotemark{a} \\
    \hline
    10 & Convolutional & No. of Filters: 256 $\vert$ Filter Size: 3 $\vert$ Strides: 1 & ReLU\tablenotemark{a} \\
    \hline
    11 & Max-Pooling & Kernel Size: 3 $\vert$ Strides: 2 & -- \\
    \hline
    12 & Local Response Normalization & -- & -- \\
    \hline
    13 & Fully Connected & No. of neurons: 4096 & tanh \\
    \hline
    14 & Dropout & Dropout probability: 50$\%$ & -- \\
    \hline
    15 & Fully Connected & No. of neurons: 4096 & tanh \\
    \hline
    16 & Dropout & Dropout probability: 50$\%$ & -- \\
    \hline
    17 & Fully Connected & No. of neurons: 3 & softmax \\
    \hline
\enddata
\tablenotetext{a}{Rectified Linear Unit}
\tablecomments{The various layers of \gamornet{} along with the important parameters of each layer and the corresponding activation functions are shown in the table above. The architecture of \gamornet{} is based on AlexNet and, broadly speaking, consists of five convolutional layers followed by three fully connected layers. The source code for \gamornet{} is made public as described in Appendix\,\ref{sec:ap:gamornet_source_code}}
\end{deluxetable*}

The artificial neural network that we use for this work is a Convolutional Neural Network (CNN; \citealp{fukushima_80, lecun_98}). This is a type of deep artificial neural network that has become extremely popular for image processing in recent years. The input to the network is the two-dimensional vector representation of an image, and in a convolutional layer, each unit receives input from a local image patch of the previous layer known as the receptive field. Convolution involves taking a filter of a particular size and repeatedly applying it (by moving it with a specific stride) to each part of the input image, resulting in a two-dimensional output map of activations called a feature map. The different units in the feature map share the same weight matrix, and hence, each feature map can be interpreted as trying to locate a particular feature at different locations in the image. Each convolutional layer is typically followed by a max-pooling layer wherein the dimensionality of the feature maps are reduced by only preserving the maximum value in a small patch and thus making the network invariant to minor distortions. The convolutional and max-pooling layers are usually followed by a few fully connected layers that use the output of the convolutional layers to infer the correct output for the input image. We refer an interested reader to \citet{nielsen} for a more detailed overview of the above concepts.

\begin{figure*}[htb]
    \centering
    \includegraphics[width = \textwidth, height= 5.2cm]{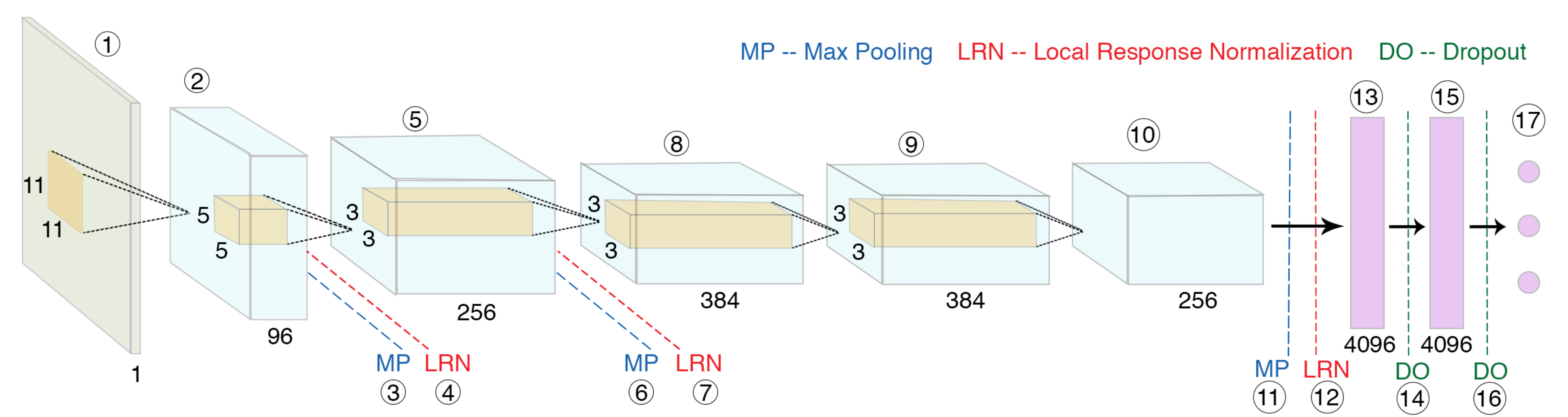}
    \caption{Schematic diagram of \gamornet{}, a CNN optimized to identify whether galaxies are bulge-dominated or disk-dominated. Its architecture, which is based on AlexNet\,\citep{alexnet}, consists of five convolutional layers and three fully connected layers. Between these layers are max-pooling, local response normalization, and dropout layers. The numbers inside the circles refer to the layer number and corresponding details for each layer can be found by looking up the corresponding layer order number in Table~\ref{tab:network_layers}. }
    \label{fig:galmnet_schematic}
\end{figure*}

The architecture of \gamornet{} is based on AlexNet \citep{alexnet}, a CNN that won the 2012 ImageNet Large Scale Visual Recognition Challenge (ILVRS), wherein different teams compete to classify about 14 million hand-annotated images. Very broadly speaking, the architecture of \gamornet{} consists of five convolutional layers and three fully connected layers. Interspersed between these are local response normalization, max-pooling and dropout layers. The dropout layers help to prevent over-fitting by randomly ignoring or ``dropping out" some number of layer outputs. The size of the input layer corresponds to the size of the images being fed-in, and the output layer corresponds to the three classes into which the galaxies are separated, which are defined in \S\,\ref{sec:initial_training}. The output layer happens to have the softmax activation function and thus, the output value of the three output neurons can be interpreted as the network's prediction probability that the input galaxy is in the corresponding category. In total, \gamornet{} has 17 layers, the details of which are summarized in Table~\ref{tab:network_layers}. Figure~\ref{fig:galmnet_schematic} shows a schematic diagram of \gamornet{}.

We implemented \gamornet{} using TFLearn\footnote{\href{http://tflearn.org}{http://tflearn.org}}\footnote{Also available in Keras now as a part of the data release}, which is a high-level Application Program Interface for TensorFlow\footnote{\href{https://tensorflow.org}{https://tensorflow.org}}, an open source library widely used for large-scale machine learning applications. We make the source code of \gamornet{} available as a part of our public data release (see Appendix\,\ref{sec:ap:gamornet_source_code} for more details).

\subsection{Initial Training} \label{sec:initial_training}
 Using the two sets of simulations corresponding to the SDSS and CANDELS data sets, we trained two different networks, both with the same structure as described in \S\,\ref{sec:network_description}. Henceforth, we refer to the networks trained on SDSS and CANDELS simulations as \gamornet{}-S and \gamornet{}-C respectively. During the training process, we trained the networks to separate galaxies into three different categories:
\begin{enumerate}[noitemsep]
\item Galaxies with $L_B/L_T < 0.45 $, i.e., disk-dominated.
\item Galaxies with $L_B/L_T > 0.55 $, i.e., bulge-dominated.
\item Galaxies with $ 0.45 \leq L_B/L_T \leq 0.55 $, i.e., indeterminate.
\end{enumerate}
Here, $L_B$ is the luminosity of the bulge component, and $L_T$ is the total luminosity of the galaxy. Since these galaxies are simulated, we used our knowledge of the actual $L_B/L_T$ for each galaxy to train the network.

Of the 100,000 galaxies simulated for each data set, we used $90\%$ for training and the rest for validation. The validation set was used to tune the different hyper-parameters in the network (like the learning rate described in \S\,\ref{sec:network_description}). We use a learning rate of 0.0001 and a batch size of 64, as these lead to $>95\%$ accuracy on the validation set and run-times of $\sim\mathcal{O}(1\mathrm{\,hour})$ on Tesla P100 GPUs. The batch size refers to the number of training samples the network works through before the model's internal parameters are updated

During the training process, we used the categorical cross-entropy loss function and minimized it using the momentum optimizer, which is a variant of SGD and accelerates SGD in the relevant direction besides dampening oscillations during the minimization process. Both SGD and the categorical cross-entropy loss function are described in \S\,\ref{sec:network_description}. 

An ``epoch" of training refers to running all of the training images through the network once. After each epoch of training, we evaluated the value of the loss function and calculated the accuracy on the validation set. The process of calculating the accuracy involves running all of the images in the validation set through the network. Since the output layer in our network is a softmax layer, the output value of each neuron can be interpreted as the network's predicted probability of the galaxy image to belong to the particular category corresponding to that neuron. A galaxy is said to belong to the $L_B/L_T$ category for which the predicted probability is the highest, and the accuracy was calculated as the number of galaxies classified correctly divided by the total number of galaxies. It is important to note here that we used an additional criterion for classifying the real images later on, as described in \S\,\ref{sec:morph_results}.

\begin{figure*}[htb]
	\begin{center}
    \subfigure[SDSS]{\label{fig:sdss_base_network}\includegraphics[width=0.45\textwidth]{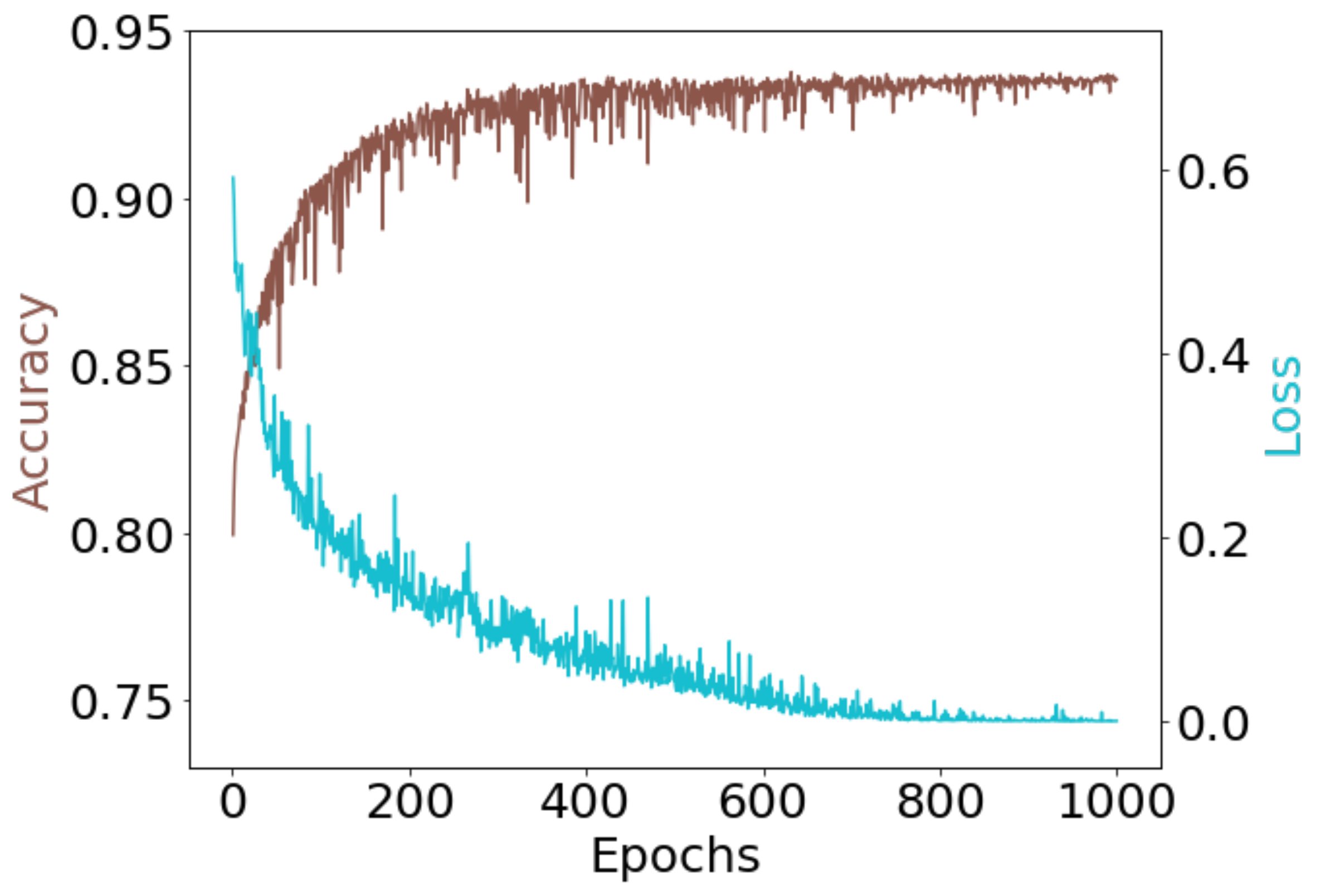}}
    \subfigure[CANDELS]{\label{fig:candels_base_network}\includegraphics[width=0.45\textwidth]{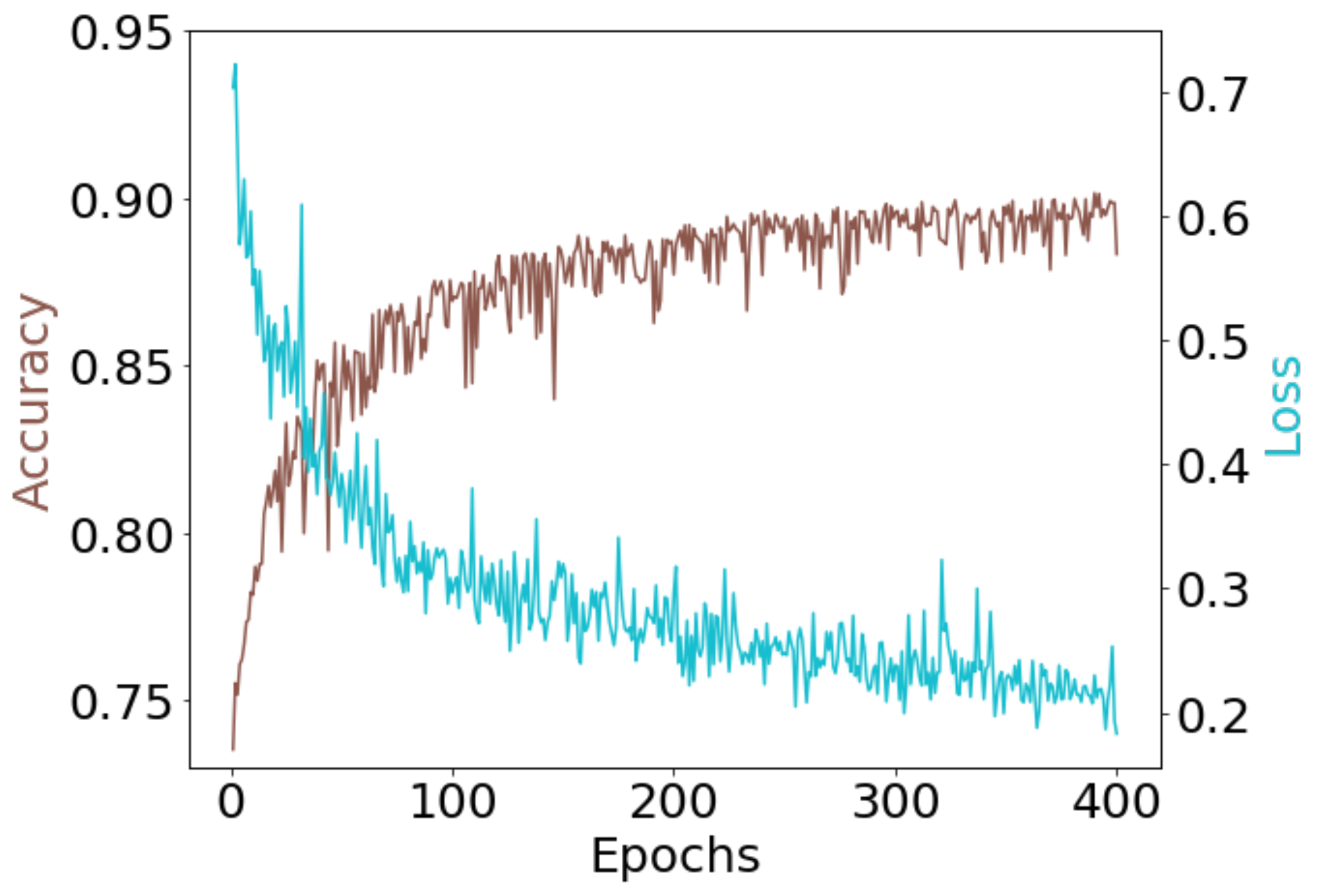}}
  \end{center}
  \caption{Learning curves for the process of training \gamornet{} on simulated galaxy images. The accuracy evaluated on the validation set (brown curves, left axes) and the value of the loss function after each epoch of training (blue curves, right axes) are shown for both \gamornet{}-S and \gamornet{}-C (left and right panels, respectively). \gamornet{}-S achieves an accuracy of $93.55\%$ after 1000 epochs of training, and \gamornet{}-C achieves an accuracy of $88.33\%$ after 400 epochs of training. For more details about the training process, see \S\,\ref{sec:initial_training}.}
  \label{fig:base_network_train}
\end{figure*}

We trained both the networks until the values of the accuracy and the loss function stabilized and a significant gain in accuracy did not seem probable with further training. This constituted training \gamornet{}-S for 1000 epochs and \gamornet{}-C for 400 epochs. Both learning curves are shown in Figure~\ref{fig:base_network_train}, which shows the accuracy as well as the value of the loss function after each epoch of training. \gamornet{}-S and \gamornet{}-C achieved net accuracies of $93.55\%$ and $88.33\%$, respectively, on the simulated images being used for validation; note that these are simulated images that the network did not ``see" during the process of training. 

\subsection{Transfer Learning} \label{sec:tf_intro}

\begin{deluxetable*}{p{2.2cm}p{3cm}p{4cm}p{4cm}p{2cm}}[htbp]
%\tablenum{4}
\tablecaption{Transfer Learning Parameters \label{tab:tl_para}}
\tablecolumns{5}
\tablehead{
\colhead{Network} & \colhead{Non-Trainable Layers
\tablenotemark{a}} & \colhead{Layers Trained from Previous Training} & \colhead{Layers Trained from Scratch} & \colhead{Learning Rate}
}
\startdata
    \hline
    \gamornet{}-S & None & All Convolutional Layers (2,5,8,9,10) & Last 3 Fully Connected Layers (13,15,17) & 0.00001 \\
    \hline
    \gamornet{}-C & First 3 Convolutional Layers (2,5,8) & Last 2 Convolutional Layers + First Fully Connected Layer (9,10,13)& Last 2 fully Connected Layers (15,17) & 0.00001 \\ 
    \hline
\enddata
\tablenotetext{a}{These layers were optimized during the initial training on simulations and then frozen at those values for the transfer learning step.}
\tablecomments{Details of the transfer learning algorithm used for both the SDSS and CANDELS networks. The numbers in parentheses refer to the layer numbers according to Table~\ref{tab:network_layers}. The above parameters were chosen by heuristically testing various options and choosing the ones that maximized accuracy, while not showing any signs of over-training.}
\end{deluxetable*}

CNNs have an extremely large number of free parameters (weights and biases) that need to be tuned during the process of training, and thus, if the size of the training set is not sufficiently large, there is a chance of ``over-fitting" after a certain number of epochs of training. That is, with further training, the accuracy of the network increases on the training data but not on the test data, and hence, the network fails to generalize. 

Transfer learning involves taking a network trained on a particular data set and optimized for a particular task, and re-tuning the weights and biases for a slightly different task or data set. The advantage here is that a much smaller training set can be used to re-tune the network than to train it from scratch. Transfer learning as a data-science concept has been around since the 1990s \citep{Pan2010}, and has been applied to a wide variety of tasks, including image classification\,\citep{zhu2011,Kulis2011,Duan2012}. As an example, transfer learning was recently applied to detect galaxy mergers\,\citep{ml_mergers}, starting from a network that could accurately identify images of everyday objects like cars, cats, dogs, etc.

In the present work, since we want to enable morphological classification even in the absence of a large training set, we use only a small fraction of the SDSS and CANDELS data sets for training. Specifically, we take the network trained on simulations and then re-train it by transfer learning on $\sim25\%$ of the real SDSS and CANDELS galaxy images.

In transfer learning, it is common to freeze the weights and biases in the initial layers of the network (i.e., those close to the input layer), while allowing variations in layers close to the output layer. The logic behind this approach is that, in a CNN, the deeper feature maps identify more complicated features while the earlier layers identify more basic features (like lines, shapes, and edges). Since transfer learning re-trains a network to do a slightly different task than it was initially trained to do, it is the last few layers that need to be re-tuned for the task at hand. At the same time, since the earlier layers correspond to more basic features, we do not expect that they will need re-tuning. We heuristically tested a combination of the various options mentioned above, and chose the one that maximized accuracy, while not showing any signs of over-training. The details of the transfer learning method used in both cases are summarized in Table~\ref{tab:tl_para}.

For the SDSS data, we have access to estimates of $L_B/L_T$ for each galaxy from \citet{simard_11}, wherein each galaxy was fitted with an $n=4$ bulge and an $n=1$ disk component. We used this as the ``ground truth'' for separating galaxies into the three categories defined in \S~\ref{sec:initial_training}. We randomly selected 10,000 galaxies from each category to make up our transfer learning training data set; this constitutes about a quarter of the full SDSS sample. We found that during transfer learning, it is important to have an equal number of galaxies from each category in the training set because otherwise, the network attempts to maximize accuracy in the category with more samples at the cost of other categories. Since both our samples have many more disk-dominated than bulge-dominated galaxies, a randomly selected training set would result in a very high accuracy in classifying disk-dominated galaxies but a very low accuracy in classifying bulge-dominated galaxies. Using the configuration given in Table~\ref{tab:tl_para}, we trained \gamornet{}-S for 300 epochs.

\begin{figure}[hbt]
    \centering
    \includegraphics[width=0.46\textwidth]{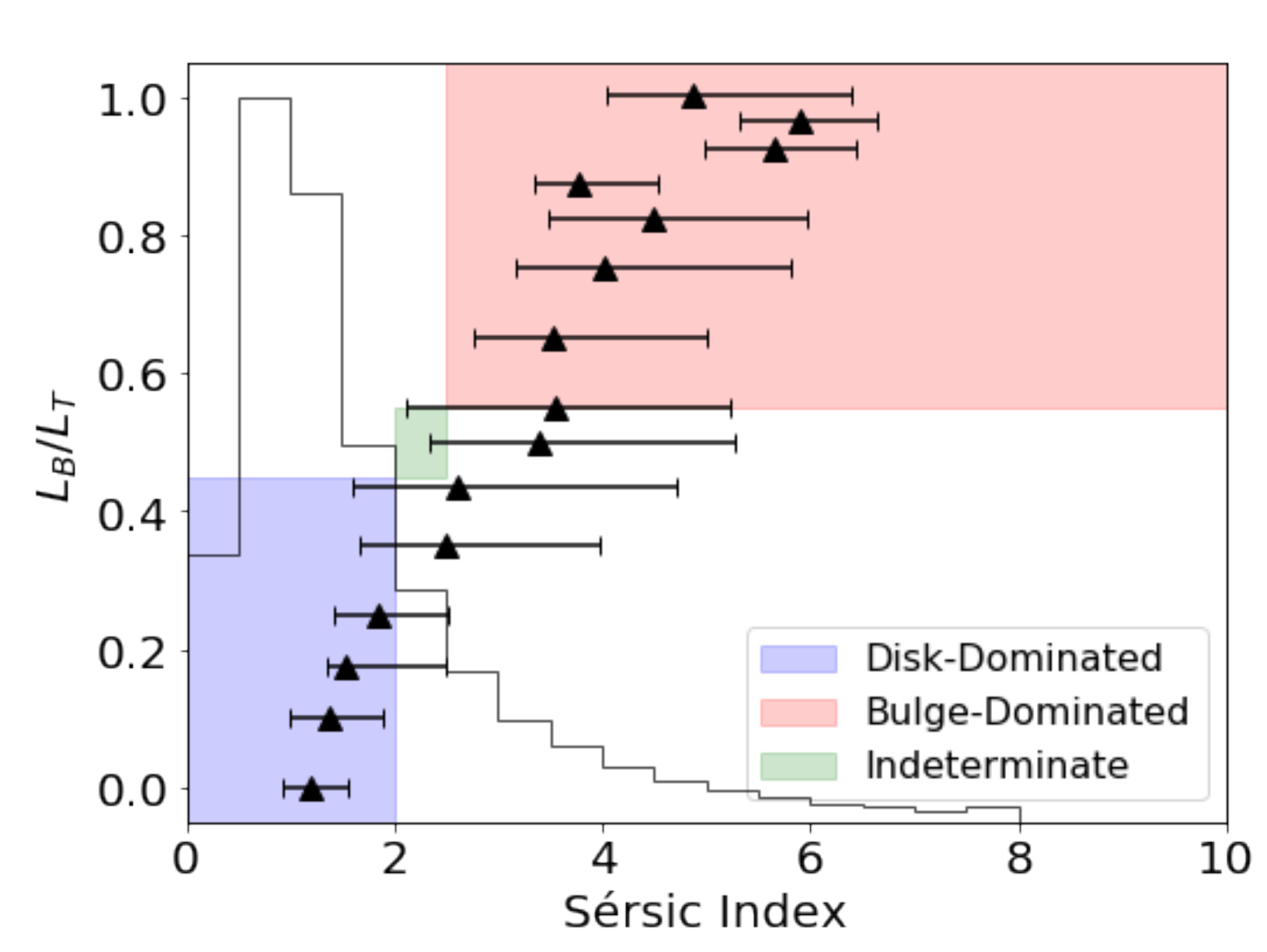}
    \caption{ The triangles show the input bulge-to-total ratio ($L_B/L_T$) versus fitted \sersic{} index for the galaxies simulated by \citeauthor{simmons_08} (\citeyear{simmons_08}; adapted from the lowest panel in their Figure 19). The plotted points are the median of each bin's distribution, and the error bars mark the central 68\% of sources in the bin. The shaded regions correspond to our definitions of the three output classes used by \gamornet{}-C. The histogram shows the distribution of the \sersic{} index for all the galaxies in our CANDELS sample, most of which are disk-dominated (see \S\,\ref{sec:data}). Clearly, all galaxies with $n<2$ are truly disk-dominated (i.e., have $L_B/L_T < 0.45$) but, because of the spread in \sersic{} indices, some disk-dominated or intermediate galaxies may get misclassified as bulge-dominated. Although a higher $n$ threshold (for, e.g., $n \sim 6$) would lead to a purer bulge-dominated sample, for reasons mentioned in \S\,\ref{sec:tf_intro}, it would make the transfer learning sample insufficiently small. Note that readers can choose different bin boundaries, doing their own transfer learning step on the simulation-trained network made available via \S\,\ref{sec:ap:gamornet_trained_models}}
    \label{fig:si_lb_lt}
\end{figure}{}
    
For the CANDELS data, no two-component bulge-disk decompositions were available in the literature. Thus, we translated the \sersic{} indices from \citet{vdw_12} into the three classifications used by \gamornet{} using results from \citet{simmons_08}, who analyzed CANDELS-depth HST ACS simulations of bulge+disk galaxies. The authors fitted single \sersic{} profiles to their simulations in order to find the correspondence between \sersic{} index and actual $L_B/L_T$. Guided by their result in the redshift bin $z=1.075$ (see their Fig.~19), appropriate for the CANDELS galaxies we wish to classify, we define galaxies with $n < 2.0$ as disk-dominated, $n > 2.5$ as bulge-dominated, and $ 2.0 \leq n \leq 2.5$ as indeterminate.
    
To illustrate these choices, we reproduce in Figure \ref{fig:si_lb_lt} the \citet{simmons_08} results, specifically, the range in \sersic{} index corresponding to different $L_B/L_T$ values for the simulated galaxies. The three broad classifications assigned by \gamornet{}-C --- disk-dominated, indeterminate, and bulge-dominated --- are shown as shaded regions. There is no unique or perfect way to go from \sersic{} index to $L_B/L_T$; although, the choice of $n<2$ is pretty clean, i.e., all such galaxies have $L_B/L_T < 0.45$ and are disk-dominated. For $n>2.5$, most galaxies are bulge-dominated (i.e., have $L_B/L_T > 0.55$) as is evident from the top-right portion of the figure; although, a few disk-dominated galaxies with $L_B/L_T \sim 0.4$ may be incorrectly included in that category. 
    
A higher $n$ threshold (for, eg., $n\sim6$) leads to a purer bulge-dominated sample, but drastically reduces the number of bulge-dominated galaxies available for transfer learning, as is evident from the histogram shown in Figure \ref{fig:si_lb_lt}. As mentioned previously, we need roughly equal numbers of galaxies in each bin for the training process during transfer learning, and thus, the upper limit on the total number of galaxies available for training is set by the size of the least populous bin. The above choices ensure a sufficient number of galaxies in each category (needed for the transfer learning step) and produce statistically acceptable classifications. Readers can set these boundaries differently, as appropriate to their science goals, using \gamornet{}-C models trained only on the bulge + disk simulations made publicly available via \S\,\ref{sec:ap:gamornet_trained_models}. Instructions on how to train these models for transfer learning are available in the GitHub repository.
    
Using the above definitions of the three classes, we re-trained the simulation-trained \gamornet{}-C for 75 epochs using the Transfer Learning configuration in Table~\ref{tab:tl_para}; note that only the weights and biases in the last two of the total five convolutional layers are adjusted during the transfer learning step. For this process, we used 2400 galaxies from each of the three morphological categories, or about a quarter of the total CANDELS sample.

\begin{deluxetable*}{ccccccc}[htbp]
%\tablenum{5}
\tablecaption{Classification Probabilities for 82,547 SDSS Galaxies}
\tablecolumns{7}
\tablehead{
\colhead{ObjID\tablenotemark{a}} & \colhead{R.A.} & \colhead{Decl.} & \colhead{Disk Prob.} & \colhead{Bulge Prob.} & \colhead{Indeterminate Prob.} & \colhead{Classification}
}
\startdata
587722953304440846 & 237.4210352 & 0.2367580 & 0.1356 & 0.4439 & 0.4205 & indeterminate \\
587722981750014081 & 202.6811651 & -1.0804622 & 1.0000 & 0.0000 & 0.0000 & disk-dominated\\
587722982831161384 & 219.5687676 & -0.3497467 & 0.9384 & 0.0001 & 0.0615 & disk-dominated\\
587722983365279858 & 213.2606435 & 0.1463757 & 0.9977 & 0.0000 & 0.0023 & disk-dominated\\
587722983366721714 & 216.5747982 & 0.1543351 & 0.0590 & 0.7170 & 0.2240 & indeterminate\\
\vdots & \vdots &\vdots &\vdots &\vdots &\vdots &\vdots\\
\enddata
\tablenotetext{a}{These are pre-DR8 ObjIDs}
\tablecomments{\gamornet{}-S classification probabilities (of being disk-dominated, bulge-dominated, or indeterminate) and final classification for all the galaxies in our SDSS test sample. This table is published in its entirety as a part of the public data release (Appendix \ref{sec:ap:prob_tables}). The first five entries are shown here for guidance regarding its form and content.}
\label{tab:lt_sdss}
\end{deluxetable*}

\begin{deluxetable*}{|>{\centering\arraybackslash}p{3cm}|>{\centering\arraybackslash}p{2.25cm}|>{\centering\arraybackslash}p{2.0cm}|>{\centering\arraybackslash}p{2.0cm}|}[htbp]
%\tablenum{6}
\tablecaption{Classification Summary for 82,547 SDSS Galaxies \label{tab:sdss_results}}
\tablecolumns{4}
\tablehead{ \multicolumn{4}{|c|}{Predictions:- Disks: 47,656 $|$ Bulges: 7963 $|$ Indeterminate: 26,928}}
\startdata
    \hhline{|====|}
    \multicolumn{4}{|c|}{Numbers} \\
    \hline 
     \multicolumn{2}{|c|}{} & \multicolumn{2}{c|}{\gamornet{}-S classifications} \\
     \cline{3-4}
     \multicolumn{2}{|c|}{} & Disks & Bulges \\
     \hline
     \citet{simard_11}   & Disks & 47,526 & 329\\ \cline{2-4} classifications & Bulges & 94 & 7552 \\
     \hhline{|====|}
     \multicolumn{4}{|c|}{Percentages} \\
     \hline
     \multicolumn{3}{|c|}{\gamornet{}-classified disks that SC\tablenotemark{a} also classified as disks} & 99.73\%\\
     \multicolumn{3}{|c|}{\gamornet{}-classified disks that SC\tablenotemark{a} classified as bulges} & 0.20\%\\
     \hline
     \multicolumn{3}{|c|}{\gamornet{}-classified bulges that SC\tablenotemark{a} also classified as bulges} & 94.84\%\\
     \multicolumn{3}{|c|}{\gamornet{}-classified bulges that SC\tablenotemark{a} classified as disks} & 4.13\%\\
     \hline
     \multicolumn{3}{|c|}{Total percentage of galaxies misclassified} & 0.7\%\\
\enddata
\tablenotetext{a}{\citet{simard_11} Classifications}
\tablecomments{Results of running the entire SDSS test set of 82,547 galaxies through \gamornet{}-S. Values in the top section refer to the number of galaxies in each category as predicted by \gamornet{}-S with respect to the \citet{simard_11} classifications. For example, the top-left cell value of 47,526 means that out of the 47,656 predicted disks, 47,526 are also classified as disks by \citet{simard_11}.}
\end{deluxetable*}

\section{Results} \label{sec:results}
\subsection{Morphology Results}\label{sec:morph_results}
After using about a quarter of the images for transfer learning, the remaining 82,547 galaxies in the SDSS sample were used as our test set. Since \gamornet{}'s output layer consists of three softmax neurons whose output values sum to $1$, each value can be interpreted as the probability that a galaxy belongs to that $L_B/L_T$ category. These probability values are the primary output of \gamornet{}. However, in order to compare our results with previous classifications and keeping in mind situations that necessitate rigid classifications, we transform the probability values into classifications.

After some experimentation, we arrived at this decision tree for classification:
\begin{enumerate}[noitemsep]
\item Disk-dominated when \gamornet{}-S reports $\geq 80\%$ probability that $L_B/L_T < 0.45$.
\item Bulge-dominated when \gamornet{}-S reports $\geq 80\%$ probability that $L_B/L_T > 0.55$.
\item Otherwise, indeterminate.
\end{enumerate}
This is slightly different than the criterion we used for the initial training, as those galaxies were idealized, and the classifications were unambiguous. For the real galaxies, simply taking the highest probability neuron, including probabilities below 80\%, made the classifications far less accurate. Requiring a threshold of 80\% greatly improved the classification accuracy at the expense of increasing the number of indeterminate galaxies.

For each galaxy, we have access to its bulge-to-total ratio, i.e., $L_B/L_T$ value from \citet{simard_11}, which we consider to be the ``true" value. For mapping $L_B/L_T$ to a classification of being bulge- or disk-dominated, we used the same criterion as during the initial training, outlined at the beginning of \S\,\ref{sec:initial_training}.

Individual morphological classifications by \gamornet{}-S are reported in Table~\ref{tab:lt_sdss} and Table~\ref{tab:sdss_results} compares the \gamornet{}-S and \citet{simard_11} classifications of SDSS galaxies. Assuming the latter are ``true", for disk-dominated galaxies, we achieved an accuracy of 99.7\% and for bulge-dominated galaxies, we achieved an accuracy of 94.8\%, resulting in a net misclassification rate of 0.7\%. A total of 26,928 galaxies, or $\sim 32$\% of the SDSS test set, were found to have indeterminate morphologies.

\begin{deluxetable*}{cccccccc}[htbp]
%\tablenum{7}
\tablecaption{Classification Probabilities for 21,746 CANDELS galaxies \label{tab:lt_candels}}
\tablecolumns{8}
\tablehead{
\colhead{Field} & \colhead{ID\tablenotemark{a}} & \colhead{R.A.} & \colhead{Decl.} & \colhead{Disk Prob.} & \colhead{Bulge Prob.} & \colhead{Indeterminate Prob.} & \colhead{Classification}
}
\startdata
GOODSN & 19 & 189.1464840 & 62.0957640 & 0.3356 & 0.3372 & 0.3272 & indeterminate\\
GOODSN & 32 & 189.1314850 & 62.0973280 & 0.3762 & 0.2722 & 0.3516 & disk-dominated\\
GOODSN & 63 & 189.1174320 & 62.1017230 & 0.3709 & 0.2877 & 0.3414 & disk-dominated\\
GOODSN & 68 & 189.1499790 & 62.1017680 & 0.4039 & 0.1798 & 0.4163 & indeterminate\\
GOODSN & 72 & 189.1432950 & 62.1022950 & 0.3006 & 0.3312 & 0.3683 & indeterminate\\
\vdots & \vdots &\vdots &\vdots &\vdots &\vdots &\vdots &\vdots\\
\enddata
\tablenotetext{a}{ID refers to the IDs assigned by the CANDELS team\,\citep{candels_1,candels_2}}
\tablecomments{\gamornet{}-C classification probabilities (of being disk-dominated, bulge-dominated, or indeterminate) and final classification for all the galaxies in our CANDELS test sample. This table is published in its entirety as a part of the public data release (Appendix \ref{sec:ap:prob_tables}). The first five entries are shown here for guidance regarding its form and content.}
\end{deluxetable*}

\begin{deluxetable*}{|>{\centering\arraybackslash}p{4cm}|>{\centering\arraybackslash}p{2cm}|>{\centering\arraybackslash}p{2.0cm}|>{\centering\arraybackslash}p{2.0cm}|}[htbp]
%\tablenum{8}
\tablecaption{Classification Summary for 21,746 CANDELS Galaxies \label{table:candels_results}}
\tablecolumns{4}
\tablehead{ \multicolumn{4}{|c|}{Predictions:- Disks: 12,549 $|$ Bulges: 580 $|$ Indeterminate: 8617}}
\startdata
    \hhline{|====|}
    \multicolumn{4}{|c|}{Numbers} \\
    \hline 
     \multicolumn{2}{|c|}{} & \multicolumn{2}{c|}{\gamornet{}-C classifications} \\
     \cline{3-4}
     \multicolumn{2}{|c|}{} & Disks & Bulges \\
     \hline
     \citet{vdw_12}  & Disks & 11,524 & 121 \\ \cline{2-4} classifications & Bulges & 992 & 456 \\
     \hhline{|====|}
     \multicolumn{4}{|c|}{Percentages} \\
     \hline
     \multicolumn{3}{|c|}{\gamornet{}-classified disks that VdwC\tablenotemark{a} also classified as disks } & 91.83\%\\
     %\hline
     \multicolumn{3}{|c|}{\gamornet{}-classified disks that VdwC\tablenotemark{a} classified as bulges} & 7.90\%\\
     \hline
     \multicolumn{3}{|c|}{\gamornet{}-classified bulges that VdwC\tablenotemark{a} also classified as bulges} & 78.62\%\\
     %\hline
     \multicolumn{3}{|c|}{\gamornet{}-classified bulges that VdwC\tablenotemark{a} classified as disks} & 20.86\%\\
     \hline
     \multicolumn{3}{|c|}{Total percentage of galaxies misclassified} & 5.3\%\\
\enddata
\tablenotetext{a}{\citet{vdw_12} Classifications}
\tablecomments{Results of running the entire CANDELS test set of 21,746 galaxies through \gamornet{}-C. Values in the top section refer to the number of galaxies in each category as predicted by \gamornet{}-C with respect to the \citet{vdw_12} classifications. For example, the top-left cell value of 11,524 means that out of the 12,549 predicted disks, 11,524 are also classified as disks by \citet{vdw_12}.}
\end{deluxetable*}

For the CANDELS data set, there were 21,746 galaxies in the test set. We classified these using \gamornet{}-C and, again, experimented with thresholds for the final neuron values in order to arrive at an acceptable balance between accuracy and fraction with indeterminate morphologies. The thresholds for the CANDELS classification (which are different from those adopted for the SDSS data) are:
\begin{enumerate}[noitemsep]
    \item Bulge-dominated if \gamornet{}-C reports $\geq 55\%$ probability that $L_B/L_T > 0.55 $.
    \item Disk-dominated if \gamornet{}-C reports $\geq 36\%$ probability that $L_B/L_T < 0.45 $ {\it and} this probability exceeds the probabilities of $L_B/L_T > 0.55 $, $0.45 \leq L_B/L_T \leq 0.55 $.
    \item Otherwise, indeterminate.
\end{enumerate}
The choice of these confidence thresholds and their impact on the results is discussed later in this section. 

Table~\ref{tab:lt_candels} reports the individual morphological classifications by \gamornet{}-C, and
Table~\ref{table:candels_results} compares these to the results of \citet{vdw_12}. From the \sersic{} index of each galaxy \citep{vdw_12}, we derive its $L_B/L_T$ following \citet{simmons_08} as described in \S\,\ref{sec:tf_intro}. Thereafter, we map these values to a classification of being bulge- or disk-dominated using the same criterion as we did during initial training, as described in \S\,\ref{sec:initial_training}. Assuming these as the ``true" classifications, \gamornet{}-C has an accuracy of 91.8\% for disk-dominated galaxies and 78.6\% for bulge-dominated galaxies, or a net misclassification rate of 5.3\%. A total of 8617 galaxies were classified in the indeterminate category, which is $\sim39$\% of the CANDELS test set. 

\begin{figure*}[htb]
	\begin{center}
    \subfigure[]{\label{fig:candels_sn}\includegraphics[height = 6cm]{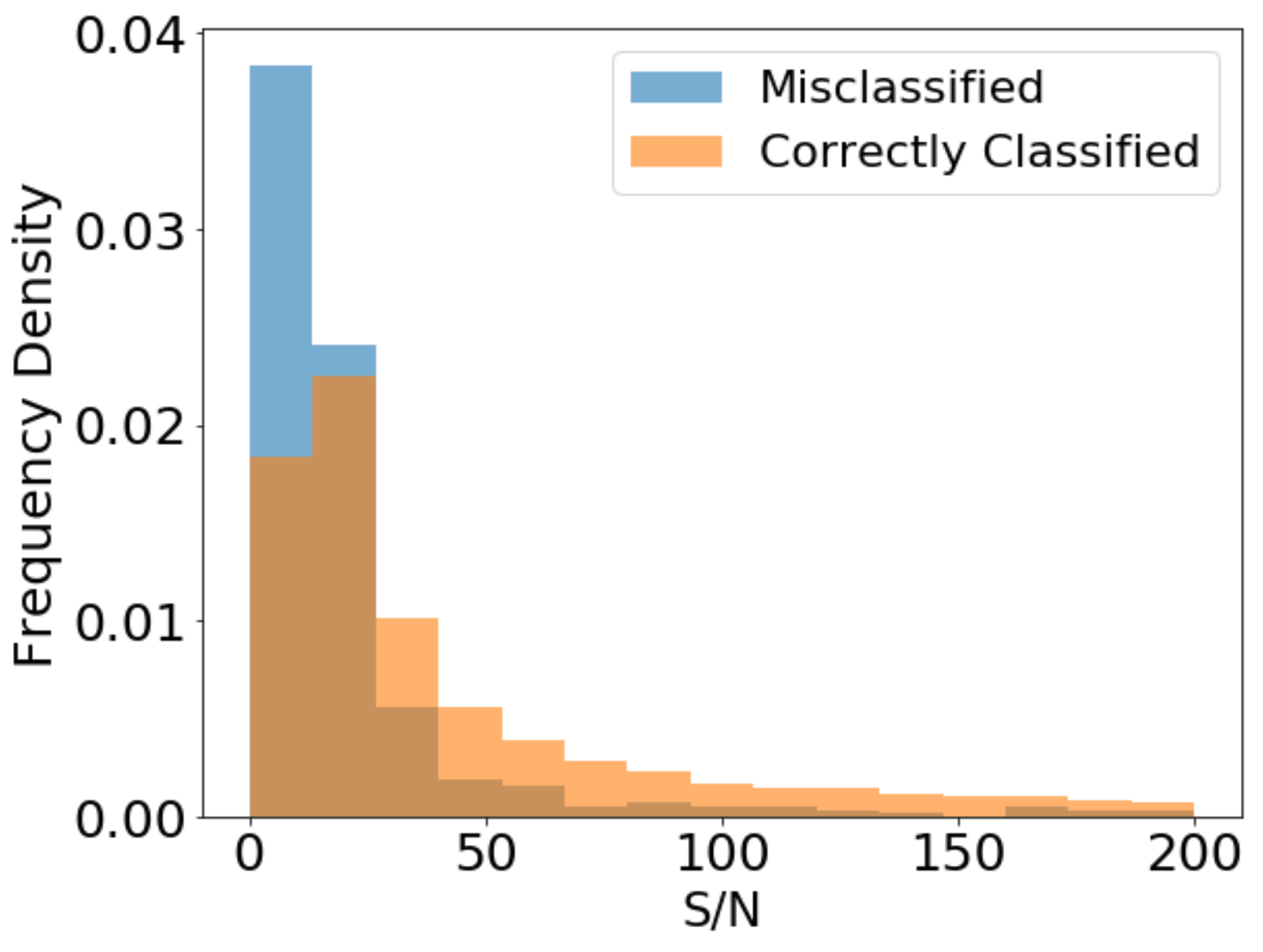}}
    \subfigure[]{\label{fig:candels_re}\includegraphics[height=6cm]{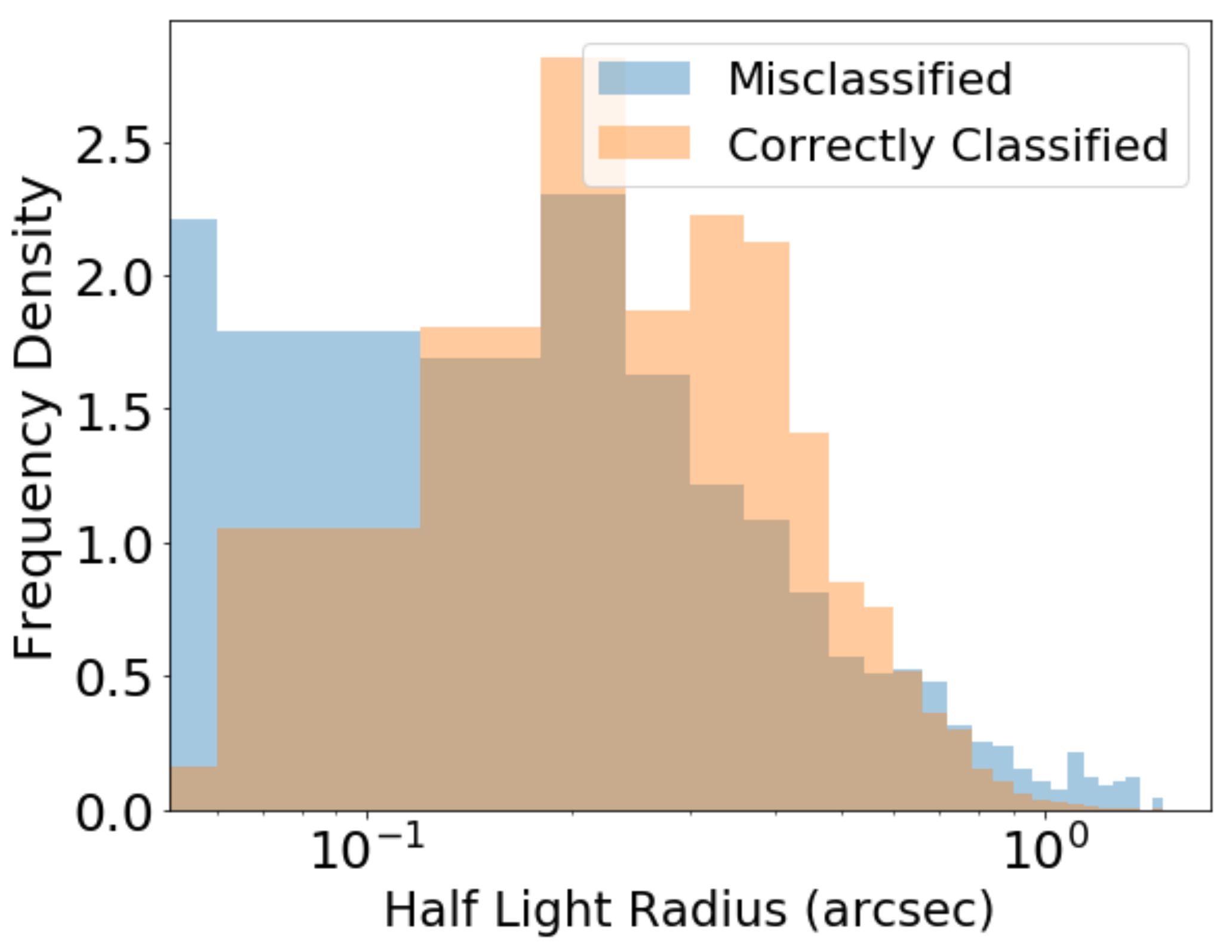}}
  \end{center}
  \caption{The normalized distribution of correctly classified and misclassifed CANDELS galaxies in the test set as a function of the signal-to-noise ratio (S/N) and half-light radius ($r_e$). Both plots show that compared to the correctly classified galaxies, a higher fraction of the misclassified galaxies have a low S/N ratio and/or small $r_e$. `Frequency density' refers to the number counts normalized to form a probability density.}
  \label{fig:candels_misclf}
\end{figure*}

The misclassification rate of CANDELS galaxies is higher than that of SDSS galaxies. To find out why, we investigated various relevant statistics for the misclassified galaxies. The two most significant variables were the S/N and the half-light radius (taken from \citealp{vdw_12}). Figure~\ref{fig:candels_misclf} shows the distribution of both the correctly classified and misclassified galaxies over these parameters. Although both the misclassified and correctly classified galaxies are distributed similarly over S/N, the misclassified population peaks much more sharply at a lower S/N, showing that a much larger fraction of the misclassifed sample has a low S/N compared to the correctly classified fraction. Similarly, a much larger fraction of the misclassified sample has low values of $r_e$ compared to the correctly classified galaxies. Therefore, we conclude that the misclassified galaxies are essentially galaxies with a small half-light radius comparable to the PSF and/or a low S/N, and thus, it is inherently difficult for \gamornet{}-C to correctly classify these galaxies. The misclassified population in the SDSS data set also peaks more sharply at a lower value of $r_e$ compared to the correctly classified galaxies; although, we have poor statistics for this, as the misclassification rate is $<1\%$.

\begin{figure*}[htb]
	\begin{center}
    \subfigure[]{\label{fig:sdss_unclass_threshold}\includegraphics[height=6cm]{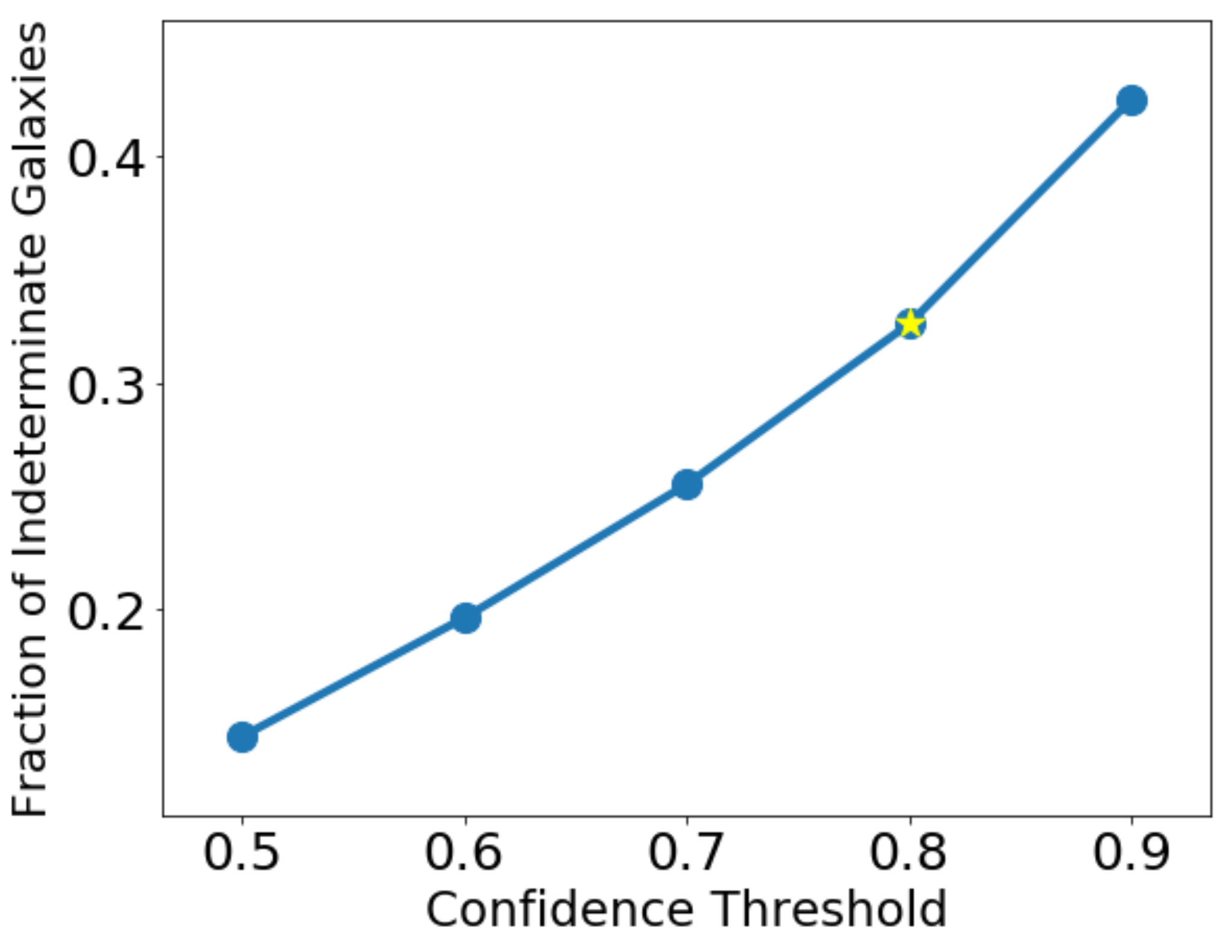}}
    \subfigure[]{\label{fig:sdss_acc_threshold}\includegraphics[height=6cm]{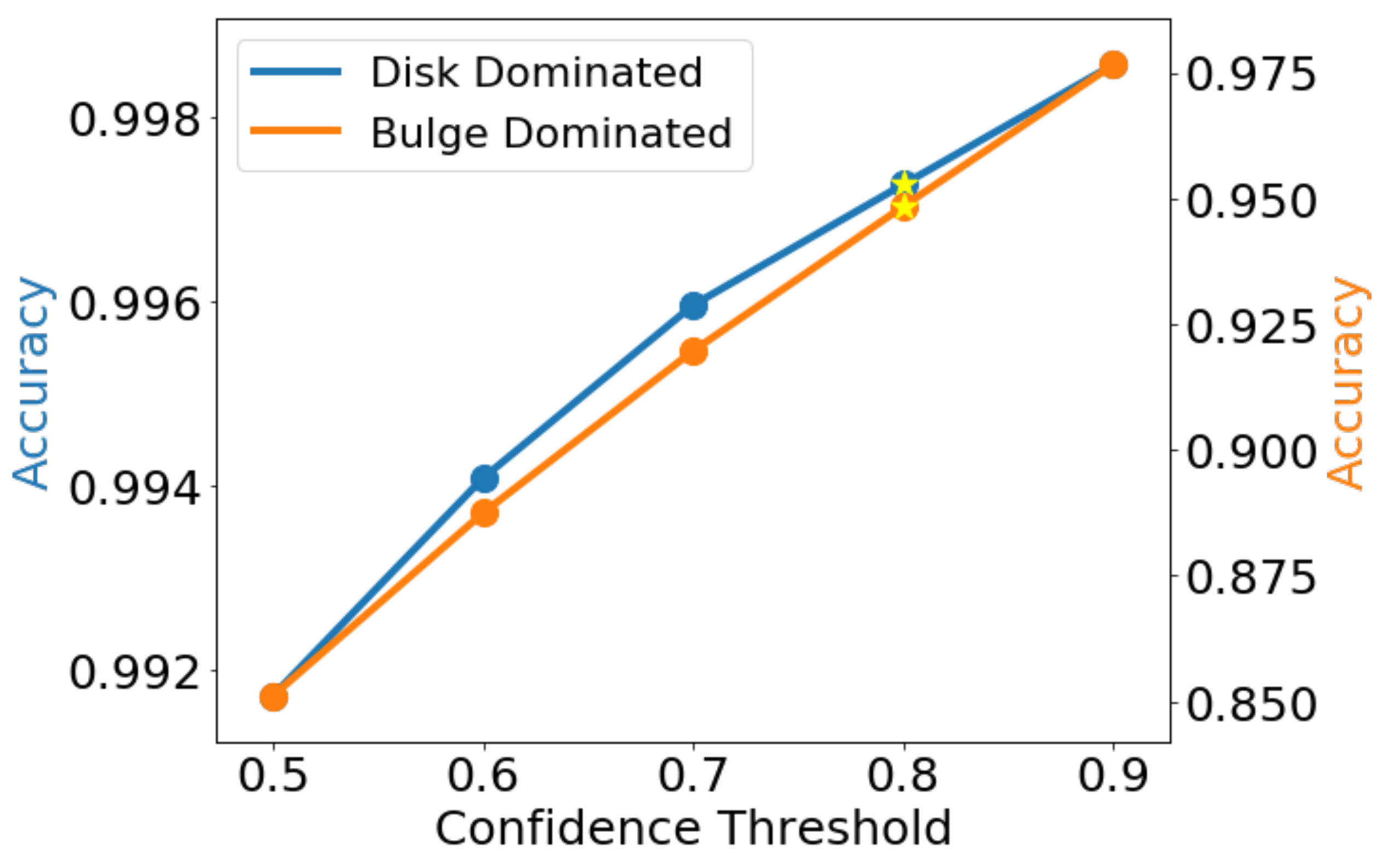}}
  \end{center}
  \caption{Relation of confidence threshold to completeness and accuracy of classification, for the SDSS data set. Left (a): The fraction of indeterminate galaxies increases with increasing confidence threshold.  Right (b): The accuracy of both disk-dominated (blue line, left axis) and bulge-dominated (orange line, right axis) classifications increases with increasing confidence threshold. We decided on a confidence threshold of 0.8 (or 80\%) for \gamornet{}-S (star in both plots) as the optimal compromise between accuracy and completeness.}
  \label{fig:sdss_threshold}
\end{figure*}

\begin{figure*}[htb]
	\begin{center}
    \subfigure[]{\label{fig:candels_acc_threshold_disk}\includegraphics[width=0.45\textwidth,height=6cm]{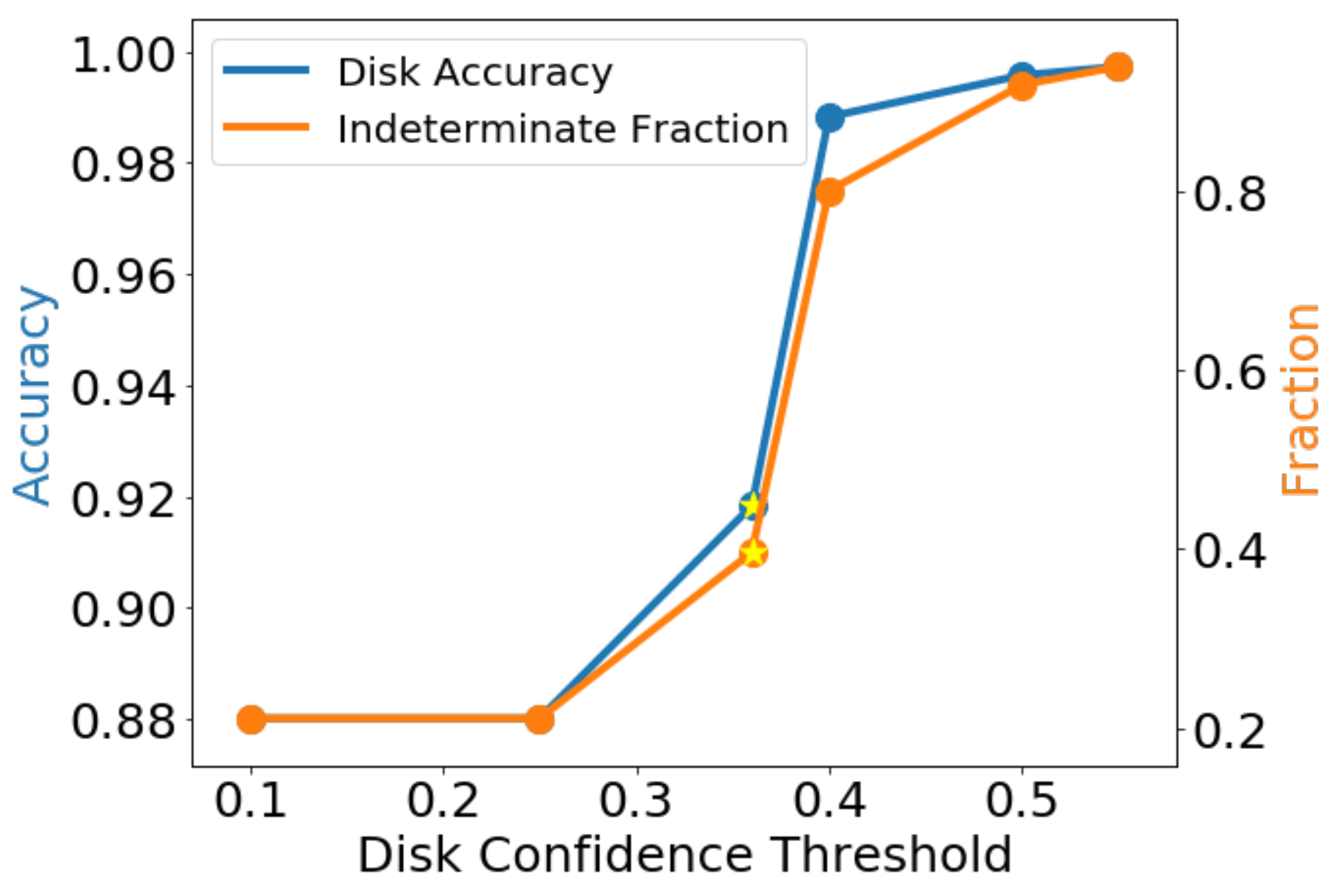}}
    \subfigure[]{\label{fig:candels_acc_threshold_ellips}\includegraphics[width=0.45\textwidth,height=6cm]{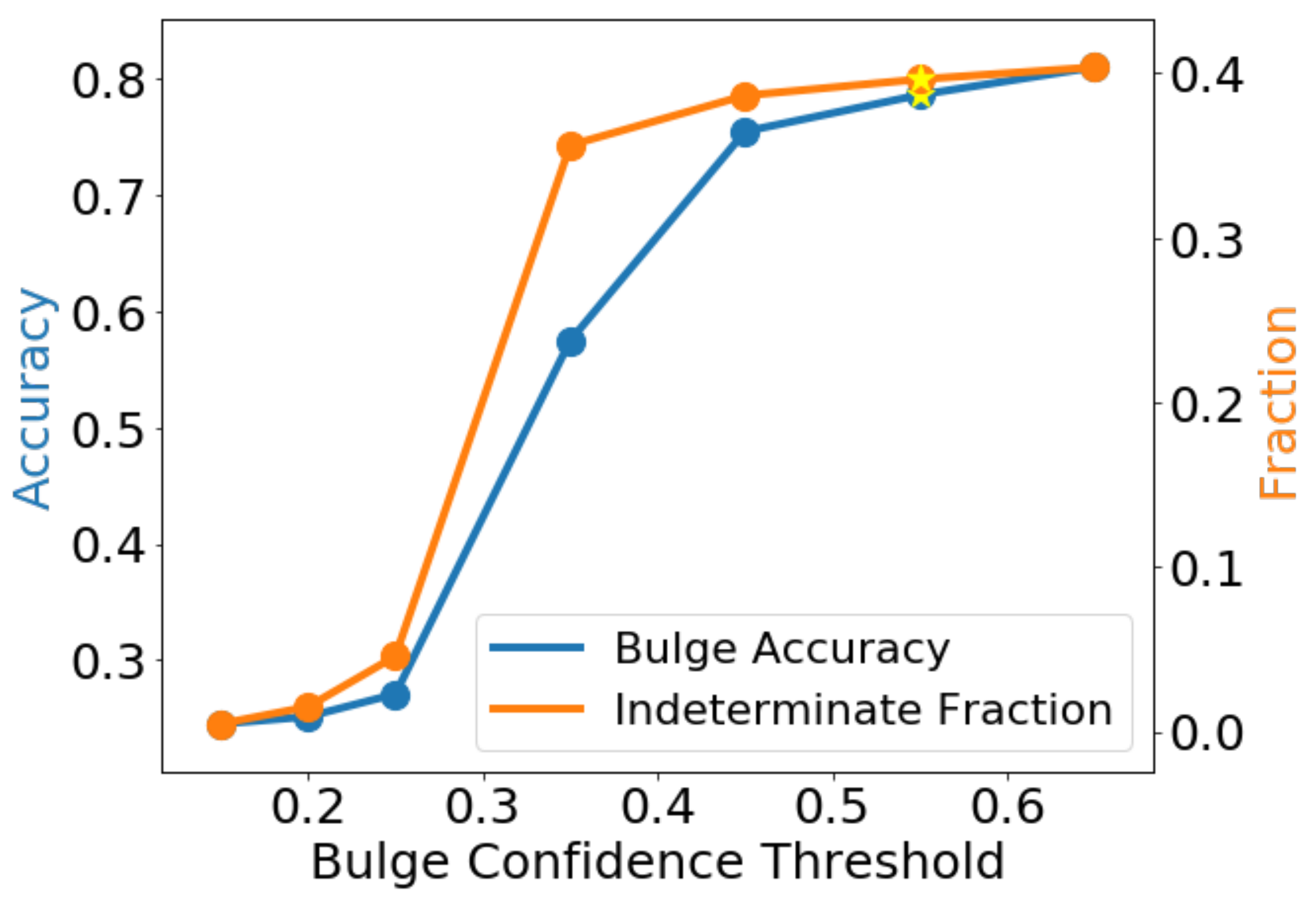}}
  \end{center}
  \caption{Relation of confidence threshold to the accuracy (blue lines, left axes) and completeness (orange lines, right axes) of \gamornet{}-C classification of the CANDELS data set. Stars denote the adopted confidence thresholds.
  Left (a): For the chosen disk confidence threshold of 0.36, provided the probability of being disk-dominated exceeds the probabilities of being bulge-dominated or indeterminate, the classification accuracy is better than $92\%$ and the indeterminate fraction $<$40\%.
  Right (b): For the chosen bulge confidence threshold of 0.55, we obtain an accuracy of $>$80\% and indeterminate fraction $<$40\%. }
  \label{fig:candels_threshold}
\end{figure*}

The choice of the confidence threshold values to classify a galaxy as bulge- or disk-dominated primarily affects two parameters: the misclassification rate and the number of indeterminate galaxies. Having a high confidence threshold results in a low misclassification rate but a high number of indeterminate galaxies, and vice-versa. We show in Figure~\ref{fig:sdss_threshold} how changing the value of the confidence threshold affects the number of indeterminate galaxies and the accuracy of both the bulge- and disk-dominated galaxies for the SDSS sample. We chose a threshold value of 0.8 or $80\%$ but as the figure shows, even with a threshold of 60\%, it is possible to get $> 85\%$ accuracy for both bulge- and disk-dominated galaxies with an indeterminate fraction as low as $\sim 20\%$.

For the CANDELS data set, setting a common/joint threshold as high as we did for the SDSS data led to most of the data being classified as indeterminate. Thus, we use separate confidence thresholds for the disk and bulge classifications, and the variation of the indeterminate fraction and accuracy with both thresholds is shown in Figure~\ref{fig:candels_threshold}. We chose the final threshold values of 0.36 and 0.55 for the disk- and bulge-dominated galaxies, respectively, as a compromise between the two competing requirements of having a low indeterminate fraction and high accuracy. 

For our choice of confidence thresholds, the indeterminate fraction is $> 25\%$ of the test set for both SDSS and CANDELS. This indeterminate fraction consists of two kinds of galaxies: those with intermediate bulge-to-total ratios (i.e., $0.45 \leq L_B/L_T \leq 0.55$) and those for which the network is not confident enough to make a prediction, because of low S/Ns and/or small sizes. For comparison, \citet{powell_17} used GALFIT to do single \sersic{} fits to 4479 GOODS-S and GOODS-N galaxies; they found that $\sim 38\%$ of the population could not be classified due to poor fits ($\chi^2 > 1.5$) or galaxies having $2.0 < n < 2.5$. Similarly, large fractions of Galaxy Zoo classifications have $\lesssim 80\%$ agreement among classifiers \citep{land_2008}. Thus, even with stringent confidence threshold values, \gamornet{} is able to match the indeterminate fraction of traditional studies. 

The choice of the confidence threshold is arbitrary and should be chosen appropriately for the particular task at hand. Toward this end, Figures \ref{fig:sdss_threshold} and \ref{fig:candels_threshold} can be used to asses the trade-off between accuracy and completeness for both the samples. We have emphasized accuracy over completeness, since we have very large samples already and can show that the misclassified objects simply have lower S/Ns and/or are too compact to classify accurately.

\vspace{1.0cm}

\subsection{Color - Mass Results}\label{sec:cm_results}
In this section, we study the quenching of star formation in  $z\sim0$ (SDSS) and $z\sim1$ (CANDELS) galaxies by examining their color-mass diagrams constructed using the morphological classifications obtained in \S\,\ref{sec:morph_results}. Refer to \S\,\ref{sec:data} for details about the calculation of colors, masses, and sSFR for both samples. 

\begin{figure*}[htbp]
	\begin{center}
	\begin{tabular}{ll}
    \subfigure[Disk-dominated Galaxies]{\label{fig:sdss_disk_scatter}\includegraphics[width=8cm,height=6cm]{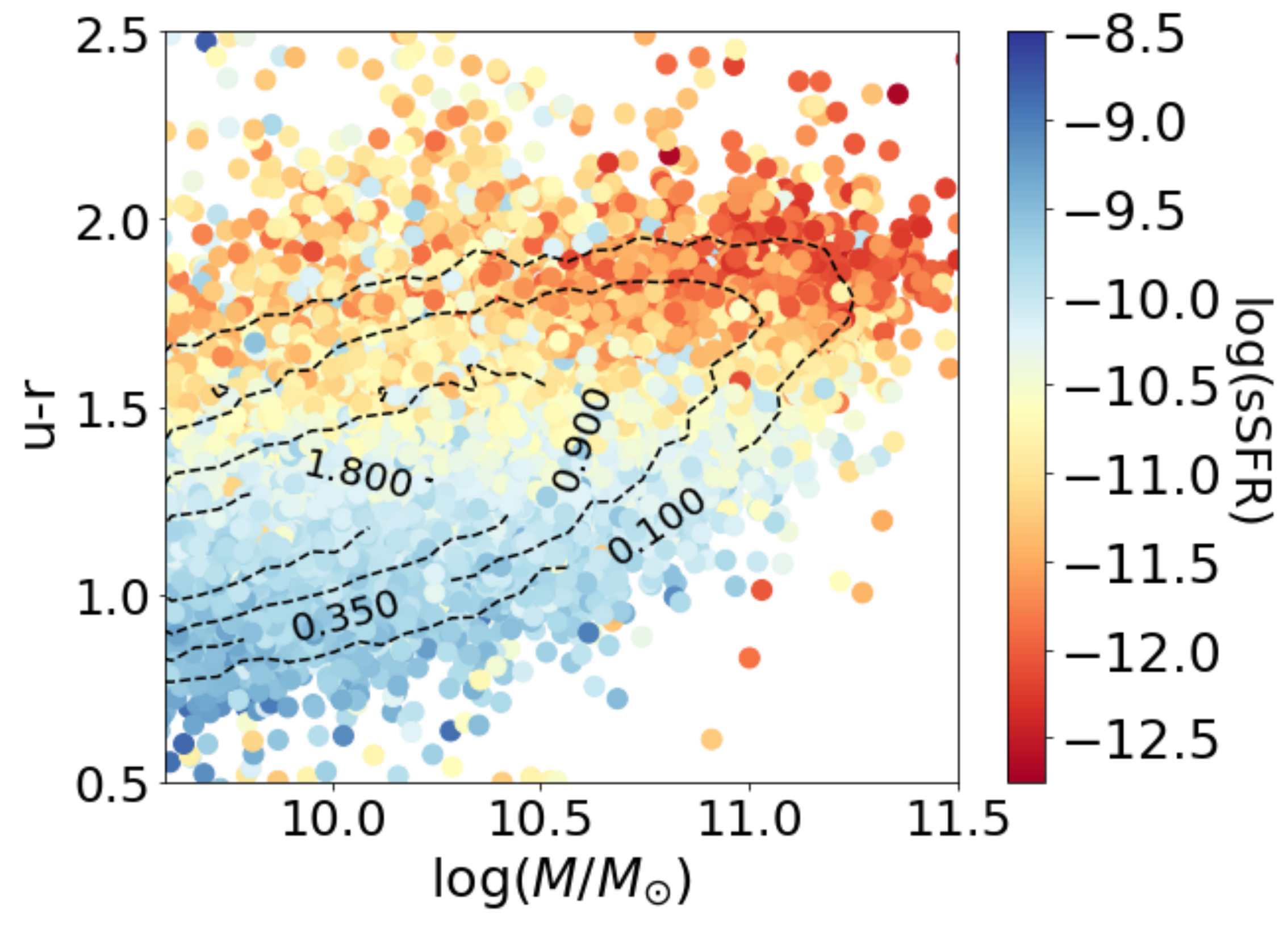}} &
    \subfigure[Bulge-dominated Galaxies]{\label{fig:sdss_ellips_scatter}\includegraphics[width=8cm,height=6cm]{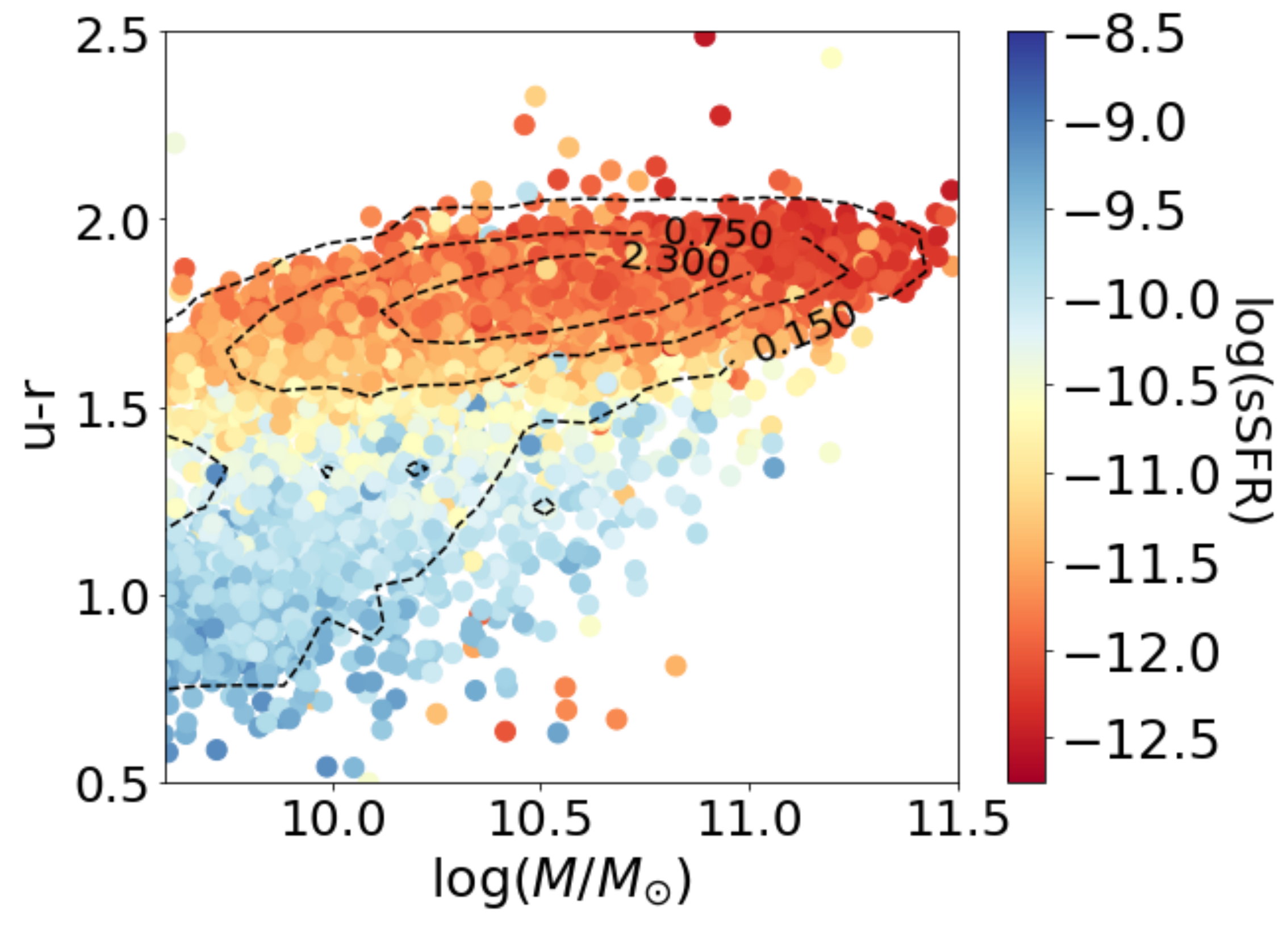}} \\
    \subfigure[Disk-dominated Galaxies]{\label{fig:sdss_disk_density}\includegraphics[width=7.6cm,height=6cm]{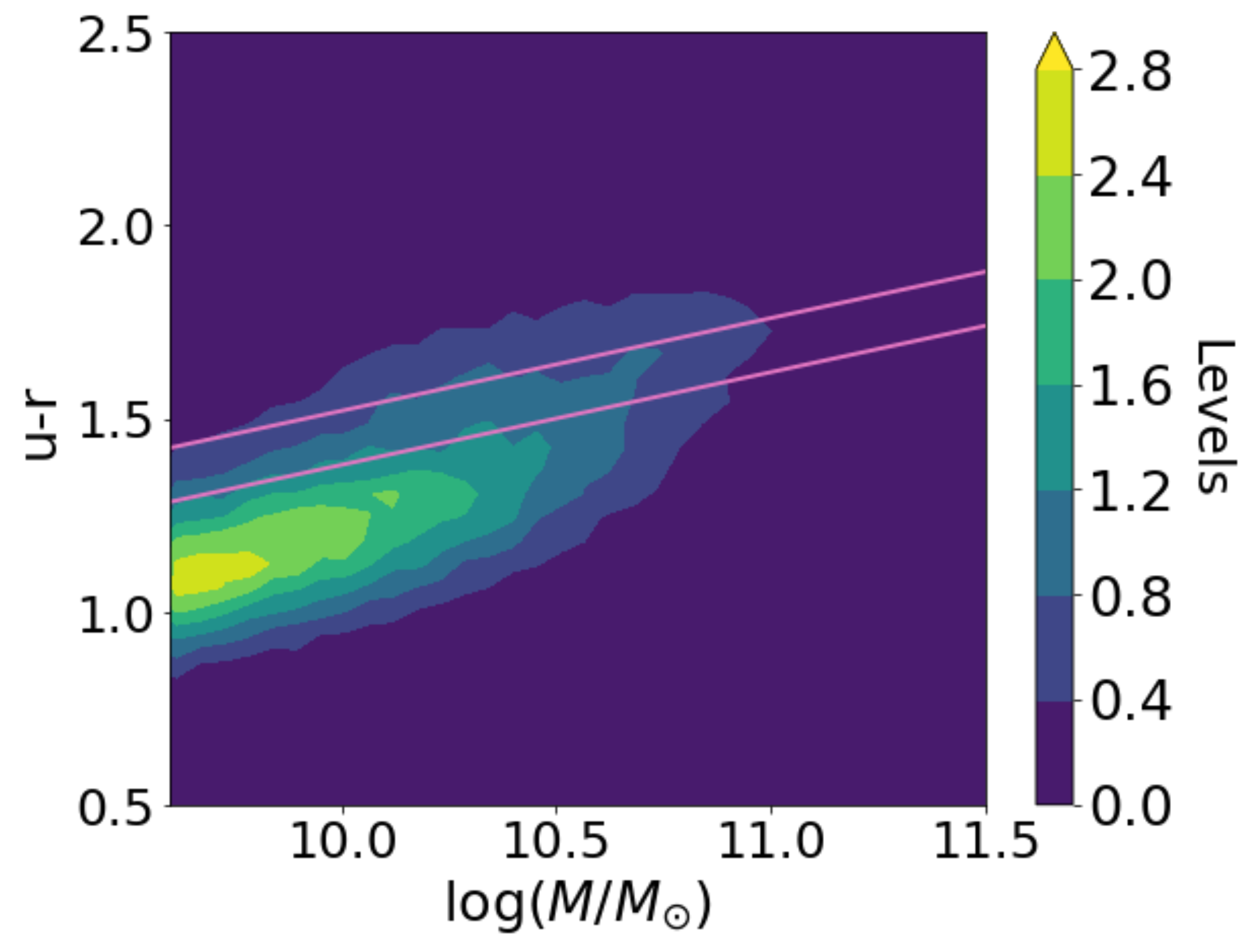}} &
    \subfigure[Bulge-dominated Galaxies]{\label{fig:sdss_ellips_density}\includegraphics[width=7.6cm,height=6cm]{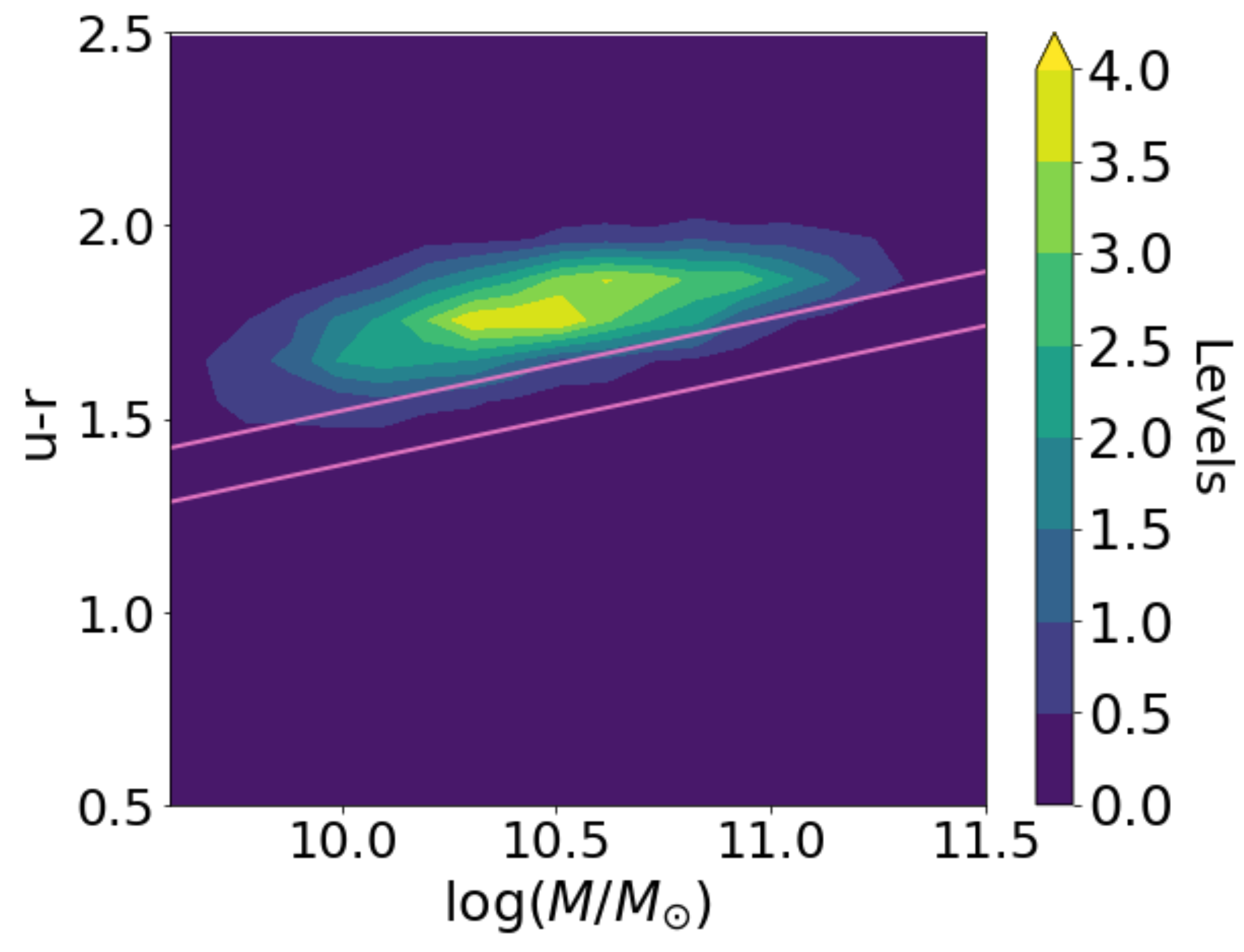}} \\
    \end{tabular}
  \end{center}
  \caption{Color-mass diagrams for the galaxies in the SDSS test set, separated by morphology. Disk-dominated galaxies (panels (a) and (c)) are mostly blue until they reach high masses (and presumably high halo masses), at which point they evolve to the red. In contrast, bulge-dominated galaxies (panels (b) and (d)) are predominately red, and appear to evolve rapidly from a short-lived population of rare, blue ellipticals that likely formed from major mergers of disky star-forming galaxies. Panels (a) and (b) show individual data points, with color indicating the specific star formation rates (sSFR) for each galaxy in units of yr$^{-1}$. Contours show the linear density of galaxies in this plot, and the numbers refer to the levels of the contours. Panels (c) and (d) are the same data plotted in terms of galaxy density. The lines mark the position of the green valley.}
  \label{fig:sdss_cmd}
\end{figure*}

\begin{figure*}[htbp]
	\begin{center}
	\begin{tabular}{ll}
    \subfigure[Disk-dominated Galaxies ]{\label{fig:candels_disk_scatter}\includegraphics[width=8.1cm,height=6cm]{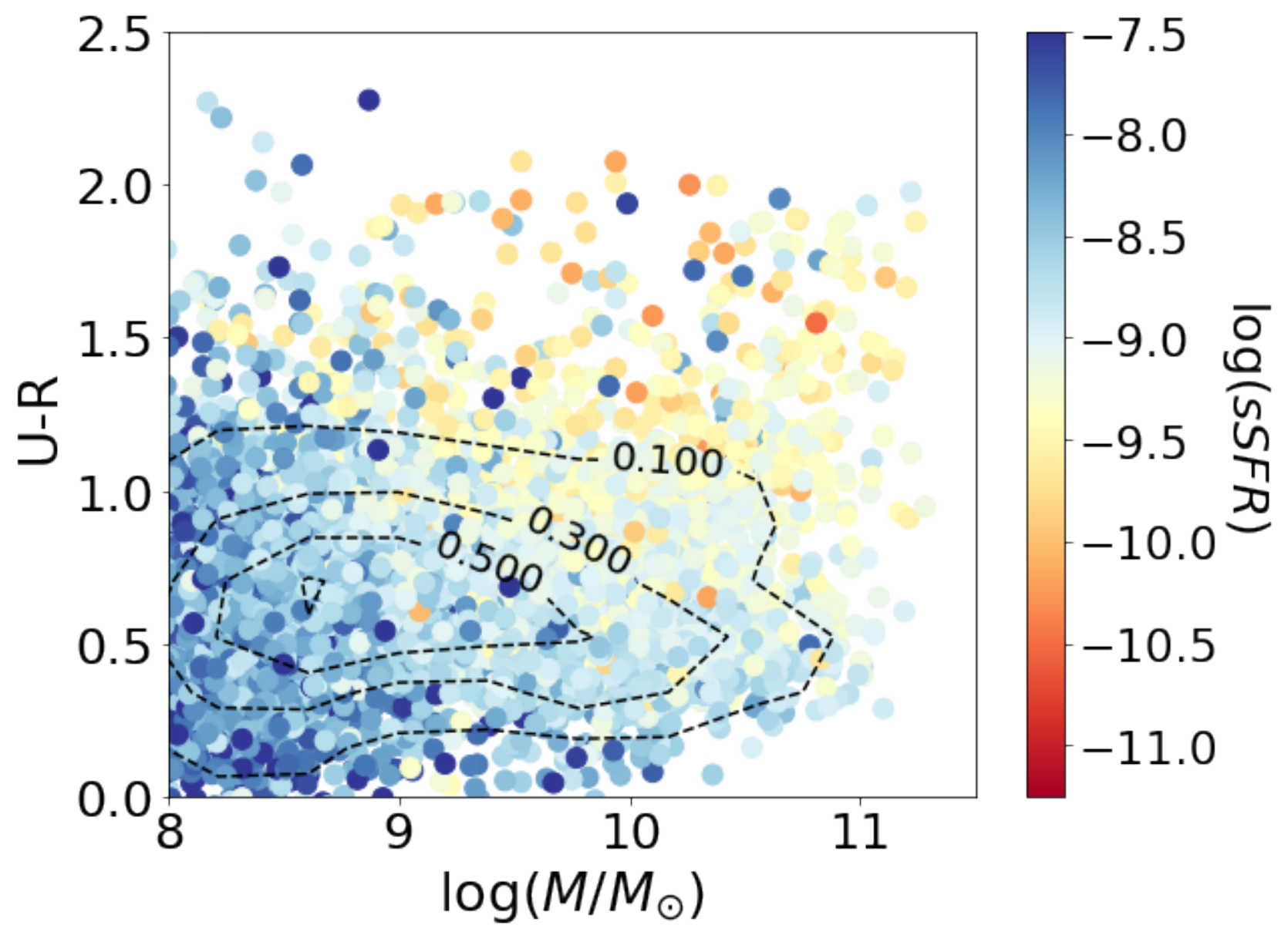}} &
    \subfigure[Bulge-dominated Galaxies ]{\label{fig:candels_ellips_scatter}\includegraphics[width=8.1cm,height=6cm]{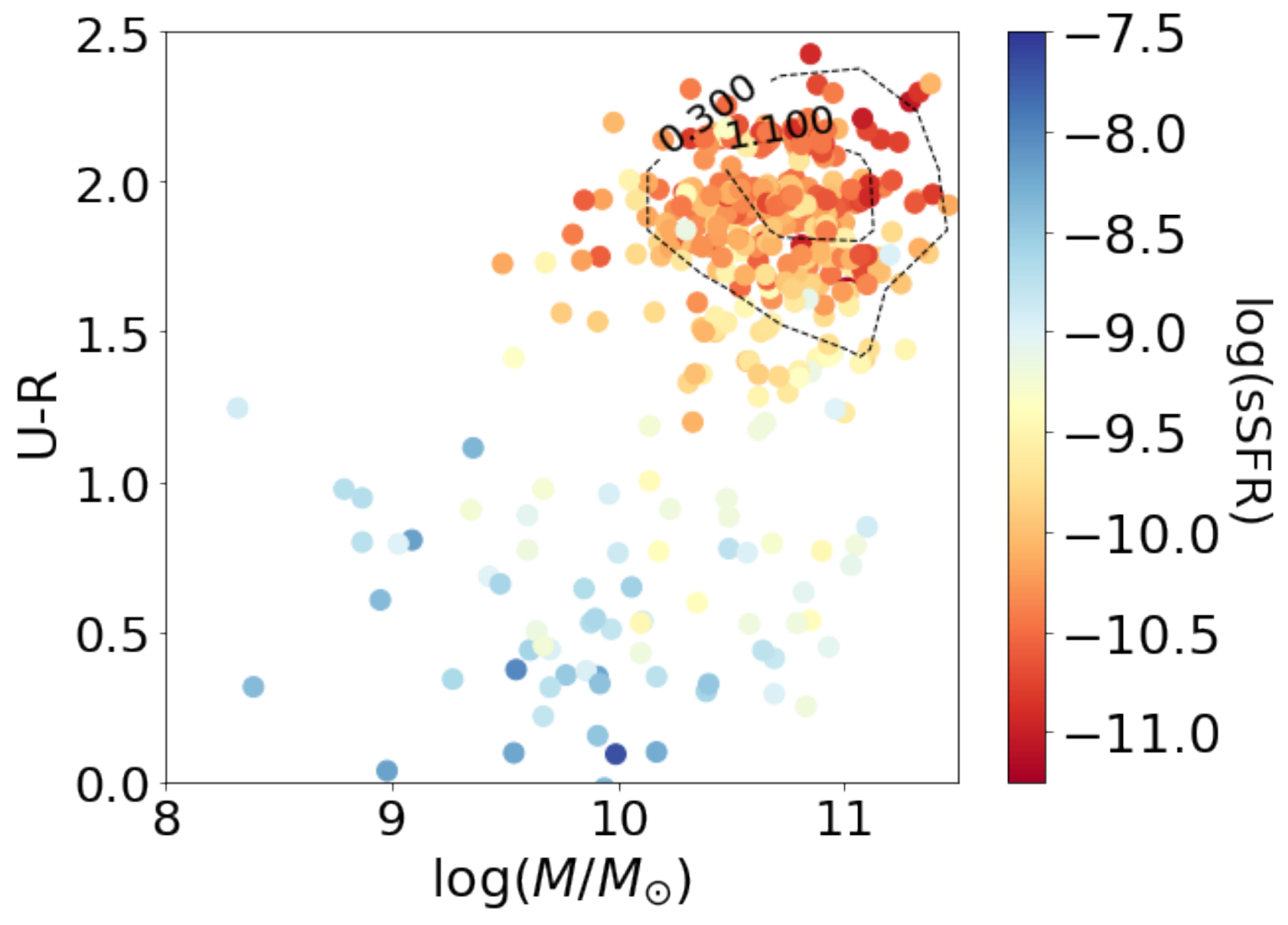}} \\
    \subfigure[Disk-dominated Galaxies]{\label{fig:candels_disk_density}\includegraphics[width=8cm,height=6cm]{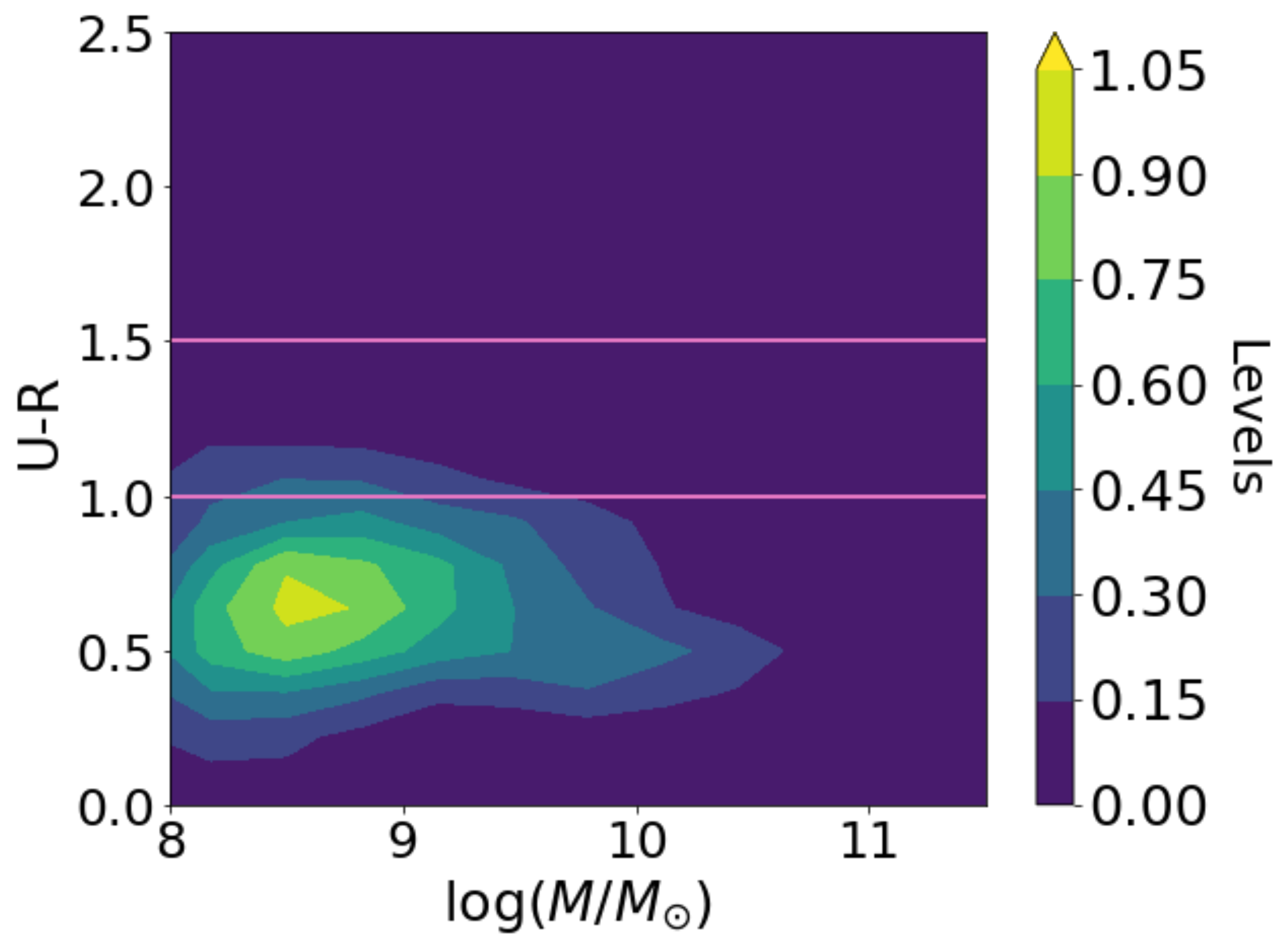}} &
    \subfigure[Bulge-dominated Galaxies  ]{\label{fig:candels_ellips_density}\includegraphics[width=8cm,height=6cm]{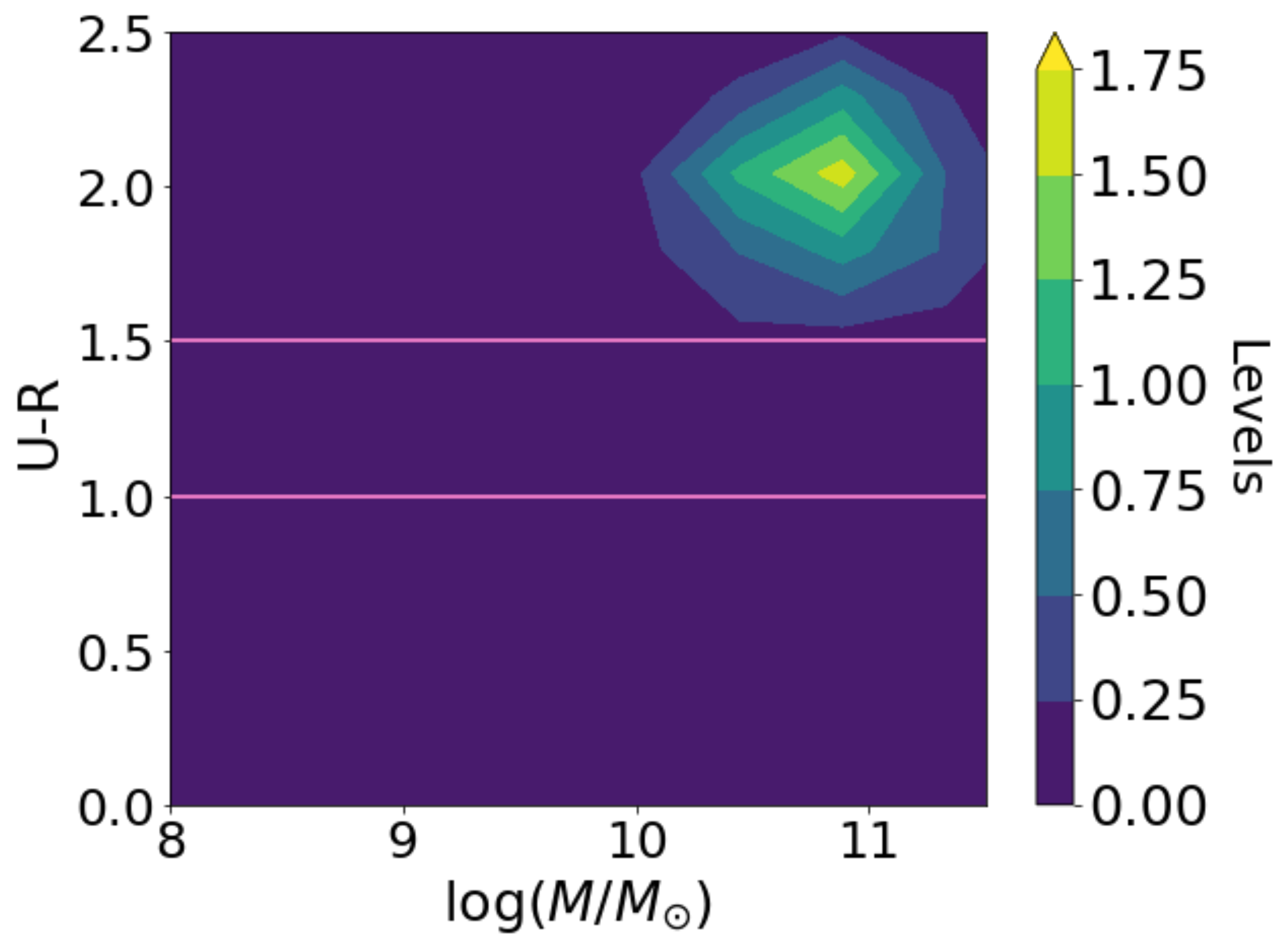}} \\
    \end{tabular}
  \end{center}
  \caption{Color-mass diagrams for the galaxies in the CANDELS test set, separated by morphology. Similar to Fig. \ref{fig:sdss_cmd}, disk-dominated galaxies (panels (a) and (c)) show signs of secular evolution, while bulge-dominated galaxies (panels (b) and (d)) appear to evolve rapidly from a short-lived population of rare, blue ellipticals. Panels (a) and (b) show individual data points, with color indicating the specific star formation rates (sSFR) for each galaxy in units of yr$^{-1}$. Contours show the linear density of galaxies in this plot and the numbers refer to the levels of the contours. Panels (c) and (d) are the same data plotted in terms of galaxy density. The lines mark the position of the green valley.}
  \label{fig:candels_cmd}
\end{figure*}

\begin{deluxetable*}{c|c|cc|cc}[htbp]
%\tablenum{9}
\tablecaption{ Statistics of the Color-Mass Diagrams \label{tab:cmd_stats}}
\tablecolumns{6}
\tablehead{\multicolumn{2}{c}{} & \multicolumn{2}{|c|}{SDSS} & \multicolumn{2}{c}{CANDELS}}
\startdata
\hline
    \multicolumn{2}{c|}{Galaxy Sample} & Number & $\%$ Population & Number & $\%$ Population\\
    \hline
    \multirow{4}{*}{Disk-dominated} & Blue Cloud & 32870 & 69.16 & 10614 & 87.10 \\ & Green Valley & 7814 & 16.44 & 1330 & 10.91 \\ & Red Sequence & 6845 & 14.40 & 242 & 1.99 \\ \cline{2-6}
    & Total & 47529 & 100 & 12186 & 100 \\ 
    \hline
    \hline
    \multirow{4}{*}{Bulge-dominated} & Blue Cloud & 995 & 12.53 & 80 & 16.19 \\ & Green Valley & 633 & 7.97 & 39 & 7.89 \\ & Red Sequence & 6313 & 79.50 & 375 & 75.91 \\ \cline{2-6}
    & Total & 7941 & 100 & 494 & 100 \\
    \hline
\enddata
\tablecomments{The demographics of SDSS and CANDELS galaxies disaggregated by morphology and color. The green valley for both samples is defined in \S\,\ref{sec:cm_results} and the three zones are shown in Figs.\,\ref{fig:sdss_cmd} and \ref{fig:candels_cmd}. We omit galaxies used in training \gamornet{} ($\sim$25\% of each sample) as well as galaxies lacking estimates for the mass or sSFR ($\sim 0.7\%$ for SDSS and $\sim 3.4\%$ for CANDELS).}
\end{deluxetable*}

Figure~\ref{fig:sdss_cmd} shows the \textit{u-r} color-mass diagram for the $z \sim 0$ SDSS test set separated by disk- and bulge-dominated morphologies. The color of each point in panels (a) and (b) refer to the specific star formation rate of each galaxy. The contours in all plots refer to the linear number density of galaxies, and the straight lines in panels (c) and (d) mark the location of the green valley, which we define to be the region between the colors mentioned below:

\begin{equation}
    u-r(M) = -1.02 + 0.24\times\log(M/M_{\odot})
\end{equation}

\begin{equation}
    u-r(M) = -0.88 + 0.24\times\log(M/M_{\odot}) .
\end{equation}

\noindent
The \textit{U-R} color-mass diagram for the $z\sim1$ CANDELS data is shown in Figure~\ref{fig:candels_cmd} and this figure is arranged in the same way as Figure~\ref{fig:sdss_cmd}. We define the green valley, in this case, as the region between \textit{U-R} colors $1.0$ and $1.5$. 

The demographics of galaxies by color and morphology for both samples is summarized in Table~\ref{tab:cmd_stats}. Note that the total number of galaxies in the table does not match that in \S\,\ref{sec:morph_results} as we have omitted galaxies that lack estimates of either mass or sSFR. The omitted fraction is $\sim 0.7\%$ and $\sim 3.4\%$ for the SDSS and CANDELS samples, respectively. 

For both the samples, we see that both bulge- and disk-dominated galaxies span the entire range of colors (i.e., we see examples of red disk-dominated galaxies as well blue bulge-dominated galaxies). As expected, the disk-dominated galaxies peak in the blue cloud while the bulge-dominated galaxies dominate the red sequence. The green valley is not a feature for either morphology; that is, there is no bimodality. Rather, the number density of galaxies declines monotonically from a red or blue peak. Thus, the green valley only arises when plotting the color-mass diagram of all galaxies together, as was first pointed out for $z\sim0$ galaxies by \citet{schawinski_14_green_herring}.

Figs.\,\ref{fig:sdss_disk_density} and \ref{fig:candels_disk_density} show that the disk-dominated galaxies peak in the blue cloud and decline gradually to the red sequence, in a unimodal way. This suggests that the disks undergo a gradual decline in star formation as opposed to being rapidly quenched through the green valley into the red sequence. At high masses, there are relatively more red disk-dominated galaxies, suggesting that high halo masses may play a role in shutting off the gas supply and quenching star formation. These conclusions agree with other studies of star formation in local galaxies \citep{tojeiro_13,schawinski_14_green_herring,lopes_16,powell_17}.

Conversely, bulge-dominated galaxies in both samples show a unimodal peak in the red sequence, with very few precursors at green and blue colors. 
This is consistent with a scenario in which bulge-dominated galaxies form from major mergers of disk-dominated blue galaxies and then are rapidly quenched through the green valley \citep{schawinski_14_green_herring}.

The morphology-sorted color-mass diagrams we obtained using \gamornet{} classifications largely agree with the previous results of \citet{schawinski_14_green_herring} at $z\sim0$ and \citet{powell_17} at $z \sim 1$; although, in the latter case, we present an order of magnitude more galaxies. For both samples, the galaxy fractions in the three zones of the color-mass diagram differ at the few percent level with respect to \citet{schawinski_14_green_herring} and \citet{powell_17}. It is important to note here that our definition of the green valley is slightly different from that used by \citet{schawinski_14_green_herring} due to their use of reddening corrected colors. Besides, \cite{schawinski_14_green_herring} and \cite{powell_17} used visual classification and GALFIT, respectively, compared to our use of \gamornet{}. Finally, our sample sizes are much larger: at $z \sim 0$, we have twice as many galaxies as \citet{schawinski_14_green_herring}, and at $z\sim1$, we have six times the galaxies analyzed by \citet{powell_17}. Larger samples are particularly important for bins with low statistics. For example, \cite{powell_17} identified only 5 bulge-dominated galaxies in the green valley, whereas we find 39, so the statistical uncertainties on that fraction are lower.

\begin{figure*}[htb]
	\begin{center}
    \subfigure[$z\sim0$ SDSS sample ]{\label{fig:ssfr_sdss}\includegraphics[height=6cm]{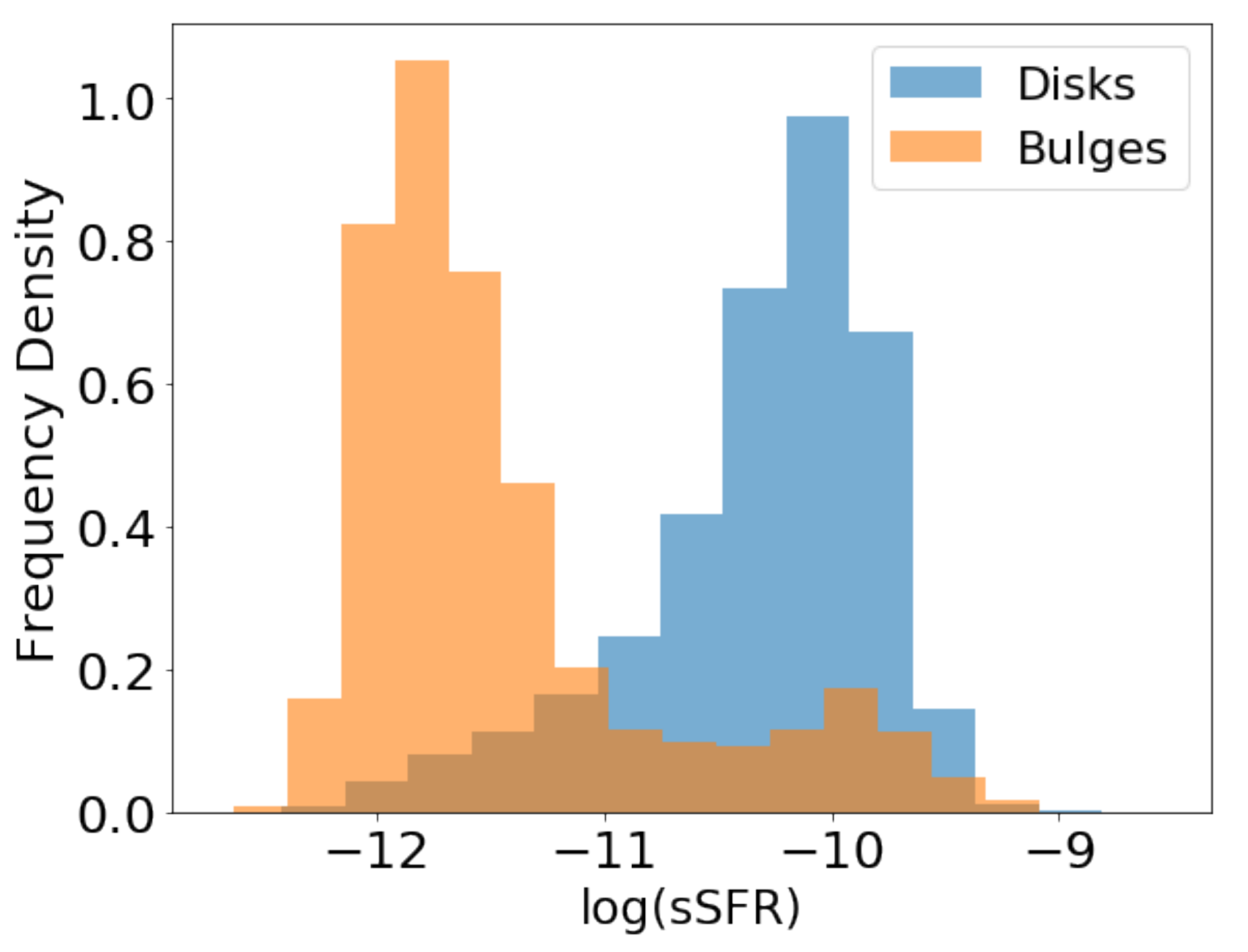}}
    \subfigure[$z\sim1$ CANDELS sample]{\label{fig:ssfr_candels}\includegraphics[height=6cm]{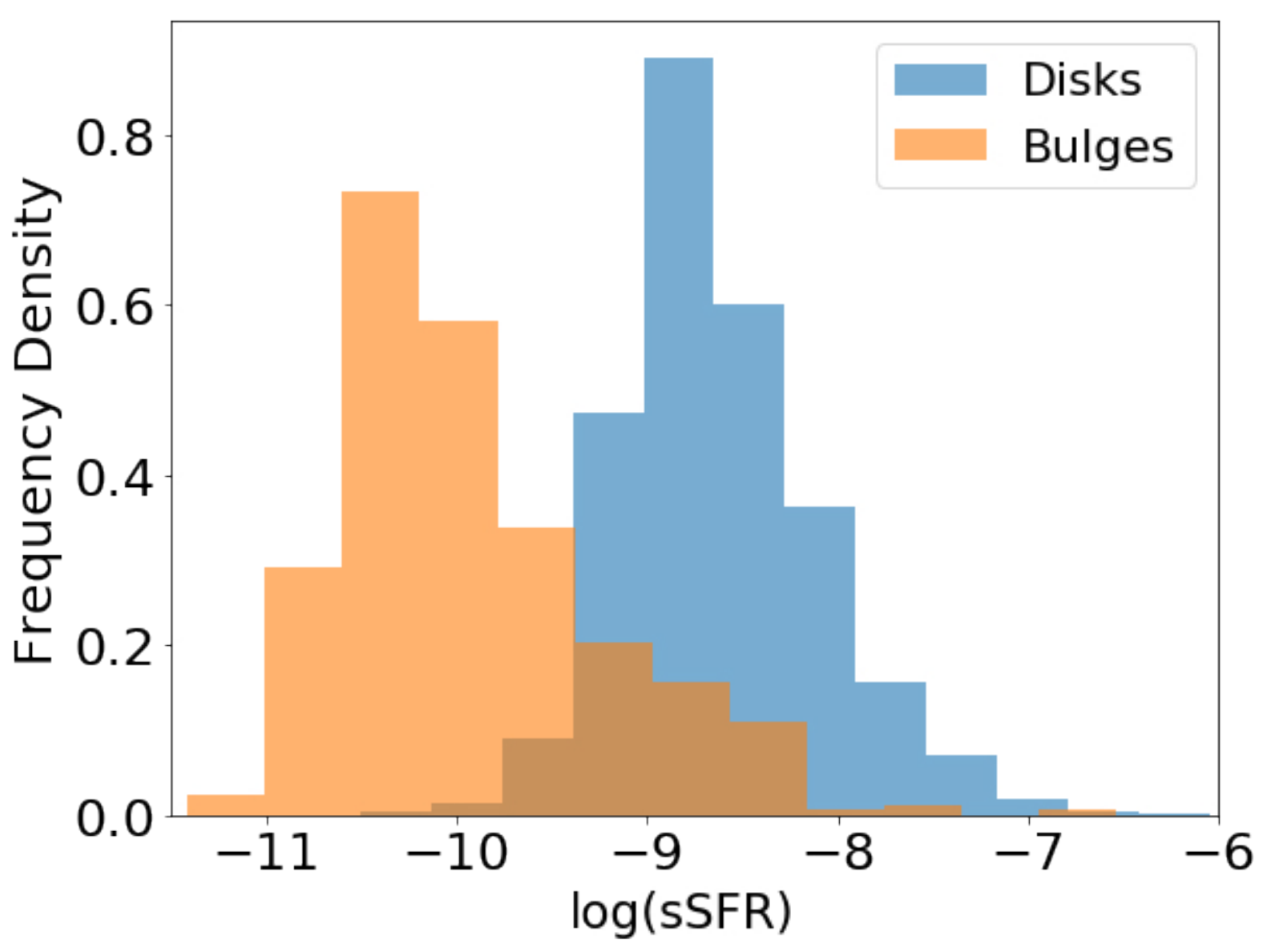}}
  \end{center}
  \caption{The normalized distribution of the specific star formation rate\,(sSFR), separated by morphology, for the SDSS and CANDELS data sets as obtained from the MPA-JHU and 3D-HST catalogs, respectively. `Frequency density' refers to the number counts normalized to form a probability density.}
  \label{fig:ssfr_distr}
\end{figure*}

Figure~\ref{fig:ssfr_distr} shows the distribution of sSFR separated by morphology. For both samples, the distribution of bulge-dominated galaxies peaks at a lower sSFR, showing the association of disk-dominated galaxies with consistent secular star formation and bulge-dominated galaxies with recent quenching. 

\section{Summary and Discussion} \label{sec:disc}
In this article, we introduced \gamornet{}, a convolutional neural network\,(CNN) that can classify galaxies morphologically. We first trained \gamornet{} on simulations of galaxies with a bulge and a disk component (\S\,\ref{sec:simulation_code}) to separate galaxies according to their bulge-to-total ratio ($L_B/L_T$). To make the network better at handling real galaxies, we then transfer learned (\S\,\ref{sec:tf_intro}) on $\sim25\%$ of both the SDSS $z\sim0$ and CANDELS $z\sim1$ samples and thereafter tested the network on the remaining $\sim75\%$ of both the samples. The net misclassification rate (calculated by weighting the disk- and bulge-dominated accuracies appropriately) achieved for both samples is $\lesssim 5\%$. For the SDSS test set of 82,547 galaxies, we achieved accuracies of $99.7\%$ for disk-dominated galaxies and $94.8\%$ for bulge-dominated galaxies. The corresponding numbers for the CANDELS test set of 21,746 galaxies are $91.8\%$ and $78.6\%$. We showed in \S\,\ref{sec:morph_results} that the misclassified CANDELS galaxies are dominated by galaxies with a half-light radius comparable to the PSF and galaxy images with low S/Ns. 

Although it has previously been shown that CNNs can be used to recover single-component \sersic{} fits of galaxies and visual morphologies \citep[eg.][]{company_15,tuccillo_18}, according to our knowledge, this is the first time it has been demonstrated that CNNs can be used to classify galaxies according to their bulge-to-total ratios.

More importantly, this work demonstrates that \gamornet{} can be applied across different data sets to perform morphological classification without the need for a large training set of real galaxies. By using a roughly 25-75 train-test split during transfer learning, we have clearly demonstrated that even when training on $25\%$ of the total sample, \gamornet{} can generalize beyond the training data and classify galaxies with high accuracy. This has very important consequences, as the applicability of CNNs to future data-intensive surveys like LSST, WFIRST, and Euclid will depend on their ability to perform without the need for a large training set of real data.

We make the source code of \gamornet{}, the trained network models, as well the morphological classifications of all the galaxies in our sample available to the public (Appendix \ref{sec:ap:public_data_release}). Although \gamornet{}-S and -C were tuned for \textit{g}-band and \textit{H}-band images, respectively, the networks should perform with comparable accuracies in other nearby bands for all SDSS $z\sim0$ and CANDELS $z\sim1$ galaxies. We also make available the weights and biases of \gamornet{} before transfer learning, i.e., after training with simulations only, so that additional data sets can be used for transfer learning. Our general prescription of training on simulations and then transfer learning should work for morphological classifications of any data set.

In \S\,\ref{sec:cm_results}, we used the morphological classifications obtained using \gamornet{} (\S\,\ref{sec:morph_results}) to study the quenching of star formation using the color-mass diagrams of our samples at $z\sim0$ and $z\sim1$\,.

For both samples, the morphology-separated color-mass diagrams do not show any bimodality. The disk-dominated galaxies peak in the blue cloud and then gradually extend to the red sequence, suggesting that quenching in disks is a secular process. Conversely, bulge-dominated galaxies in both samples peak  in  the  red  sequence,  with  very few precursors in the green valley and blue cloud. This is consistent with a scenario in which bulge-dominated galaxies form from major mergers of disk-dominated blue galaxies and then are rapidly quenched through the green valley.

Our results largely agree with previous similar studies performed at these redshifts. Our sample sizes are twice and six times as large, respectively, as those in the two previous studies done using visual classifications \citep{schawinski_14_green_herring} and using GALFIT \citep{powell_17}. The reason that we were able to use such large sample sizes is that \gamornet{}, once trained, can process large data sets very quickly and easily compared to more traditional methods.

In the future, we aim to use \gamornet{} to study the correlation of AGN with host galaxy morphology. We also plan to take \gamornet{} beyond bulge/disk classification and use it to derive different properties of AGN host galaxies. 

\acknowledgments
We would like to thank the anonymous referee for a thorough review of the manuscript and suggesting changes that greatly improved the quality and clarity of our manuscript.

This work used data from SDSS. Funding for the SDSS and SDSS-II has been provided by the Alfred P. Sloan Foundation, the Participating Institutions, the National Science Foundation, the U.S. Department of Energy, the National Aeronautics and Space Administration, the Japanese Monbukagakusho, the Max Planck Society, and the Higher Education Funding Council for England. The SDSS website is \href{http://www.sdss.org/}{http://www.sdss.org/}.

The SDSS is managed by the Astrophysical Research Consortium for the Participating Institutions. The Participating Institutions are the American Museum of Natural History, Astrophysical Institute Potsdam, University of Basel, University of Cambridge, Case Western Reserve University, University of Chicago, Drexel University, Fermilab, the Institute for Advanced Study, the Japan Participation Group, Johns Hopkins University, the Joint Institute for Nuclear Astrophysics, the Kavli Institute for Particle Astrophysics and Cosmology, the Korean Scientist Group, the Chinese Academy of Sciences (LAMOST), Los Alamos National Laboratory, the Max-Planck-Institute for Astronomy (MPIA), the Max-Planck-Institute for Astrophysics (MPA), New Mexico State University, Ohio State University, University of Pittsburgh, University of Portsmouth, Princeton University, the United States Naval Observatory, and the University of Washington.

This work is based on observations taken by the CANDELS Multi-Cycle Treasury Program with the NASA/ESA HST, which is operated by the Association of Universities for Research in Astronomy, Inc., under NASA contract NAS5-26555.

This work is based on observations taken by the 3D-HST Treasury Program with the NASA/ESA HST, which is operated by the Association of Universities for Research in Astronomy, Inc., under NASA contract NAS5-26555.

This material is based upon work supported by the National Science Foundation under grant No. 1715512

C.M.U would like to acknowledge support from National Aeronautics and Space Administration via ADAP Grant 80NSSC18K0418. 

\clearpage

\appendix

\section{Public Release of Code, Models, and Galaxy Morphological Classifications}\label{sec:ap:public_data_release}

Here, we provide an outline of all the material that we make public as a part of this work. An up-to-date record of this public data release will also be maintained at \href{http://gamornet.ghosharitra.com}{http://gamornet.ghosharitra.com} and \href{http://www.astro.yale.edu/aghosh/gamornet.html}{http://www.astro.yale.edu/aghosh/gamornet.html} in case any of the URLs below stop working over time.   

\subsection{\gamornet{} Source Code}\label{sec:ap:gamornet_source_code}

\gamornet{} was implemented using TFLearn\,\,(\href{http://tflearn.org}{http://tflearn.org}), which is a high-level Application Program Interface for TensorFlow\,\,(\href{https://tensorflow.org}{https://tensorflow.org}), an open source library widely used for large-scale machine learning applications.

The source code of \gamornet{} is maintained as a GitHub Repository and is available at \href{https://github.com/aritraghsh09/GaMorNet}{https://github.com/aritra ghsh09/GaMorNet}. Instructions for installing TFLearn and using \gamornet{} are available in the above GitHub repository. An implementation of \gamornet{} in Keras (\href{https://keras.io/}{https://keras.io/}) is also available at the above repository. 

\subsection{\gamornet{} Trained Models }\label{sec:ap:gamornet_trained_models}

Trained Models for both \gamornet{}-S and -C are being made available as a part of this data release. 

For more details about the various stages of training, refer to \S\,\ref{sec:initial_training} \& \ref{sec:tf_intro}. All of the models below are being made available via Yale Astronomy's Public FTP service \href{ftp://ftp.astro.yale.edu/pub/aghosh/gamornet/trained_models}{ftp://ftp.astro.yale.edu/pub/aghosh/gamornet/trained\textunderscore models}. 

You can copy and paste the above link into a browser window to download the files, or you can also issue the following commands from a terminal to login to the ftp server

\begin{verbatim}
ftp ftp.astro.yale.edu
\end{verbatim}

Use the username `anonymous' and keep the password field blank. After logging-in, do the following:

\begin{verbatim}
cd pub/aghosh/gamornet/<appropriate_subdirectory> 
get <file_name>
quit
\end{verbatim}

To list the files at your current location, you can use the `ls' command. 

The various subdirectories are named as follows in the list below:

\begin{enumerate}
    \item \gamornet{}-S model trained only on simulations $\xrightarrow{}$ /trained\_models/SDSS/sim\textunderscore trained/
    \item \gamornet{}-S model trained on simulations and transfer learned on real data $\xrightarrow{}$ /trained\_models/SDSS/tl/
    \item \gamornet{}-C model trained only on simulations $\xrightarrow{}$ /trained\_models/CANDELS/sim\textunderscore trained/
    \item \gamornet{}-C model trained on simulations and transfer learned on real data $\xrightarrow{}$ /trained\_models/CANDELS/tl/
\end{enumerate}

Models 2 and 4 can be applied directly to SDSS \textit{g}-band data at $z\sim0$ and CANDELS \textit{H}-band data at $z\sim1$ (or data in other nearby bands), respectively, without any further training. However, if you plan to apply \gamornet{} to data that is different from the above mentioned data sets, we recommend using any of the models above and then transfer learning on your new data. The exact nature of the data will decide which of the models above is the best starting point for the transfer learning process. 

For more information on how to load these models in TFLearn and use them, refer to the documentation of the \gamornet{} GitHub repository mentioned in Sec.~\ref{sec:ap:gamornet_source_code}.

\subsection{ Tables with predicted probabilities and classifications}\label{sec:ap:prob_tables}
                                                                          
The predicted probabilities (of being disk-dominated, bulge-dominated, or indeterminate) and the final classifications for all the galaxies in our SDSS and CANDELS test sets, as determined by \gamornet{}-S and -C, are made available below as .txt files. These tables are the full versions of Tables~\ref{tab:lt_sdss} \&~\ref{tab:lt_candels}.

Both the tables are being made available via Yale Astronomy's Public FTP service \href{ftp://ftp.astro.yale.edu/pub/aghosh/gamornet/pred_tables}{ftp://ftp.astro.yale.edu/ pub/aghosh/gamornet/pred\_tables}. Instructions for accessing the service from the command line can be found in \S\,\ref{sec:ap:gamornet_trained_models}. The two files are located according to the list below:

\begin{itemize}
    \item Full version of Table~\ref{tab:lt_sdss} corresponding to the SDSS data set $\xrightarrow{}$ /pred\textunderscore tables/pred\textunderscore table\textunderscore sdss.txt
    \item Full version of Table~\ref{tab:lt_candels} corresponding to the CANDELS data set $\xrightarrow{}$ /pred\textunderscore tables/pred\textunderscore table\textunderscore candels.txt
\end{itemize}

\subsection{ GalaxySim Source Code }\label{sec:ap:galsim}

The code that was used to simulate the galaxies described in \S\,.\ref{sec:simulation_code} is available as a GitHub repository at \href{https://github.com/aritraghsh09/GalaxySim}{https://git hub.com/aritraghsh09/GalaxySim}

This code makes use of GALFIT\,\citep{galfit} to simulate idealized double component galaxies. Since the simulations of galaxy surface brightness profiles are independent of each other, the code could be trivially parallelized. Instructions for using GalaxySim are available in the above GitHub repository. 

\bibliographystyle{aasjournal}
\bibliography{references}

\end{document}